\def\arraystretch{0.85}
\def\l{$\lambda$}
\def\mbh{$M_{\rm BH}$\/}
\def\nh{$n_{\mathrm{H}}$\/}
\def\lledd{$L/L_{\rm Edd}$}

\def\nc{$N_{\rm c}$\/}
\def\rfe{$R_\mathrm{FeII}$}

\def\feiiq{\rm Fe{\sc ii}$\lambda$4570\/}

\def\ltsima{$\; \buildrel < \over \sim \;$}
\def\ltsim{\lower.5ex\hbox{\ltsima}}  
\def\gtsima{$\; \buildrel > \over \sim \;$}

\def\gtsim{\lower.1ex\hbox{\gtsima}} 
\def\h{$H_{0}$}
\def\lya{{ Ly}$\alpha$}
\def\civfull{{\sc{Civ}}$\lambda$1549\/}
\def\civ{{\sc Civ}}

\def\civbc{{\sc{Civ}}$\lambda$1549$_{\rm BC}$\/}

\def\hb{{\sc{H}}$\beta$\/}

\def\hbbc{{\sc{H}}$\beta_{\rm BC}$\/}

\def\niv{{\sc{Niv]}}$\lambda$1486\/}
\def\nivfull{{\sc{Niv]}}$\lambda$1486\/}
\def\ciii{{\sc{Ciii]}}\/}
\def\ciiifull{{\sc{Ciii]}}$\lambda$1909\/}

\def\caii{{Ca{\sc ii}}}

\def\oiiiuv{{\sc{Oiii]}}$\lambda$1663\/}
\def\niii{{\sc{Niii]}}$\lambda$1750\/}
\def\siiii{Si{\sc iii]}\/}
\def\siiiifull{Si{\sc iii]}$\lambda$1892\/}
\def\aliii{Al{\sc  iii}}
\def\heiiuv{He{\sc{ii}}}
\def\nv{{N\sc{v}}}
\def\aliiifull{Al{\sc iii}$\lambda$1860\/}
\def\heiiuvfull{He{\sc{ii}}$\lambda$1640}

\def\feii{{Fe\sc{ii}}\/}
\def\siii{{Si\sc{ii}}$\lambda$1814\/}
\def\feiii{{Fe\sc{iii}}\/}
\def\fe{{\sc{Fe}}\/}

\def\fe76087{{\sc [Fe vii]}$\lambda$6087\/}
\def\oiii{{\sc [Oiii]}$\lambda$5007}
\def\b{$\beta$}
\def\kms{km~s$^{-1}$}

\def\ergss{erg s$^{-1}$\/}
\def\hii{H{\sc ii}\/}

\def\heii{{{\sc H}e{\sc ii}}$\lambda$1640\/}

\def\siiv{Si{\sc iv}\/}
\def\oiv{O{\sc iv]}\/}
\def\siivfull{Si{\sc iv}$\lambda$1397\/}
\def\oivfull{O{\sc iv]}$\lambda$1402\/}

\documentclass[twocolumn]{aastex63}

\def\nh{$n_\mathrm{H}$\/}
\usepackage{epsfig}
\usepackage{rotating}
\received{}
\revised{}
\accepted{}
\submitjournal{}
\shorttitle{Metallicity in highly accreting quasars}
\shortauthors{\'Sniegowska et al.}
\begin{document}
\title{High metal content of highly accreting quasars}
\correspondingauthor{Marzena \'Sniegowska}
\email{msniegowska@camk.edu.pl}
\author[0000-0003-2656-6726]{Marzena \'Sniegowska}
\affiliation{Nicolaus Copernicus Astronomical Center, Polish Academy of Sciences, Bartycka 18, 00-716 Warsaw, Poland}
\affiliation{Center for Theoretical Physics, Polish Academy of Sciences, Al. Lotnik{\'o}w 32/46, 02-668 Warsaw, Poland}

\author[0000-0002-6058-4912]{Paola Marziani}
\affiliation{Istituto Nazionale di Astrofisica (INAF), Osservatorio Astronomico di Padova, 35122 Padova, Italy }

\author[0000-0001-5848-4333]{Bo\.zena Czerny} 
\affiliation{Center for Theoretical Physics, Polish Academy of Sciences, Al. Lotnik{\'o}w 32/46, 02-668 Warsaw, Poland}

\author[0000-0002-5854-7426]{Swayamtrupta Panda}
\affiliation{Center for Theoretical Physics, Polish Academy of Sciences, Al. Lotnik{\'o}w 32/46, 02-668 Warsaw, Poland}
\affiliation{Nicolaus Copernicus Astronomical Center, Polish Academy of Sciences, Bartycka 18, 00-716 Warsaw, Poland}

\author[0000-0002-7843-7689]{Mary Loli Mart\'{\i}nez-Aldama}
\affiliation{Center for Theoretical Physics, Polish Academy of Sciences, Al. Lotnik{\'o}w 32/46, 02-668 Warsaw, Poland}

\author{Ascensi{\'o}n del Olmo}
\affiliation{Instituto de Astrof{\'\i}sica de Andaluc\'\i{a}\ (IAA- CSIC), Glorieta de Astronom\'{\i}a, E-18080 Granada, Spain}

\author{Mauro D'Onofrio}
\affiliation{Dipartmento di Fisica \& Astronomia, Universit\`a di Padova, Padova, Italy}



\begin{abstract}

We present an  analysis of UV spectra of 13 quasars believed to belong to extreme Population A (xA) quasars, aimed at the estimation of the
chemical abundances of the broad line emitting gas. Metallicity estimates for the broad line emitting gas of quasars  are subject to a number of caveats,
xA sources with the strongest \feii\ emission offer several advantages with respect to the quasar general population,
as their optical and UV emission lines can be interpreted as the sum of a low-ionization component roughly at quasar rest frame (from virialized gas),
plus a blueshifted excess (a disk wind), in  different physical conditions. Capitalizing on these results, 
we analyze  the component at rest frame and the blueshifted one, exploiting the dependence  of several intensity line ratios on metallicity $Z$.  We find that the validity of intensity line ratios as metallicity indicators depends on the physical conditions.  We apply the measured diagnostic ratios to estimate the physical properties of sources such as density, ionization, and metallicity of the gas.  Our results  confirm that the two regions (the low-ionization component and the blue-shifted excess) of different dynamical conditions also show different physical conditions and suggest metallicity values that are high, and probably the highest along the quasar main sequence, 
with  $Z \sim 20 - 50 Z_{\odot}$, if the solar abundance ratios can be assumed constant.  We found some evidence of an overabundance of Aluminium with respect to Carbon, possibly due to selective enrichment of the broad line emitting gas by supernova ejecta. 
\end{abstract}

\section{Introduction} 
\label{sec:intro} 

Thanks to large public databases as the Sloan Digital Sky Survey (SDSS) catalogs, we have unrestricted access to a large wealth of astronomical data (for example, several editions of quasar catalogues, \citealt{schneideretal10,parisetal17}, and of value-added measurements by \citealt{shenetal11}). SDSS spectra of high redshift quasars ($z \gtrsim$ 2) cover the rest frame UV spectral range. It is known since the 1970s that measurements of UV emission lines  can be used to explore the physical and chemical properties of active galactic nuclei (AGN). Landmark papers provided the basic understanding of line formation processes due to photoionization \citep[e.g.,][]{willsnetzer79,davidsonnetzer79,baldwinetal03}. 

The chemical composition of the line emitting gas is an especially intriguing problem from the point of view of the evolution of cosmic structures, but also from the technical side.  \citet{nagaoetal06} investigated BLR metallicities using various emission-line flux ratios and claimed that the typical metallicity of the gas in that region is at least super-solar, with typical $Z \sim 5 Z_\odot$. Moreover, studies of metallicity-redshift dependence \citep{nagaoetal06,juarezetal09} show a lack of metallicity evolution up to $z \approx 5$. Similar results are obtained for  \citep{nagaoetal06b}. The highest-redshift quasars ($z \gtrsim 5$ \citealt[e.g.,][]{banadosetal16,nadinietal19}) are known to show UV spectra remarkably similar to the ones observed at low-redshift, especially the ones accreting at high rate and radiating at high Eddington ratio \citep{diamond-stanicetal09,plotkinetal15,sulenticetal17}.\footnote{The effect is most likely due to a bias: for a flux limited sample, the highest radiators at a given black hole mass are the ones that remain detectable at highest $z$\ \citep{sulenticetal14}.} Perhaps surprisingly, these sources are suspected to have high metal content in their line emitting gas, due to the consistent values of several diagnostic ratios measured in quasars with similar spectral properties at low and high z \citep{martinez-aldamaetal18a}, and indicating highly super-solar metal content. 

Several techniques are applied to estimate the chemical composition in Galactic nebulae (see e.g., \citealt{feibelmanaller87} for planetary nebulae). Classical techniques used for \hii\ and other nebul\ae\ (including the Narrow Line Regions, NLRs) are unfortunately not applicable to the broad line regions of quasars. Permitted and inter-combination lines are too broad to resolve fine structure components of doublets; line profiles are composites and may  originate in regions that are spatially unresolved, and unresolved or only partially resolved in radial velocity as well. 

However, quasar emission line profiles still offer important clues in the radial velocity domain. The shape of the profile is strongly dependent on the ionization potential of the ionic species from which the line is emitted: it is expedient to subdivide  the broad lines in low- and high ionization lines (LILs and HILs). The LIL group in the spectral range under analysis (1200 \AA\ -- 2000 \AA) includes the following lines: Si{\sc ii}$\lambda$1263, \siii,  Al{\sc ii}$\lambda$1671,   \aliiifull, \siiiifull, \ciiifull. High ionization lines are \nivfull, \oivfull, \civfull, \siivfull, \oiiiuv, and \heiiuvfull\  \citep[for detailed discussion see][]{collinsouffrinetal88,collinsouffrinlasota88,gaskell00}. The \aliii, \siiii, and \ciii\ lines sometimes referred to as ``intermediate ionization lines:'' even if they are mainly produced within the fully ionized region of the emitting gas clouds \citep{negreteetal12}, the ionization potential of their ionic species is closer to the ones of the LILs, and typically $\lesssim 20 $ eV.    

The two groups of lines (HILs and LILs) do not only show different kinematic properties \citep{sulenticetal95}, but their emission is also likely to occur in fundamentally different physical conditions \citep{marzianietal10}.  The HILs  are characterized also by the evidence of strong blueshifted emission, very evident in \civ\ \citep[e.g.,][]{sulenticetal07,richardsetal11,coatmanetal16}.  Therefore, a  careful line comparison/decomposition is necessary, lest inferences may be associated with a non-existent region with inexplicable properties.  

The interpretation of two line components involves a   virialized region, of relatively low ionization (hereafter referred to the virialized, low-ionization BLR associated with a symmetric broad component, BC), possibly including emission from the accretion disk, and a region of higher ionization, associated with a disk wind or a clumpy outflow, a scenario first proposed by \citet[][ and further developed by \citealt{elvis00}]{collinsouffrinetal88}, and observationally supported by reverberation mapping \citep[e.g., ][]{petersonwandel99} and  the apparent lack of correlation between HILs and LILs in luminous quasars \citep[e.g.,][]{mejia-restrepoetal16,sulenticetal17}. Even if all lines were emitted by a wind \citep{murrayetal95,murraychiang97,proga07a}, the conditions at the base of the textcolor{wind} may strongly differ from the ones downstream in the outflow. 

While each UV metal line contains information related to composition \citep{hamannferland92}, not all of the lines listed above can be used in practice.  For instance, the \nv\ and Si{\sc ii}$\lambda$1263 lines are strongly affected by blending with Ly$\alpha$; other lines such as \siii\ and \niv\ are usually weak and require high S/N to be properly measured. The choice of diagnostic ratios used for metallicity estimates will be a compromise between S/N, easiness of deblending, and straightforwardness of physical interpretation. In practice, apart from \lya, only the strongest broad features will be considered as potential metallicity estimators in this work (Sect. \ref{sec:methods}). The ratio (\siiv+\oiv)/\civ\ has been widely used in past studies \citep[][and references therein]{hamannferland99}; this ratio is relatively easy to measure and  seems to be the most stable ratio against distribution of gas densities and ionization parameter in the BLR \citep{nagaoetal06}. The ratios involving NV$\lambda1240$, like \nv/\civ\ are   sensitive to ionization parameter and    to   nitrogen abundance \cite[e.g][]{dietrichetal03,wangetal12}.  We will rediscuss the use of these ratios in the context of the xA quasar spectral properties (Sect. \ref{quasar_evolution}). 


Both physical conditions and chemical abundances vary along the quasar main sequence \citep[see e.g.,][]{sulenticetal00c,kuraszkiewiczetal09,shenho14,widyczerny19,pandaetal19a}. Solar and even slightly subsolar values are possible toward the extreme Population B, where \feii\ emission is often undetectable above noise \citep[e.g.,][]{hamannetal02,punslyetal18a}. At the other extreme, where \feii\ is most prominent, estimates suggest $Z \gtrsim 10 Z_\odot$\ \citep{pandaetal18,pandaetal19}. \citet{baldwinetal03} derived $Z \approx 15 Z_\odot$, although in the particular case of a ``nitrogen-loud'' quasars. 
Apart from the extremes, it is not obvious whether there is a continuous systematic trend along the sequence. Previous estimates consistently suggest super-solar metallicity  up to $Z \lesssim 10$ $Z_\odot$ \citep{warneretal04}. Other landmark  studies consistently found super-solar metallicity: \citet{hamannferland92}\ derived $Z$ up to $\lesssim 15 Z_\odot$; \citet{nagaoetal06} found   typical values $Z \approx 5 Z_\odot$, with $Z\sim 10 Z_\odot$\ for the most luminous quasars  from the (\siiv+\oiv)/\civ\ ratio. \citet{sulenticetal14} inferred a large dispersion with the largest value in excess of $10 Z_\odot$.   Similar results were reached by \citet{shinetal13} whose \siiv+\oiv/\civ\ ratio measurements suggested $Z \gtrsim 10 Z_\odot$. 

Most interesting along the quasar main {sequence  are} the high accretors. They are selected according to empirical criteria \citep[e.g.,][]{wangetal13,marzianisulentic14,wangetal14,duetal16a}, and defined by having \rfe$ >$ 1, that is with the \feiiq\ blend on the blue side of \hb\ (as defined by \citealt{borosongreen92}) flux exceeding the flux of \hb.  In the optical diagram of the quasar main sequence \citep{sulenticetal00c,shenho14} they are at the extreme tip in terms of \feii\ prominence, and identified as extreme Population A (hereafter xA), following \citet{sulenticetal02}.  Depending on  redshift, we look for high accretors using different criteria. In case of z $\gtrsim$ 1, it is expedient to  use a criterion based on two UV line intensity ratios: 
 
 \begin{itemize}
 \phantomsection
\label{the_label}
\item  \aliii/\siiii $>$ 0.5 
\item \ciii/\siiii $<$ 1.0,
\end{itemize}
 \vfill
  following  \citep{marzianisulentic14}.  These criteria are met by the sources identified as {xA
  Population by Sulentic and collaborators}.  xA quasars are radiating at the highest luminosity per unit mass, and,  at low $z$\ they are characterized by relatively low black hole masses for their luminosities and  high Eddington ratios \citep{mathur00,sulenticetal00a}. There is evidence that xA sources tend to have high-metallicity \citep{shemmeretal04,martinez-aldamaetal18a}. Similar properties have been identified as characteristic of narrow-line Seyfert 1 galaxies (NLSy1s) with strong \feii\ emission. 
  {NLS1s also} have unusually high metallicities for their luminosities. \citet{shemmernetzer02} have shown that NLSy1s  deviate significantly from the nominal relationship between metallicity and luminosity in AGN. As several studies distinguish between NLSy1s and ``broader-lined'' AGN, we remark here that all \feii\ strong NLSy1s meeting the selection criterion \rfe $> 1$\ are {\em } extreme Pop. A sources.\footnote{NLSy1s are identified by the line width of the \hb\ broad component being FWHM(\hbbc)$\le 2000$  \kms \citep{osterbrockpogge85}, Pop. A sources are  identified FWHM(\hbbc)$\le 4000$  \kms\ \citep{sulenticetal00a}. Imposing a fixed limit on line FWHM, although very convenient observationally, has no direct physical meaning, and its interpretation might be sample dependent. See \citet{marzianietal18} for a discussion of the issue.}

 


The aim of this work is to investigate the metallicity-sensitive diagnostic ratios of the UV spectral range for extreme Population A quasars i.e., {for highly accreting quasars}.  Section \ref{sec:sample}  defines  our sample, and provides some basic information.  In Sect. \ref{sec:methods} we define the diagnostic ratios, and describe the basic observational results. In Section \ref{sec:results} we compare measured diagnostic ratios and we compare them  with the ones obtained from photoionization simulations.  In Sect. \ref{Discussion} we discuss our results in terms of method caveats, metal enrichment, accretion parameters and their implications on the nature of xA sources. We show  the  UV spectra in Appendix \ref{app:spec}  along  with the multicomponent fit analysis of the emission blends, and in the Appendix \ref{sec:iso} we show the  trend of $Z$-sensitive ratios as a function of ionization parameter, density, and metallicity.


\section{Sample}
\label{sec:sample}
\subsection{Sample definition}

Qualitatively, extreme Pop. A objects show prominent \aliii\ and weak or absent \ciii\ emission lines. In general, they show low emission line equivalent widths ($\approx \frac{1}{2}$\ of them meet the $W$(\civ)$\lesssim$ 10 \AA, and qualify as weak-lined quasars following \citealt{diamond-stanicetal09}),\footnote{Weak-lined quasars are mostly xA sources, judging from their location along the MS \citep{marzianietal16a}, and that the limit at $W \approx 10$ \AA\ separates the low-$W$\ side of a continuous distribution of the xA \civ\ equivalent width peaked right at around 10 \AA\ \citep{martinez-aldamaetal18a}.} and a spectrum that is easily recognizable even by a visual inspection, also because of the 
``trapezoidal" shape of the \civ\ profile and the intensity of the $\lambda$1400 blend, comparable to the one of \civ\  \citep{martinez-aldamaetal18}.

xA sources were selected according to the criteria given in Sect. \ref{sec:intro}, using line measurements  automatically obtained by the {\tt splot} task with a cursor script within the IRAF data reduction package. We focus on the spectral range from $\approx$ 1200 \AA\ to 2100 \AA, where (1) UV lines used for xA identification are present; (2) the strongest emission features helpful for metallicity diagnostics are also located. The \lya\ + \nv\ blend is usually too heavily compromised by absorptions which make it impossible to reconstruct the emission components especially for \lya.  We will make some consideration on the mean strength of the \nv\ with respect to \civ\ and \heii\ (Sect. \ref{nv}), but will not consider \nv\ as a diagnostics. We selected SDSS DR12\footnote{https://www.sdss.org/dr12/} spectra in the redshift range 2.15 $<$ z $<$ 2.40, relatively bright ($r < 19$) to ensure moderate-to-high S/N in the continua (in all cases S/N $\gtrsim 5$ in the continuum, and the wide majority with S/N $\gtrsim 10$), and of low declination $\delta < 10$.  The redshift range was chosen to allow for the possibility of \hb\ coverage in the $H$ band by eventual near-IR spectroscopic observations. The DR12 sample selected with these criteria is $\approx $ 500\ sources strong. xA sources were selected out of this sample with an automated procedure, inspected  to avoid broad absorption   lines, and further vetted for obtained a small pilot sample of $\sim 10$ sources. A larger sample of xA sources will be considered in a subsequent work (Garnica et al., in preparation). The final selection includes 13 sources.   With the adopted selection  criteria in flux and redshift, we expect a small dispersion in the accretion parameters (especially luminosity; Sect. \ref{acc}).  Indeed, the selected  sources are  rather homogeneous in terms of spectral appearance, with a few sources {included in our sample} that however show borderline criteria.   {They will be considered is Sect.~\ref{intruders} in terms of their individual $U$, \nh}.


\subsection{Sample properties}

Table \ref{tab:general} provides basic information for the 13 sources of our sample: SDSS name, redshift from the SDSS, the difference between our redshift estimation using \aliii\  (described in \ref{redshift}) and the SDSS redshift $\delta z $ = $z - z_\mathrm{SDSS}$, the $g$-band magnitude provided by \citet{adelman-mccarthyetal08}, the $g-r$\ color index, the rest-frame-specific continuum flux at 1700 \AA\ and 1350 \AA\ measured on the rest frame, the S/N at 1450 \AA. All other sources were covered by the FIRST \citep{beckeretal95}, but undetected. Considering that the typical rms scatter of FIRST radio maps is $\approx$0.15 Jy, and the typical fluxes of in the $g$\ band, we have upper limits $\lesssim 5$ in the  radio-to-optical ratio, qualifying the  sample sources  {as} radio quiet. Distances were computed using the formula  provided by \citet[][their Eq. B.5]{sulenticetal06}, and $\Lambda$CDM cosmology ($\Omega_{\Lambda} =0.7, \Omega_\mathrm{M} =0.3, H_{0} = 70$ \kms\ Mpc$^{-1}$).  The bolometric luminosity is around $\sim 10^{47}$ \ergss, assuming a bolometric correction  B.C.$_{1350}$ = 3.5 \citep{richardsetal06}.  The sample rms is just $\approx 0.2$ dex: all sources are in a narrow range of distances and have observed fluxes within a factor $\approx$ 2 from their average. This is, in principle, an advantage for the estimation of the physical parameters such as \lledd, considering the large uncertainty and {serious} biases associated with the estimation of \mbh\ from UV high-ionization lines. Accretion parameters will be discussed  in Sect. \ref{acc}.

 \begin{deluxetable*}{lcccccccc}
\tabletypesize{\scriptsize}
\tablecaption{Source identification and basic properties}
\tablehead{ 
\colhead{SDSS NAME  }     & \colhead{$z_\mathrm{SDSS}$} & \colhead{$\delta z$} &  \colhead{${g}$} & \colhead{$g-r$}   &   \colhead{$f_\lambda$(1700 \AA)} & \colhead{$f_\lambda$(1350 \AA)}     & \colhead{S/N }  \\
\colhead{(1)}& \colhead{(2)}& \colhead{(3)}&\colhead{(4)}& \colhead{(5)}& \colhead{(6)}&\colhead{(7)}& \colhead{(8)}& }
\startdata
J010657.94-085500.1 & 2.355 & 0.006 & 18.18  & 0.095 & 662 & 951  & 20                 \\
J082936.30+080140.6 & 2.189 & 0.008 & 18.366 & 0.302 & 672 & 939  & 11                  \\
J084525.84+072222.3 & 2.269 & 0.017 & 18.204 & 0.331 & 668 & 989  & 13                 \\
J084719.12+094323.4 & 2.295 & 0.004 & 18.940 & 0.234 & 368 & 511  & 17                 \\
J085856.00+015219.4 & 2.160 & 0.002 & 17.916 & 0.255 & 709 & 1204 & 21                 \\
J092641.41+013506.6 & 2.181 & 0.004 & 18.591 & 0.337 & 377 & 670  & 21                 \\
J094637.83-012411.5 & 2.212 & 0.002 & 18.561 & 0.178 & 385 & 595  & 18                 \\
J102421.32+024520.2 & 2.319 & 0.008 & 18.49  & 0.177 & 478 & 694  & 23                 \\
J102606.67+011459.0 & 2.253 & 0.003 & 18.982 & 0.206 & 428 & 525  & 13                 \\
J114557.84+080029.0 & 2.338 & 0.009 & 18.545 & 0.369 & 243 & 360  & 5                  \\
J150959.16+074450.1 & 2.255 & 0.008 & 18.938 & 0.278 & 223 & 346  & 9            \\
J151929.45+072328.7 & 2.394 & 0.008 & 18.662 & 0.171 & 405 & 507  & 19                 \\
J211651.48+044123.7 & 2.352 & 0.000 & 18.825 & 0.220 & 404 & 573  & 32                 \\
  \enddata
\tablecomments{Columns are as follows: (1) SDSS coordinate name; (2) SDSS redshift; (3) correction to redshift estimated in the present work ($\delta z $ = $z - z_\mathrm{SDSS}$); (4) $g$-band magnitude from \citet{2008ApJS..175..297A}; (5) color index $g-r$; (6) continuum flux measured at 1700 \AA\ in units of 10$^{-17}$ erg s $^{-1}$ cm $^{-2}$ \AA $^{-1}$; (7) continuum flux measured at 1350 \AA\ in the same units;
(8) S/N measured at continuum level at 1450 \AA.}
\label{tab:general}
 \end{deluxetable*}
 
\section{Methods}
\label{sec:methods}

\subsection{Redshift determination}
\label{redshift}

The estimate of the quasar systemic redshift in the UV is not trivial, as there are no low-ionization narrow lines available in the spectral range \citep{vandenberketal01}. In practice, one can resort to the broad LIL. \citet{negreteetal14} and \citet{martinez-aldamaetal18a} consider the Si{\sc ii}$\lambda$1263 and O{\sc i}$\lambda$1302 lines to obtain a first estimate.  A re-adjustment is then made from the wavelength of the \aliii\ doublet which is found, in almost all cases, to have a consistent redshift.  To determine the \aliii\ shift those authors used multicomponent fits with all the lines in the region of the blend $\lambda$1900 included.  The peak of \aliii\  is clearly visible in the spectra of our sample, since in high accretors emission of \aliii\ is strong with respect to the other lines in the blend at $\lambda$1900\AA.  We decided to use only this  method for redshift estimation (in Tab. \ref{tab:general}) and to measure the peak we use single Gaussian fitting from the {\tt splot} task of the \aliii\ doublet and/or of the \siiii\ line, depending on which feature is sharper. The obtained values are usually $\ge z_\mathrm{SDSS}$\  (Table \ref{tab:general}). This is not a surprise as   $z_\mathrm{SDSS}$\ is based on lines that are mainly blueshifted in xA sources, and hence is a systematic underestimation of the unbiased redshift.


\subsection{Diagnostic ratios sensitive to $U$, density, $Z$}
\label{ratios}


Line ratios are sensitive to different parameters. In the UV range, three groups of diagnostic ratios are defined in the literature
\citep[e.g.][]{negreteetal12,martinez-aldamaetal18}. 

\begin{itemize}
\item \civ/\siiv+\oiv, \civ/\heiiuv\ have been widely applied as metallicity indicators \citep[e.g.,][]{shinetal13}. In principle, \civ/\-\heiiuv\ and \siiv/\-\heiiuv\ should be sensitive to C and Si abundance because the He abundance relative to Hydrogen can be considered constant. The ionization potentials of C$^{2+}$ and He$^{+}$   are similar. The main difference is that the  \heiiuv\ line is a recombination line, equivalent to H{\sc i}  H$\alpha$, and the regions where they are formed are not coincident (see Fig. \ref{fig:iontz}).  
\item Ratios involving \nv, \nv/\civ\ and \nv/\heiiuv\ have been also widely used in past work, after it was noted that the \nv\ line was stronger than expected in a photoionization scenario \citep[e.g.,][]{osmersmith76}.  A selective enhancement of nitrogen \citep{shields76} is expected  due to  secondary production of N by massive and intermediate mass stars, yielding [N/H]$\propto Z^{2}$\ \citep{vila-costasedmunds93,izotovthuan99}. This process might be especially important at the high metallicities inferred for the quasar BLR. Therefore estimates based on \nv\ may differ in a systematic way from estimates based on other metal lines \citep[e.g.,][]{matsuokaetal11}.  In the present sample of quasars, contamination by narrow and semi-broad absorption is severe, and even 
{if we model precisely the high ionization lines,}
it might be impossible to reconstruct the unabsorbed profile of the red wing of \lya. In addition, S/N is not sufficient to allow for a careful measurement of \niv\ and \niii\ lines.  We defer the systematic analysis of nitrogen  lines to a subsequent work, while discussing the consistency of the \nv\ measures in a high-$Z$\ scenario (Sect. \ref{nv}).   
\item The ratios \aliii/\-\siiii\ and \siiii/\-\ciii\ are sensitive to density, as the ratios involve intercombination lines with a well defined critical density ($n_\mathrm{c}\sim 10^{10}$ cm$^{-3}$ for \ciii\ \citep{hamannetal02} and $n_\mathrm{c}\sim$ 10$^{11}$ cm$^{-3}$ for  \siiii\ \citep{negreteetal10}). 
\item \siiii/\siiv, \siii/\siiii, and \siii/\siiv\ are sensitive to the ionization parameters and insensitive to $Z$, as they are different ionic species of the same element. 
\end{itemize}

Other intensity ratios entail a dependence on metallicity $Z$, but also on ionization parameter $U$\ and density \nh\ \citep{marzianietal20}. 


\subsection{Line interpretation and diagnostic ratios}

The comparison between LILs and HILs has provided insightful information  over a broad range of redshift and luminosity \ \citep{corbinboroson96,marzianietal96,marzianietal10,sulenticetal17,bisognietal17,shen16,vietrietal18}.  A LIL-BLR appears to remain basically virialized \citep{marzianietal09,sulenticetal17}, as  the \hb\ profile remains (almost) symmetric and unshifted with respect to rest frame even if \civ\ blueshifts can reach several thousands of \kms. In Population A, the lines have been decomposed into two components:

\begin{itemize}
\item  The broad component (BC), also known as   the intermediate  component, the core component or the central broad component  following various authors \citep[e.g.,][]{brothertonetal94a,popovicetal02,kovacevicdojcinovicpopopvic15,adhikarietal16}. The BC is {modeled} by a symmetric and unshifted profile  (Lorentzian for  Pop. A; \citealt{veroncettyetal01,sulenticetal02,zhouetal06}), {and is believed} to be associated with a virialized BLR subsystem.   
\item The blue shifted component (BLUE).  A strong  blue excess in Pop. A \civ\ profiles is {obvious, as in some \civ\ profiles -- like the one of the xA  prototype I Zw1 or high luminosity quasars -- BLUE  dominates} the total emission line flux \citep{marzianietal96,leighlymoore04,sulenticetal17}.  For BLUE, there is no evidence of a regular  profile,  and the fit attempts to empirically reproduce the observed excess emission. BLUE is detected in a LIL such as \hb\ at a very low level, and is not strongly affecting FWHM measurements \citep{negreteetal18}.
\end{itemize}


\subsubsection{Broad component}

Diagnostic ratios are not equally well measurable for the BC and the BLUE. For the BC, the following constraints and caveats apply: 

\paragraph{\civ/, \siiv/, \aliii/ over \heiiuv} \heiiuv\ is weak but  measurable in most of the objects. Ratios such as \civ/\-\heii, \siiv/\-\heii, \aliii/\-\heii\ ($U$-dependent) offer $Z$ \ indicators. 
Especially for the low-ionization conditions of the BC emitting gas, these ratios are well-behaved (Sect. \ref{photoion} and \ref{explo}) and will form the basis of the $Z$\ estimates presented in this paper.  

\paragraph{\siiv/\civ}   There are problems in estimating the \siiv\ line intensity: an overestimation might be possible because of difficult continuum placement (see, for example, the case of  \object{SDSSJ085856.00+015219.4}  in Appendix \ref{app:spec}). The relative contribution of \siiv\ to the blend at $\lambda$1400 is unclear \citep{willsnetzer79}.  A strong BC contribution of \oiv\  is unlikely, as this line has a critical density $n_\mathrm{c}\sim 10^{10}$ cm$^{-3}$ \citep[][see also the isophotal contour of \siiv/\oiv\ in Appendix \ref{sec:iso}]{zheng88}.  Our measurements are nonetheless compared to \siiv +  total O{\sc iv]}  {CLOUDY} prediction. 

\paragraph{\aliii/\siiii} This ratio is sensitive to density in the low-ionization BLR domain \citep{negreteetal12}. Values \aliii/\siiii$>$1 are possible  if density is higher than 10$^{11}$ cm$^{-3}$, the critical density of \siiii. We will not use this parameter as a metallicity estimator, although, in principle, for fixed physical conditions (setting  \nh\ and $U$) the \aliii/\siiii\ and \siiii/\ciii\ ratios may become dependent mainly on electron temperature and so on metallicity (Sect. \ref{photoion}). The ratio of the  total emission in the 1900 blend \aliii+\siiii\ +\ciii\ over \civ\ has been used as a metallicity estimator \citep{sulenticetal14}. Considering the uncertain contribution of \feiii\ emission and especially of the \feiii\ $\lambda$1914 line in the xA spectra, we will not use the total intensity of the $\lambda$1900 blend as a diagnostic.

\paragraph{\civ/\aliii} Biases might be associated with the estimate of the \civbc\ especially when BLUE is so prominent that \civbc\  contributes to a minority fraction.


\subsubsection{BLUE component}

\paragraph{\civ/\heii}The \heii\ BLUE is well-visible merging smoothly with the red wing of \civ. The ratio \civ/\heii\ might be  affected by the decomposition of the blend, leading to an overestimate of the \heiiuv\ emission.   This ratio is in principle sensitive to metallicity. However, the increase is not monotonic at relatively high $U$\ (see the panel for \civ/\heii\ in Fig. \ref{fig:civheiifixed}). The resulting effect is that the \civ/\heii\ ratio within the uncertainties leaves the $Z$\ unconstrained between 0.1 and 100 solar. 

\paragraph{ \civ/(\oiv\ + \siiv)} The blueshifted excess  at $\lambda$1400 is ascribed to O{\sc iv} + \siiv\ emission. A significant  contribution can be associated with \oiv\ and several transitions of O{\sc iv} that are computed by { CLOUDY} \citep[see e.g.,][]{keenanetal02} are especially relevant at high $U$\ values and moderately low \nh\ ($\sim 10^8$ cm$^{-3}$). The  blue side of the line is relatively straightforward to measure  for computing  \civ/$\lambda1400$\ with a multicomponent fit, although difficult continuum placement, narrow absorption lines, and blending on the blue side make it difficult to obtain a very precise measurement. A total  {$\lambda$1400\ BLUE} emission exceeding \civ\ is possible if, assuming $\log U \gtrsim 0$, $\log $ \nh $\gtrsim 9$ [cm$^{-3}$], the metallicity value is very high $20 \lesssim Z \lesssim 100 Z_\odot$, (Sect. \ref{explo}).

\paragraph{(\oiv +\siiv)/\heii} By the same token, the \heii\ overestimation may lead to a lower  (\oiv +\siiv)/\heii\ ratio. 


\subsection{Analysis via multicomponent fits}
\label{fitting}

We analyze 13 objects using the {\tt specfit} task from IRAF \citep{kriss94}. The use of the $\chi^{2}$ minimization is  aimed to  provide a heuristic separation between the broad component (BC) and the blue component (BLUE) of the emission lines. After redshift correction following the method  described in Sect. \ref{redshift}, for each source of our sample we perform a detailed modeling  using various components as described below, including computation of  asymmetric errors (Sect. \ref{errors}).  As mentioned in Sect. \ref{ratios}, in our analysis we consider five diagnostic ratios for the BC: \civ/$\lambda1400$, \civ/\-\heii, \aliii/\-\heii, $\lambda$1400/\heii, $\lambda1400$/\aliii,  and three for the BLUE:  \civ/$\lambda1400$, \civ/\-\heii, $\lambda1400$/\heii.  The \civ/\heii\ is used with  care, as  it may yield poor constraints. In addition, it is important to stress that, of the five ratios measured on the BC, only three (the ones dividing by the intensity of \heii\ BC) are independent. We compare the fit results with arrays of CLOUDY \citep{ferlandetal13} simulations  for various metallicities and physical conditions (Sect. \ref{photoion}). 

For each source we perform the multicomponent fitting in three ranges described below. The best fit is identified by the model with the lowest $\chi^2$\ i.e., with minimized difference between the observed and the model spectrum.
Following the data analysis by \citet{negreteetal12}, we use the following  components:
 \paragraph{The continuum } was modeled as a power-law, and we use the line-free windows around 1300 and 1700 \AA\ (two small ranges where there are no strong emission lines) to scale it. If needed, we divide the continuum into three parts (corresponding to the three regions mentioned below). Assumed continua  are shown in the Figures of Appendix \ref{app:spec}.

 \paragraph{\feii\ emission} usually does not contribute significantly in the studied spectral ranges. We consider the \feii\ template which is based on CLOUDY simulations of \citet{bruhweilerverner08} when necessary. In practice the contamination by the blended \feii\ emission yielding a pseudo-continuum is negligible.  Some \feii\  emission lines were detectable in only a few objects and around  $\approx$ 1715\ \AA, at 1785 \ \AA, and at 2020 \ \AA. In these cases, we model them using single Gaussians.
\paragraph{\feiii\ emission} affects more the 1900\ \AA \ region and seems to be strong when AlIII $\lambda$1860 is strong as well \citep{hartigbaldwin86}.
To model these lines we use the template of \cite{vestergaardwilkes01}.
\paragraph{Region 1300 - 1450\AA \ }  is dominated by
the \siiv + \oiv\  high ionization blend with strong blueshifted component. The fainter lines as Si{\sc ii}$\lambda$1306, O{\sc i}$\lambda$1304 and C{\sc ii}$\lambda$1335 are also detectable.  For the broad and blueshifted components we use the same model as in case of \civ\ and \heii. This spectral range is often strongly affected by absorption.
\paragraph{Region 1450 - 1700 \AA \ }  is dominated by \civ\ emission line which we model as a fixed in the rest-frame wavelength
Lorentzian profile representing the BC and two blueshifted asymmetric Gaussian profiles vary freely. The same model is used for \heii.
\paragraph{Region 1700 - 2200 \AA\ }  is dominated by \aliii, \siiii\ and \feiii\  intermediate - ionization lines. We model \aliii\ and \siiii\ using Lorentzian profiles, following \cite{negreteetal12}. \ciii\ emission is also included in the fit, although the dominant contribution around $\lambda1900$\ is to be ascribed to \feiii\ \citep[][and references therein]{martinez-aldamaetal18a}. We use the template of \citet{vestergaardwilkes01} to model \feiii\ emission.  No BLUE is ascribed to these intermediate ionization lines. 
\paragraph{Absorption lines} are modeled by Gaussians, and included whenever necessary to obtain a good fit.
\medskip

The fits to the observed spectral ranges  are shown in the Figures of Appendix \ref{app:spec}.

\subsubsection{Error estimation on line fluxes}
\label{errors}

\begin{figure}[ht]
\includegraphics[scale=0.75]{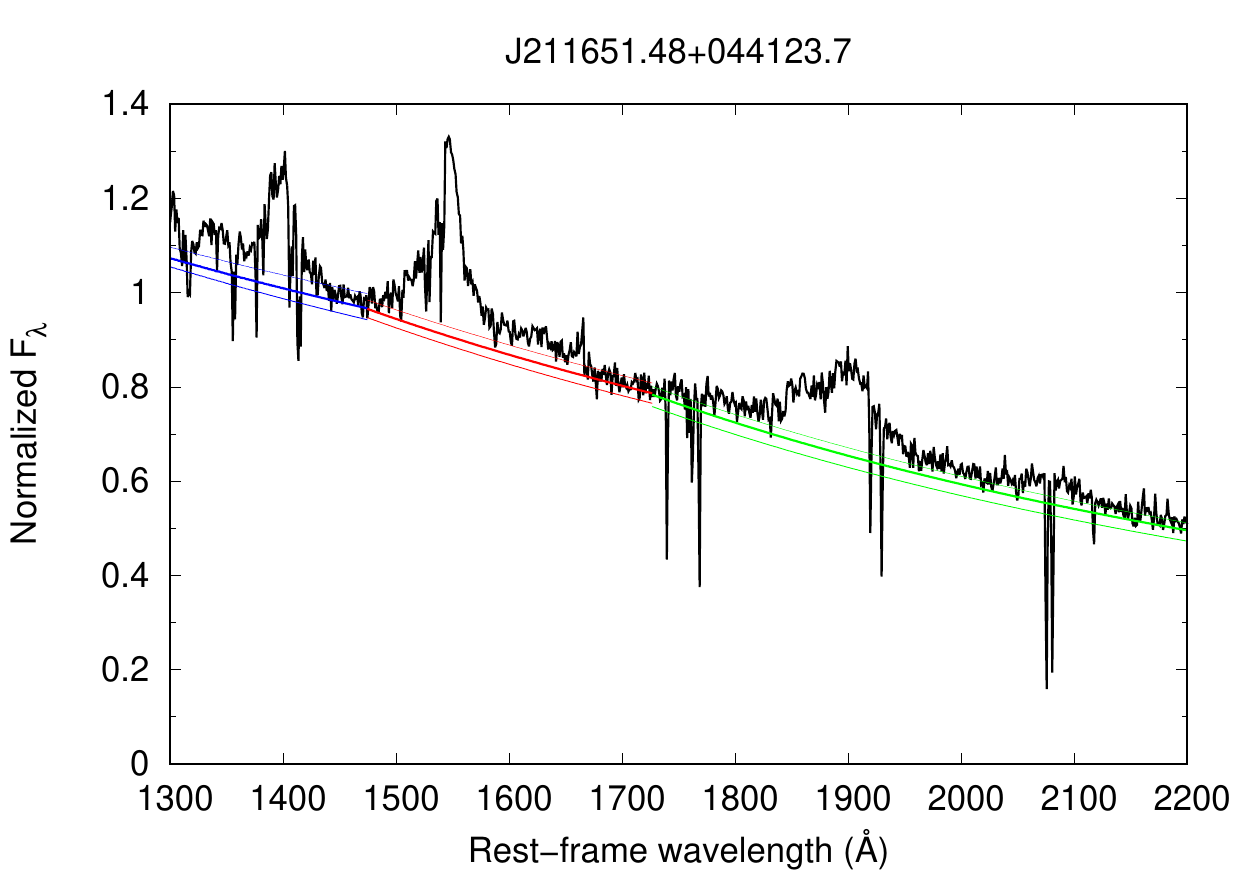}
  \caption{{Continuum estimation for J211651.48+044123.7 from our sample. Range 1300 - 1450 \AA\ is shown in blue, 1450 - 1700 \AA\ in red and 1750 - 2200 \AA\ in green. The continuum lines in each range represent from the top: the maximum, the best and the minimum continuum placement.}\label{fig:min_max_cont}}
  \end{figure}

The choice of the continuum placement is the main source of uncertainty in the measurement of the emission line intensities. The fits in Appendix \ref{app:spec} show that, in the   majority of cases, the FWHM of the \aliii\ and \siiii\ lines (assumed equal) satisfy the condition FWHM(\aliii) $\sim$ FWHM(\civ$_\mathrm{BC} \sim$ FWHM(\siiv$_\mathrm{BC}$).   Figure \ref{fig:min_max_cont}  shows the best fit, maximum and minimum placement of the continuum, which we choose empirically. With this approach the continua of Figure \ref{fig:min_max_cont} should provide the continuum uncertainty at a  $\pm 3 \sigma$\ confidence. 

The continuum placement strongly affects the measurement of an extended feature such as the \feiii\ blends and the \heii\ emission.  Figure \ref{fig:min_max_cont}  makes it evident that errors on fluxes   are asymmetric. The thick line shows the continuum best fit, and the thinner the minimum and maximum plausible continua. Even if the minimum and maximum are displaced by the same difference in the intensity with respect to the best fit continuum, assuming the minimum continuum would yield an increase in line flux larger than the flux decrease assuming the maximum continuum level. In other words, a symmetric uncertainty in the continuum specific flux translates into an asymmetric uncertainty in the line fluxes. To manage asymmetric uncertainties, we assume that the distribution of errors follows  the triangular distribution  \citep{dagostini03}. This method assumes linear decreasing in either side of maximum of the distribution (which is the best fit in our case) to the values obtained for maximum and minimum contributions of the continuum. {We motivate using the triangular error distribution as a relatively easy analytical method to deal with asymmetric errors.} For each line measurement we calculate the variance using the following formula for the triangular distribution:
\begin{equation}
    \sigma^2(X) = \frac{\Delta^2 x_+ + \Delta^2 x_- + \Delta x_+ + \Delta x_-}{18}
\end{equation}
{where $\Delta x_+$ and $\Delta x_-$ are differences between measurement with maximum and best continuum and with best and minimum continuum respectively. To analyze error of diagnostic ratios we propagate uncertainties using standard formulas of error propagation.}

\subsection{Photoionization modeling}
\label{photoion}

To interpret our fitting results we compare the line intensity ration for BC and BLUE  with the ones predicted by CLOUDY     13.05 and 17.02  simulations  \citep{ferlandetal13,ferlandetal17}.\footnote{ The arrays  were computed over several years with {\tt CLOUDY 13.05}, in large part before {\tt CLOUDY 17.02} became available. The computations with the two versions of the code are in agreement as far as the trends with $U$, \nh, and $Z$\ are concerned, although the derived $Z$\ values are a factor $\approx$ 2 systematically lower with the 17.02 release of {\tt CLOUDY}. In this paper we present the CLOUDY 17.02 for all estimates of metallicity and physical parameters $U$\ and \nh. } An array of simulations is used as reference for comparison with the observed line intensity ratios. It was computed under the assumption that  (1) column density is $N_\mathrm{c}$\ = 10$^{23}$ cm$^{-2}$; (2) the continuum is represented  by the model continuum of \citet{mathewsferland87} which is believed to be appropriate for Population A quasars, and (3)  microturbulence is negligible. The simulation arrays cover the Hydrogen  density range 7.00$\leq \log(n_\mathrm{H}) \leq$14.00 and the ionization parameter $-4.5 \leq \log(U) \leq$ 1.00, in intervals of 0.25 dex. They are repeated for values of metallicities in a range encompassing five orders of magnitude: 0.01, 0.1, 1, 2, 5, 10, 20, 50, 100, 200, 500 and 1000 $Z_{\odot}$. Extremely high metallicity $Z \gtrsim 100 Z_{\odot}$\ is considered physically  unrealistic ($Z \approx 100 Z_\odot$ implies that more than half of the gas mass is made up by metals!), unless the enrichment is provided in situ within the disk \citep{cantielloetal20}. 
The behavior of diagnostic line ratios as a function of $U$ and \nh\ for selected values of $Z$ is shown in Fig. \ref{fig:isoph1} of Appendix \ref{sec:iso}.  

\subsubsection{Basic Interpretation}
\label{basic}

The line emissivity $\epsilon_\mathrm{coll}$\ (${\rm ergs\ cm}^{-3}\ {\rm s}^{-1}$) of a collisionally excited line emitted from an element $X$\ in its $i-$th ionization stage  has a strong temperature dependence. In the high density limit 
\begin{eqnarray}
\epsilon_\mathrm{X^{i},coll}\ & = &\  n_\mathrm{X^{i},l}\beta A_\mathrm{X^{i}, ul} h\nu_\mathrm{0} \,\frac{g_\mathrm{X^{i},l}}{g_\mathrm{X^{i},u}}\, \exp{\left(-\frac{h\nu_0}{kT_\mathrm{e}}\right)}\\   \nonumber
& \propto  & n_\mathrm{X^{i},l} \exp{\left(-\frac{h\nu_0}{kT_\mathrm{e}}\right)}\\  \nonumber
\end{eqnarray}
the line is said to  be ``thermalized,'' as its strength depends only on the atomic level population and not on the transition strength \citep{hamannferland99}. $\beta$\ is the photon escape probability and $A_\mathrm{X^{i}, ul}$\ is the spontaneous decay coefficient.
At low densities we have,
\begin{equation}
\epsilon_\mathrm{X^{i},coll}\ =\  n_\mathrm{X^{i},l} n_\mathrm{e} q_\mathrm{X^{i},lu} h\nu_0\ \propto\  n_\mathrm{X^{i}}^{2} T_\mathrm{e}^{-1/2}\,\exp{\left(-\frac{h\nu_0}{kT_\mathrm{e}}\right)}
\end{equation}

The recombination lines considered  in our analysis are \hb\ and \heiiuvfull, for which the emissivity (with an approximate dependence of radiative recombination coefficient $\alpha$\ on electron temperature, \citealt{osterbrockferland06}) becomes: 
\begin{equation}
\epsilon_\mathrm{Y^{j}, rec}\ =\ n_\mathrm{Y^{j}}\,n_\mathrm{e}\,\alpha\,h\nu_0\ \propto\ n_\mathrm{Y^{j}}^{2} T_\mathrm{e}^{-1} \ \ \ \
\end{equation}
and $n_\mathrm{Y^{j}}$ is the number density of the parent ion. 

Under these simplifying, illustrative assumption we can write:

\begin{equation}
\frac{\epsilon_\mathrm{X^{i},coll}}{\epsilon_\mathrm{Y^{j}, rec}} \propto \left(\frac {n_\mathrm{X^{i}}}{n_\mathrm{Y^{j}}}\right)^{2} T_\mathrm{e}^{\frac{1}{2}} \exp{\left(-\frac{h\nu_0}{kT_\mathrm{e}}\right)}
\end{equation}
for the low-density case, and 
\begin{equation}
\frac{\epsilon_\mathrm{X^{i},coll}}{\epsilon_\mathrm{Y^{j}, rec}} \propto \frac {n_\mathrm{X^{i}}}{n_\mathrm{Y^{j}}^{2}} T_\mathrm{e} \exp{\left(-\frac{h\nu_0}{kT_\mathrm{e}}\right)}
\end{equation} 
for the high density case. 

Similarly, for the ratio of two collisionally excited lines at frequencies $\nu_{0}$\ and $\nu_{1}$, 

\begin{equation}
\frac{\epsilon_\mathrm{X^{i},coll}}{\epsilon_\mathrm{Y^{j}, coll}} \propto \left(\frac {n_\mathrm{X^{i}}}{n_\mathrm{Y^{j}}}\right)^{\kappa}  \exp{\left(-\frac{h(\nu_0-\nu_{1})}{kT_\mathrm{e}}\right)}
\end{equation}

where $\kappa =1,2$\ in the high- and low-density case respectively.

Connecting the relative chemical abundance to the line emissivity ratios in the previous equation requires the reconstruction of the ionic stage distribution for each element, i.e., the computation of the ionic equilibrium, as well as the consideration of the extension of the emitting region within the gas clouds i.e., that the line emission is not cospatial, and possible differences in optical depth effects. This is achieved by the CLOUDY simulations. However, we can see that the main variable parameter for a given relative emissivity is $T_\mathrm{e}$. In other words, electron temperature is the main parameter connected to metallicity. This is especially true for fixed physical condition ($U$, \nh, $N_\mathrm{c} = 10^{23}$, SED given). This is most likely the case of xA sources: the spectral similarity implies that the scatter in physical properties is modest. We further investigate this issue in Sect.  \ref{ind}. 

 The electron temperature is also the dominating factor affecting the strength of  the \heiiuvfull\ line, for a given density. The \heiiuv\ \lya\ line at 304 \AA\ ionizes Hydrogen atoms and other ionic species with ionization potential up to 3 Ryd. Being absorbed by  different ionic species, \heii\ \lya\  cannot sustain a population of excited electrons at the level $n=2$ of \heii. This is markedly different from Hydrogen \lya\ that in case B is assumed to scatter many times and to sustain a population of Hydrogen atoms at level $n=2$. The \heiiuv\ line is therefore produced almost only by recombination, and no collisional excitation from level $n=2$ or radiative transfer effects are expected, unlike the case of the Hydrogen Balmer lines \citep{marzianietal20}. The prediction of the \heiiuv\ line is relatively simple once the electron temperature and the density are known by assumption or computation. The additional advantage in the use of \heiiuv\ is  that there is no  significant  enhancement of the  He abundance over the entire lifetime of the Universe \citep{peimbertetal01,peimbert08}. The normalization to the \heiiuv\ line flux of the flux of metal lines should yield robust $Z$ estimates.  This is shown by the isophotal contours of Appendix \ref{sec:iso} ( Fig. \ref{fig:isoph1}), tracing the behavior of the diagnostic ratios as a function of $Z$ and $U$: the (\siiv+\oiv)/\heiiuv\ and \aliii/\heiiuv\ ratios monotonically increase with $Z$\ over a large range of $U$; for \civ/\heii\ the behavior is monotonic at low $U$, but more complex at $\log U \sim -1 - 0$. Ratios involving pairs of metal lines   yield more complex trends in the plane  $Z$ -- $U$. At low \nh, the \civ/\aliii\ ratio is a good $Z$ estimator, although of limited usefulness since \aliii\ is weak; at high \nh, its sensitivity is greatly reduced (Appendix \ref{sec:iso}; Fig. \ref{fig:isoph1}). The \civ/(\siiv + \oiv) does not appear to be especially sensitive to $Z$. The diagnostic ratios change as a function of $Z$, although the behavior as a function of \nh\ and $U$\ is roughly preserved (Appendix \ref{sec:iso}; Fig. \ref{fig:isoph2}).

\subsection{Explorative analysis of photoionization trends at fixed ionization parameter and density}
\label{explo}

One of the main results of previous investigations is the systematic differences in ionization between BLUE and BC \citep{marzianietal10,negreteetal12,sulenticetal17}.  Previous inferences suggest very low ionization ($U \sim 10^{-2.5}$), also because of the relatively low \civ/\hb\ ratio for the BC emitting part of the BLR, and high density. A robust lower limit to density \nh $\sim 10^{11.5} $ cm$^{-3}$\   has been obtained from the analysis of the CaII triplet emission \citep{matsuokaetal07,pandaetal20}. Less constrained are the physical conditions for BLUE emission.  Apart from \civ/\hb$\gg1$ and Ly$\alpha$/\hb\ and \civ/\ciii\ also $\gg 1$, little constraints exist on density and column density. This result hardly comes  as a surprise considering the difference in dynamical status associated with the two components. While it is expected that the BC is emitted in a region of high column density $\log$\nc $\gtrsim 23 $ [cm$^{{-2}}$], not last because radiation forces are proportional to the inverse of $N_\mathrm{c}$\ \citep[][see also \citealt{ferlandetal09}]{netzermarziani10}. More explicitly, the equation of motion for a gas cloud under the combined effect of gravitation and radiation forces contains an acceleration term  due to radiation that is inversely proportional to $N_\mathrm{c}$. The high \nc\ region is expected to be relatively stable (at rest frame, with no sign of systematic, large shifts in Population A) and presumably devoid of low-density gas (considering the weakness of \ciii, \citealt{negreteetal12}). The same cannot be assumed for BLUE. BLUE is associated with a high radial velocity outflow, probably with the outflowing streams creating BAL features when intercepted by the line-of-sight \citep[e.g.,][]{elvis00}. 

\begin{figure*}[htp!]
\includegraphics[scale=0.28]{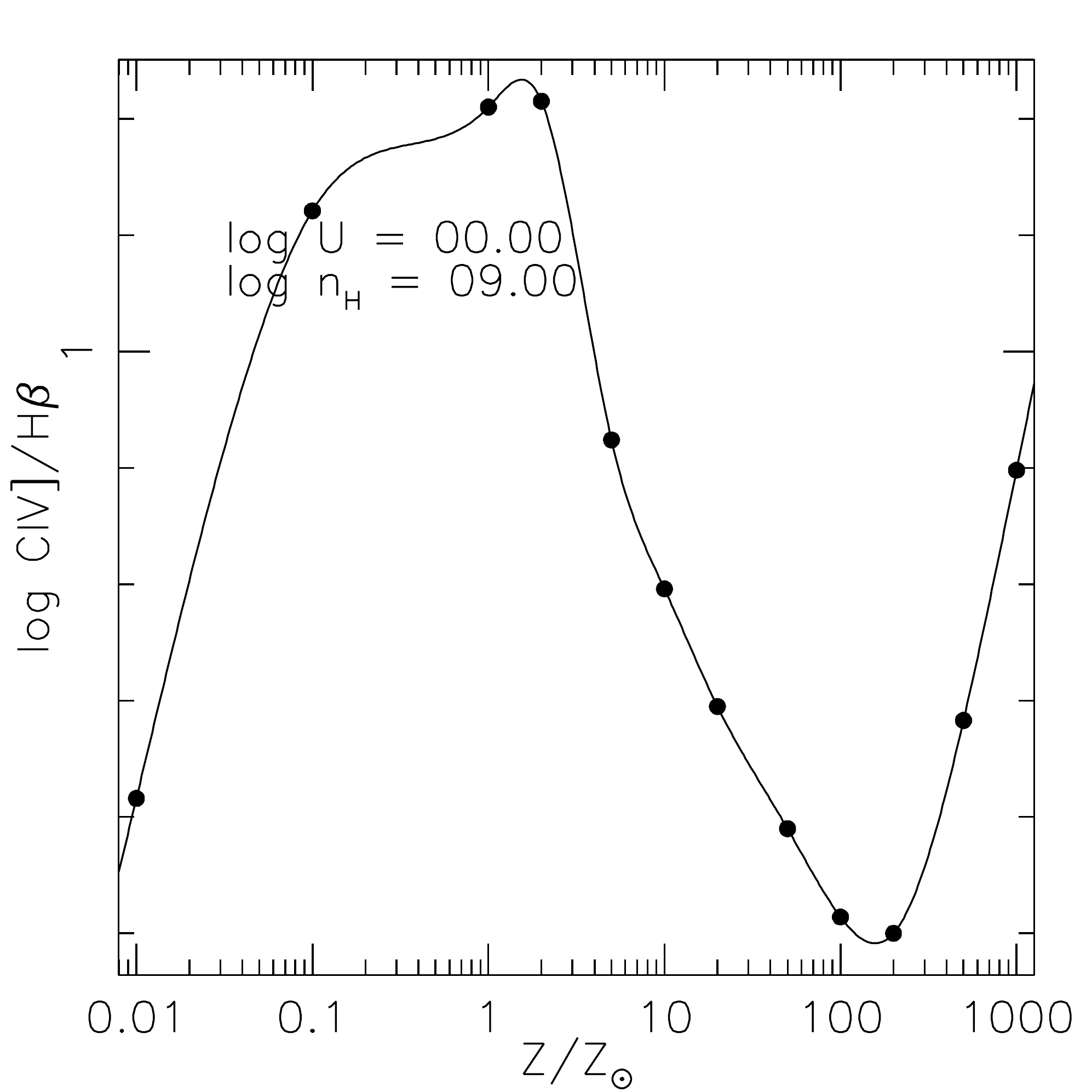}\hspace{-0.2cm}
\includegraphics[scale=0.28]{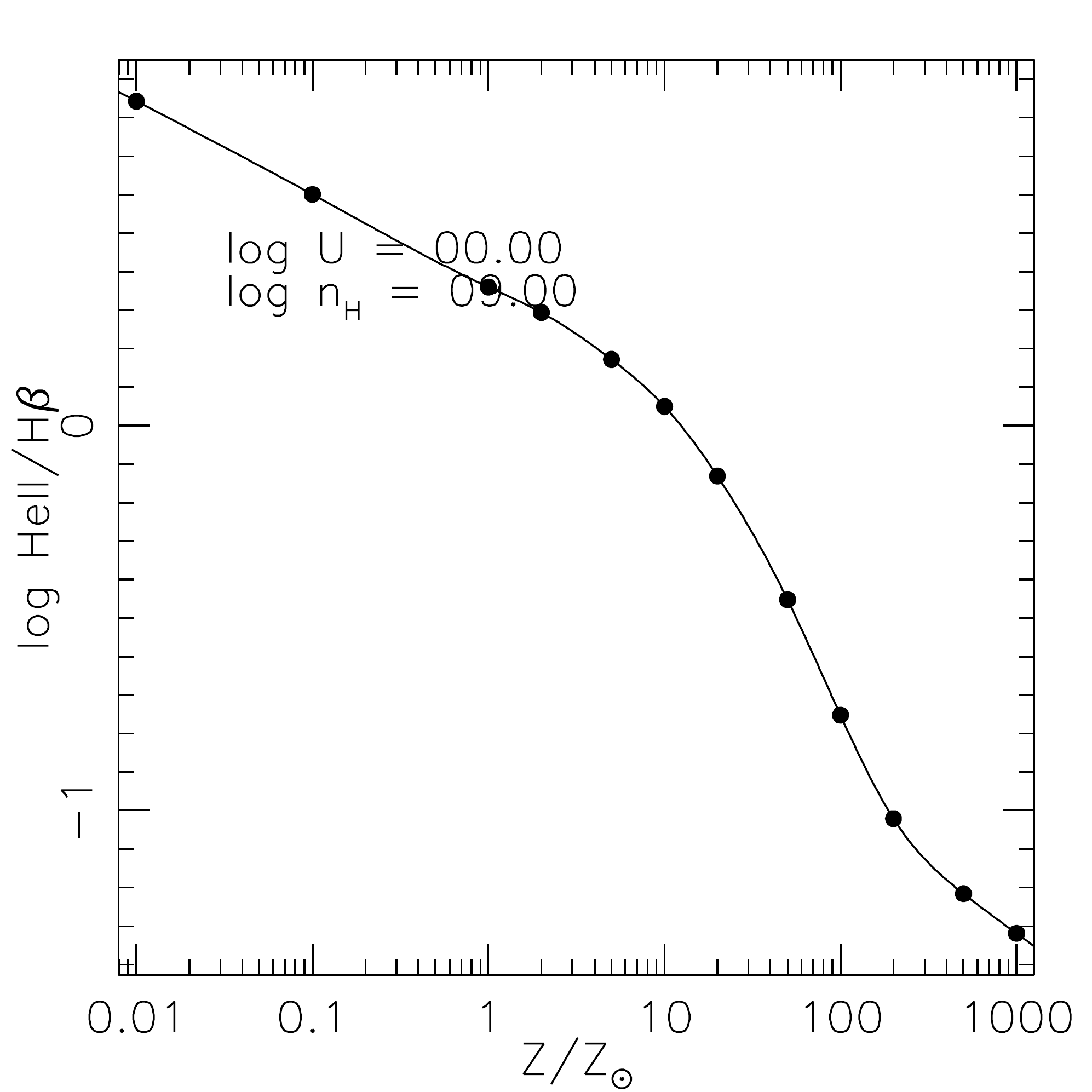}\hspace{-0.2cm}
\includegraphics[scale=0.28]{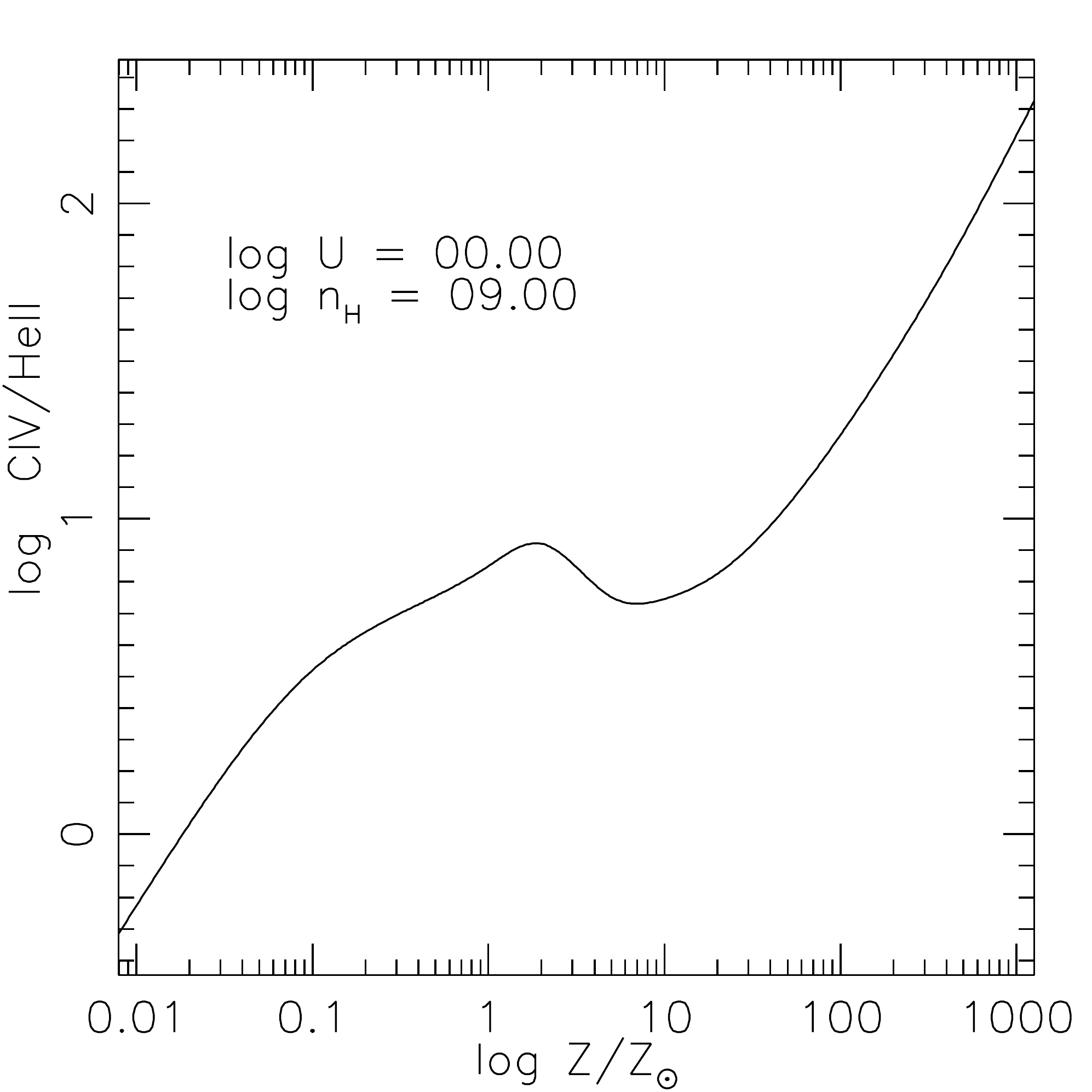}\\
\includegraphics[scale=0.28]{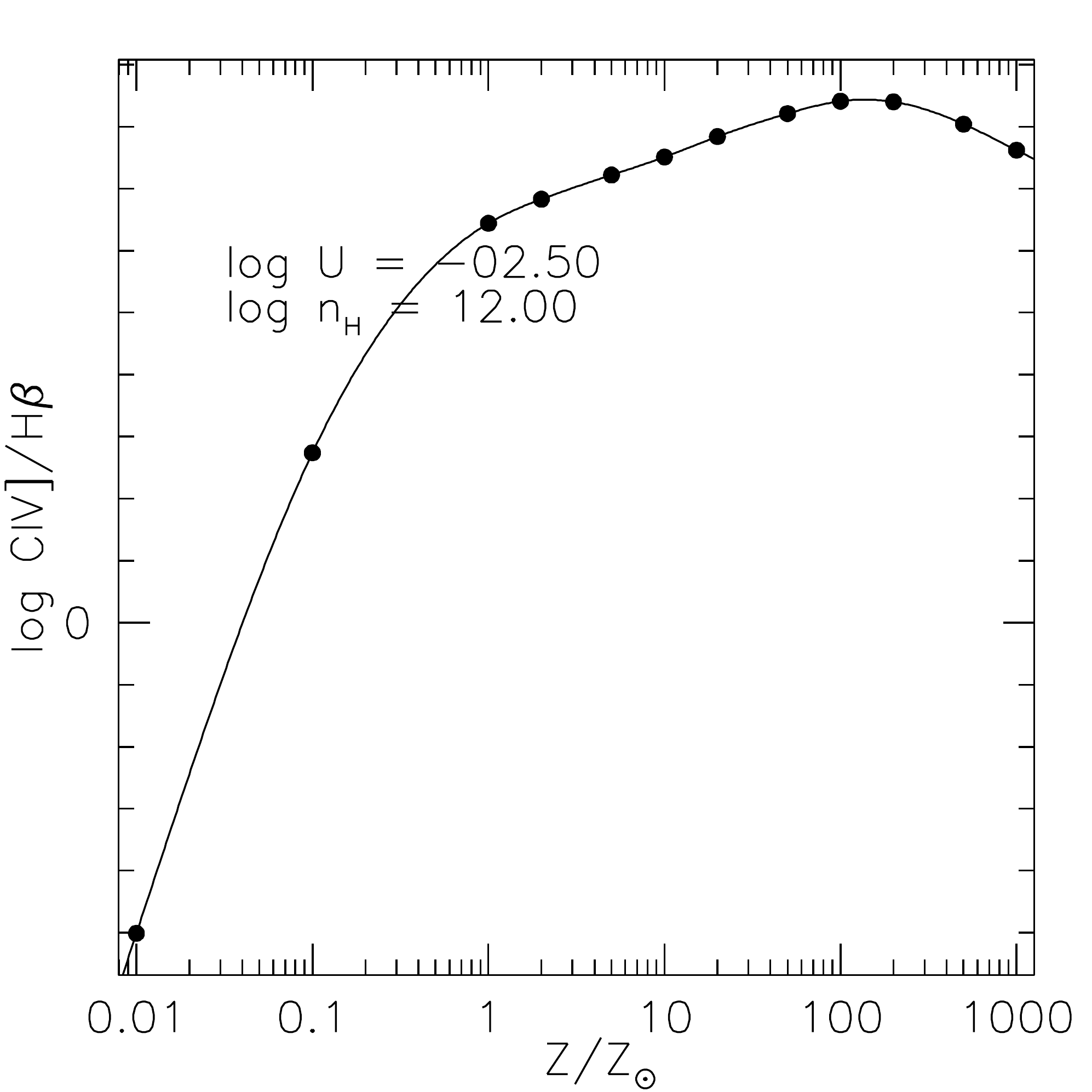} \hspace{-0.2cm}
\includegraphics[scale=0.28]{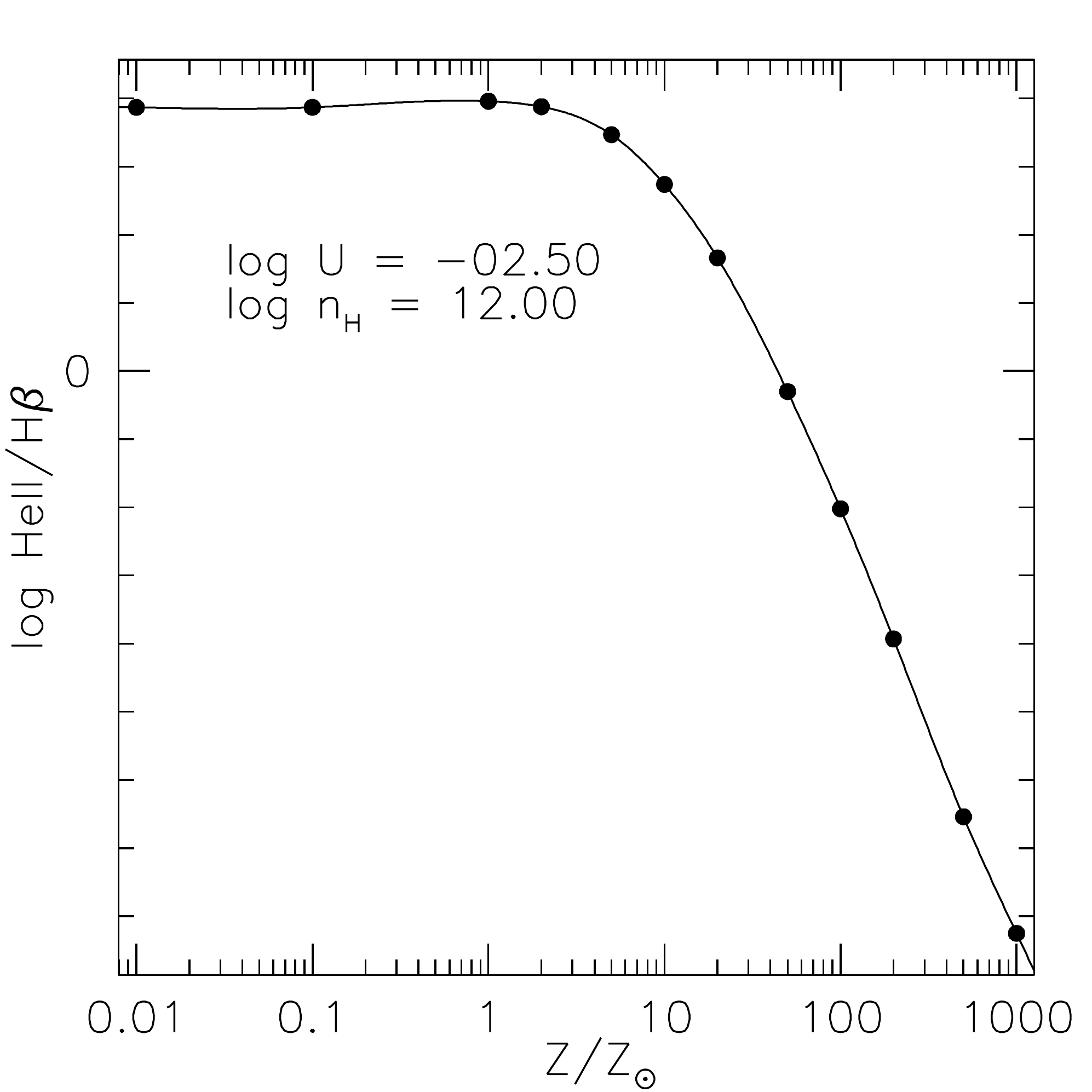}\hspace{-0.2cm} \includegraphics[scale=0.28]{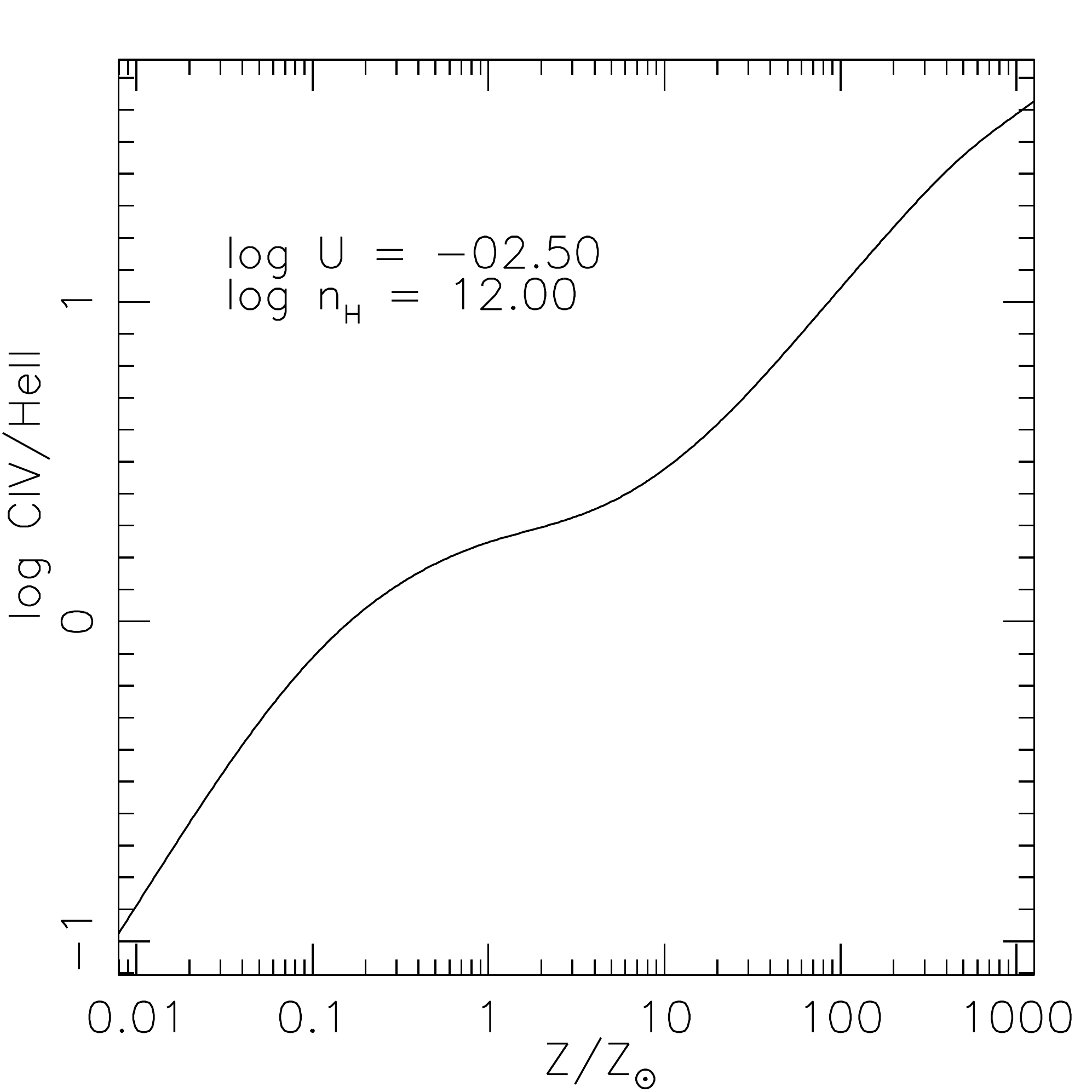}
\caption{Computed intensity ratios involving \civ\ and \heii\ as a function of metallicity, for physical parameters $U$\ and \nh\ fixed: ($\log U$, $\log$ \nh) = (-1,9) (top) and ($\log U$, $\log$ \nh) = (-2.5,12) (bottom). Columns from left to right show \civ/\hb, \heii/\hb, \civ/\heii.  \label{fig:civheiifixed}}
  \end{figure*}

Here we consider $\log U =-2.5$, $\log $ \nh=12 (-2.5,12), and $\log U=0, \log $\nh=9\ (0,9) as representative of the low and high- ionization emitting gas.  Fig. \ref{fig:civheiifixed} illustrates the behavior of the \civ/\hb, \heii/\hb\ and \civ/\heii\ in the high and low-ionization cases as a function of metallicity. The \civ\ intensity with respect to \hb\ has a steep drop around $Z \gtrsim 1 Z_{\odot}$, after a steady increase for sub-solar $Z$. The \heii/\hb\ ratio decreases steadily, with  a steepening at round solar value. Physically, this behavior is due  to the high value of the ionization parameter (assumed constant), while the electron temperature decreases with metallicity, implying a much lower collisional excitation rate for \civ\ production. The dominant effect for the \heii\ decrease is likely the ``ionization competition'' between \civ\ and \heii\ parent ionic species \citep{hamannferland99}.  As a consequence, the ratio \civ/\heii\ has a non-monotonic behavior with a local maximum around solar metallicity. At low ionization and high density, the behavior is more regular, as the steady increase in \civ/\hb\ is followed by a saturation to a maximum \civ/\hb. The \heii/\hb\ ratio is constant up to solar, and steadily decreases above solar, where the ionization competition with triply-ionized carbon sets on. The result is a smooth, steady increase in the \civ/\heii\ ratio. 
 
\begin{figure*}[htp!]
\includegraphics[scale=0.28]{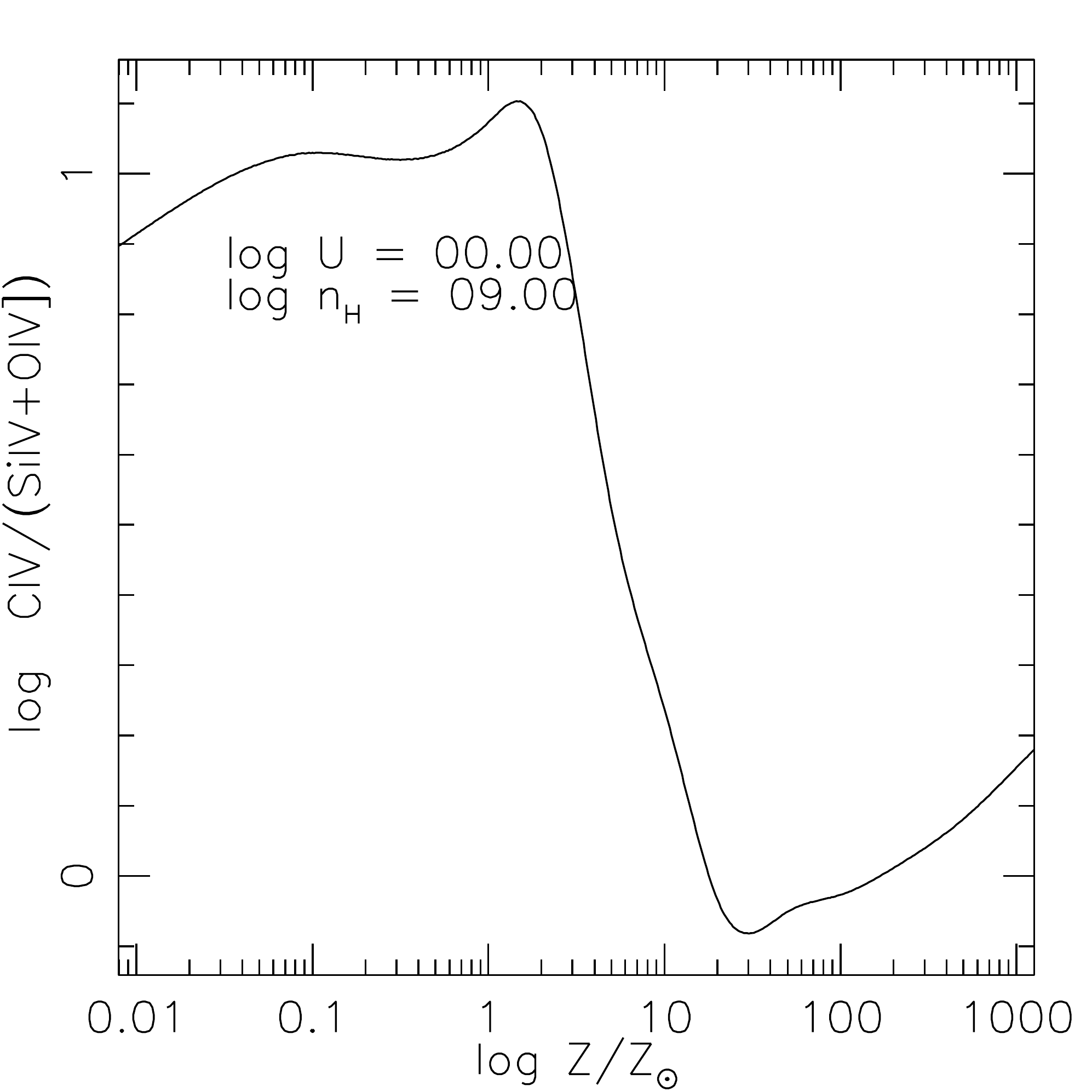}\hspace{-0.2cm}
\includegraphics[scale=0.28]{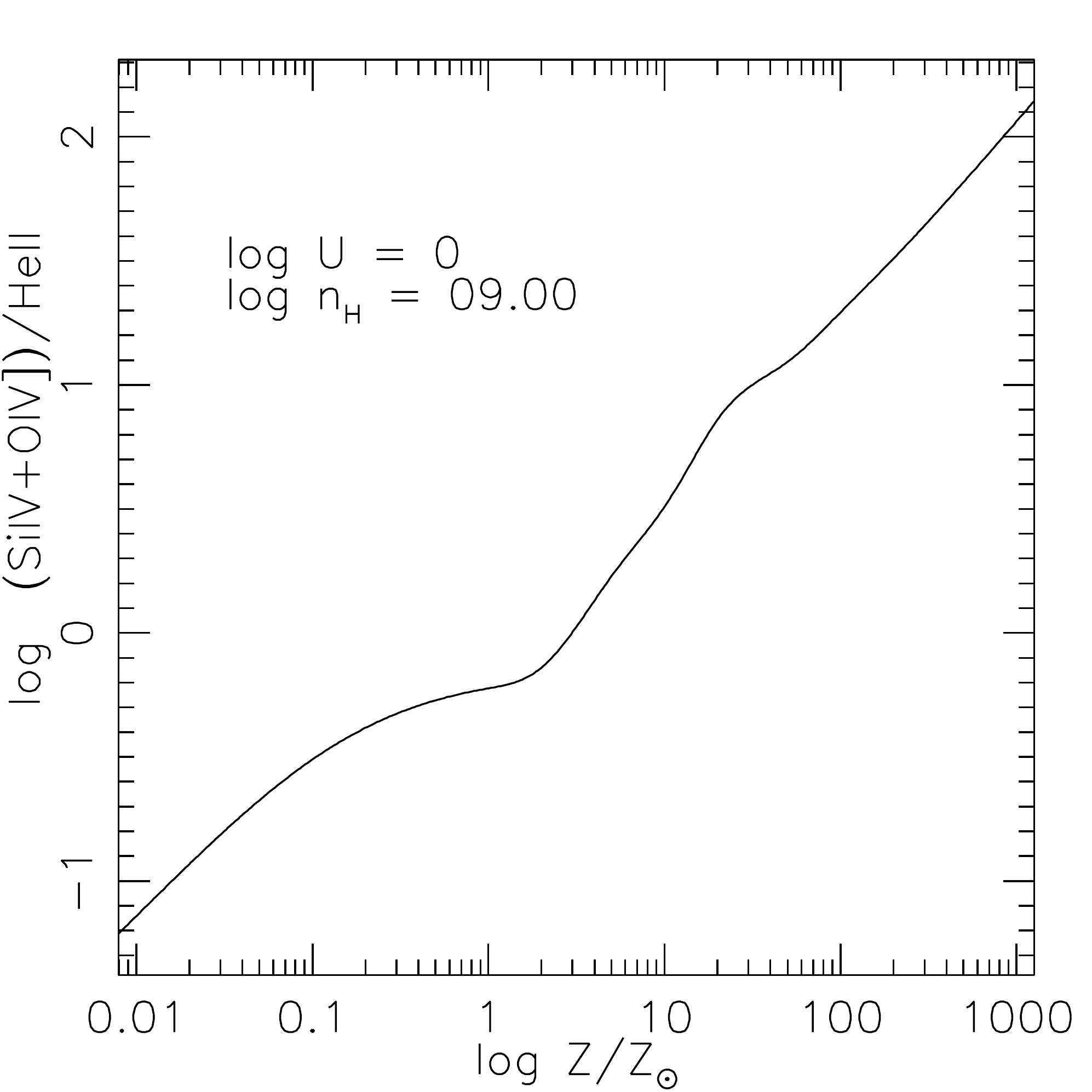}\hspace{-0.2cm}
\includegraphics[scale=0.28]{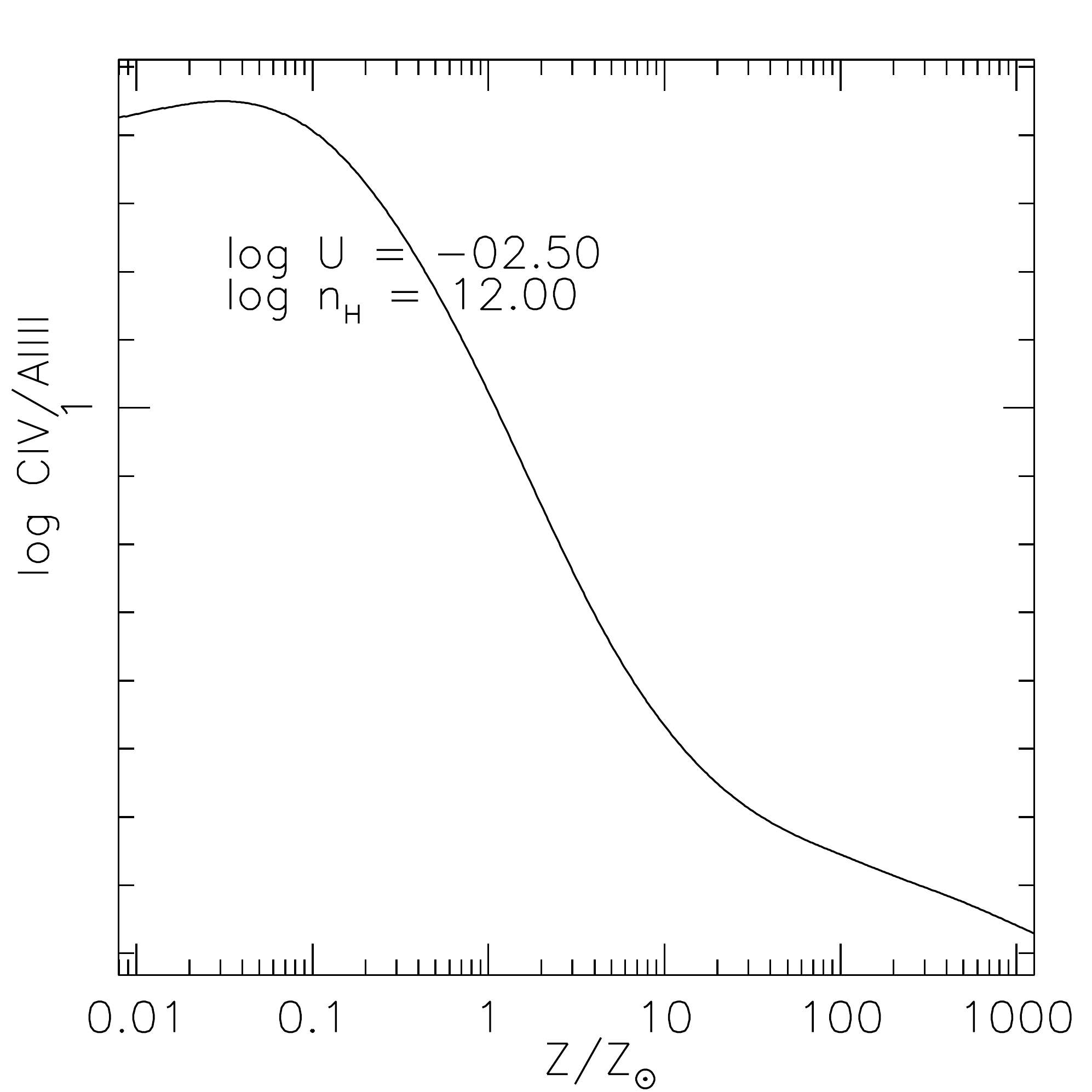} \hspace{-0.2cm}\\
\includegraphics[scale=0.28]{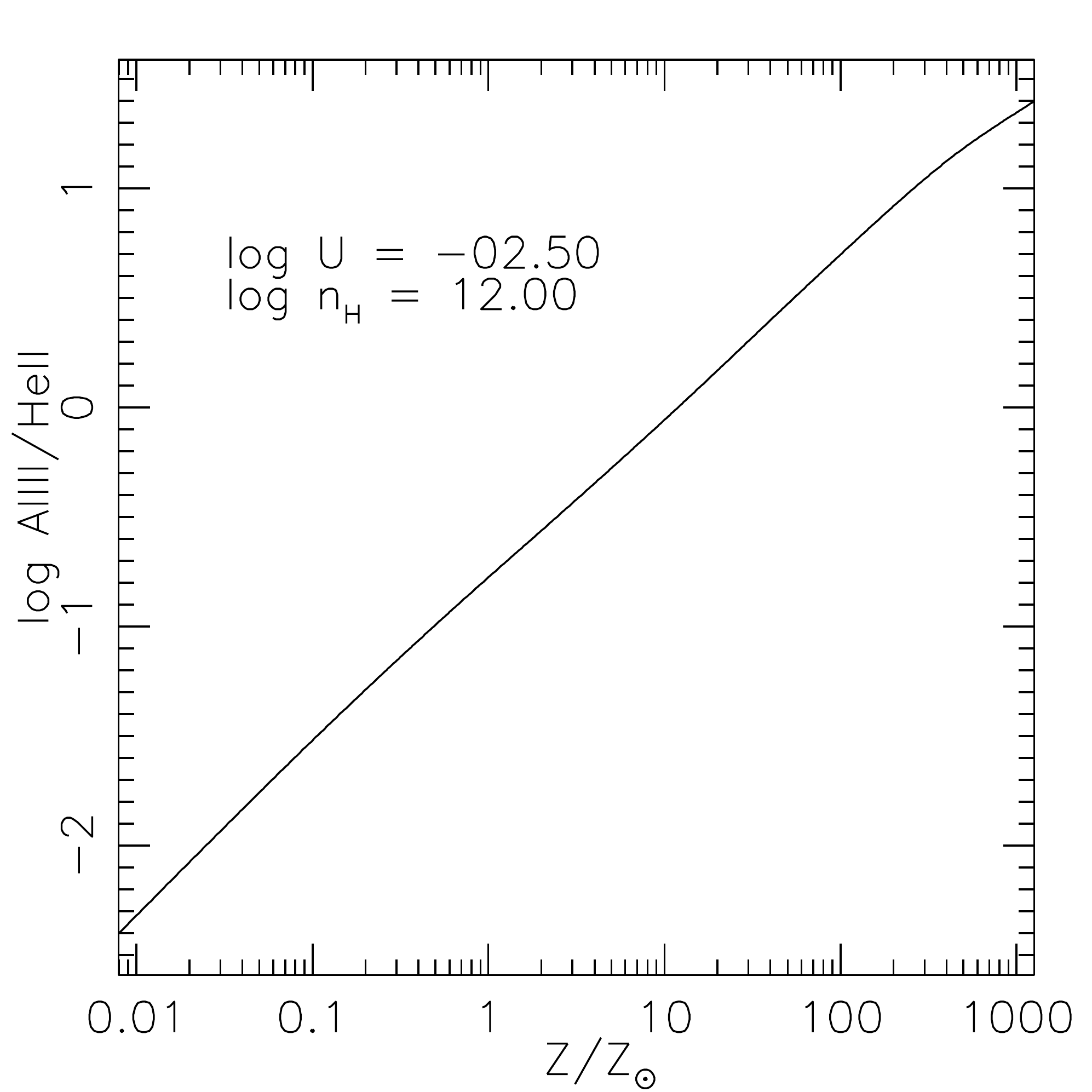}\hspace{-0.2cm}
\includegraphics[scale=0.28]{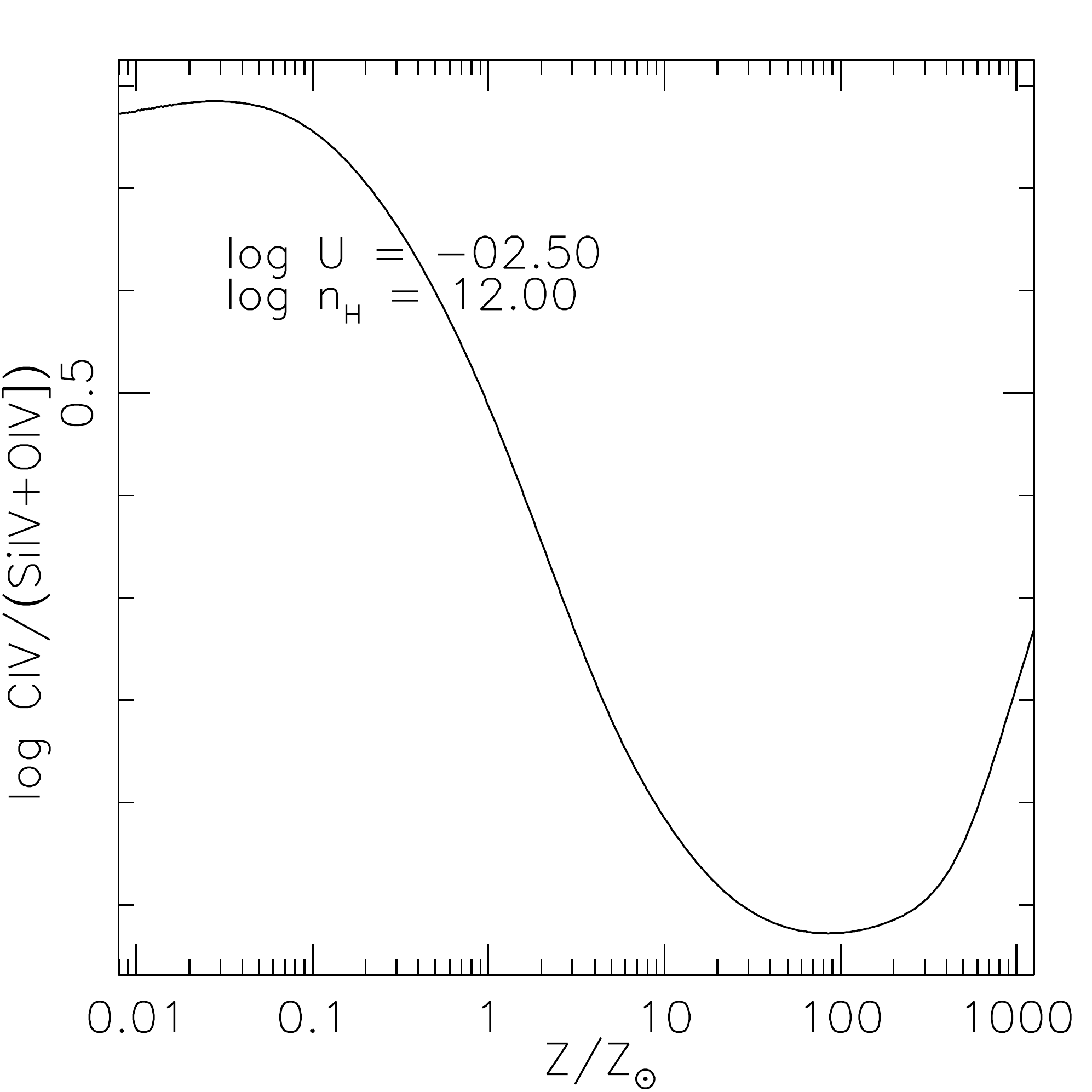}\hspace{-0.2cm}
\includegraphics[scale=0.28]{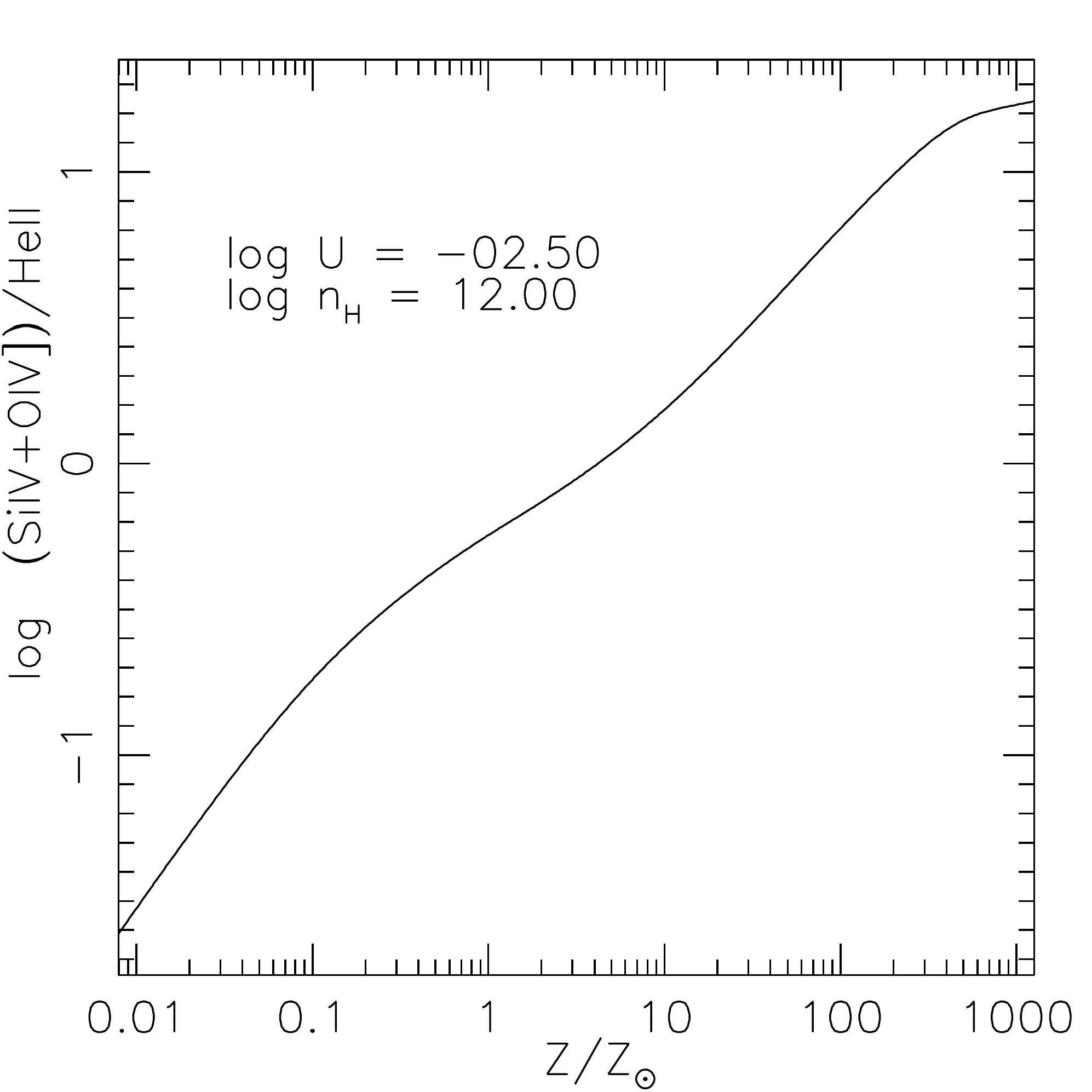}\\
 \caption{Behavior of the intensity ratios employed in this work (with the exception of \civ/\heii\ shown in the previous Figure), as a function of metallicity, for physical parameters $U$\ and \nh\ fixed: 
($\log U$, $\log$ \nh) = (-1,9)   and ($\log U$, $\log$ \nh) = (-2.5,12). Top panels, from left to right: \civ/\siiv+\oiv,(\siiv+\oiv)/\heii, \civ/\aliii. Bottom panes, from left to right: \aliii/\heii, \civ/(\siiv+\oiv), (\siiv+\oiv)/\heii.   \label{fig:fixed}}
  \end{figure*}

Fig. \ref{fig:fixed} shows the behavior of the other intensity ratios used as metallicity diagnostics, for BLUE and BC. \siiv+\oiv/\civ\ and \siiv+\oiv/\heii\ saturate above 100 $Z_{\odot}$. Only around $Z \sim 10 Z_{\odot}$ values \civ/\siiv+\oiv $\lesssim$ 1 are possible, but the behavior is not monotonic and the ratio rises again at $Z \gtrsim 30 Z_\odot$, with the unpleasant consequence that a ratio \civ/\siiv+\oiv $\approx 1.6$ might imply 10 $Z_\odot$ as well as 1000 $Z_\odot$. The ratios usable for the BC also show regular behavior.  The \civ/\aliii\ ratio remains almost constant up $Z \sim 0.1 Z_{\odot}$, and the starts a regular decrease with increasing $Z$, due to the decrease of $T_\mathrm{e}$\ with $Z$\ (\civ\ is affected more strongly than \aliii).  Interestingly, \aliii/\heii\ shows the opposite trend, due to the steady decrease of the \heii\ prominence with $Z$. Especially of interest is however the behavior of  ratio \aliii/\heii\ that shows a monotonic, very linear behavior in the log-log diagram. As for the high-ionization case, values (\siiv+\oiv)/\civ$\gtrsim$ 1 are   possible only at very high metallicity, although the non monotonic behavior (around the minimum at $Z \approx 200 Z_\odot$)  complicates the interpretation of the observed emission line ratios.

\begin{figure*}[ht]
\includegraphics[scale=0.38]{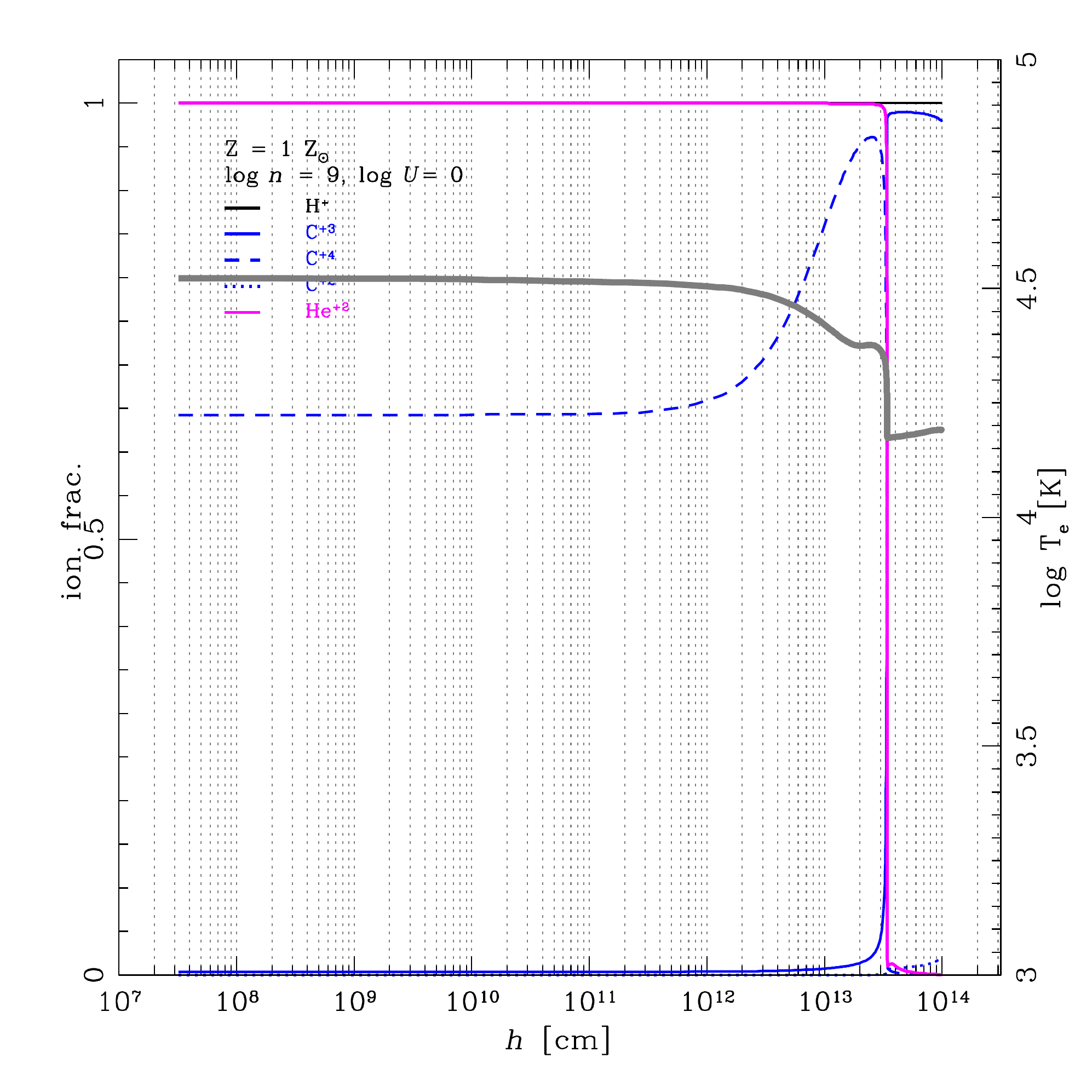}
\includegraphics[scale=0.38]{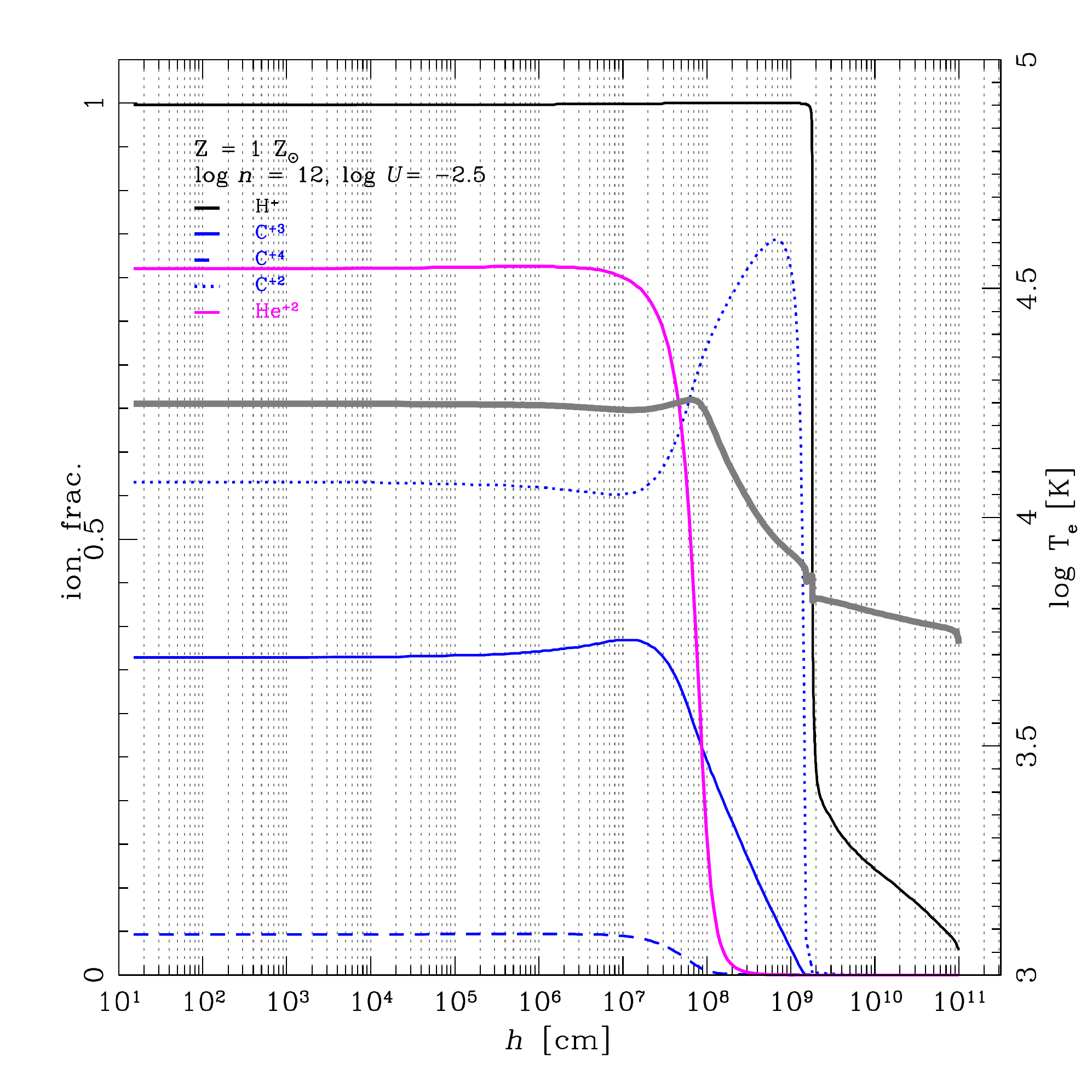}\\
\includegraphics[scale=0.38]{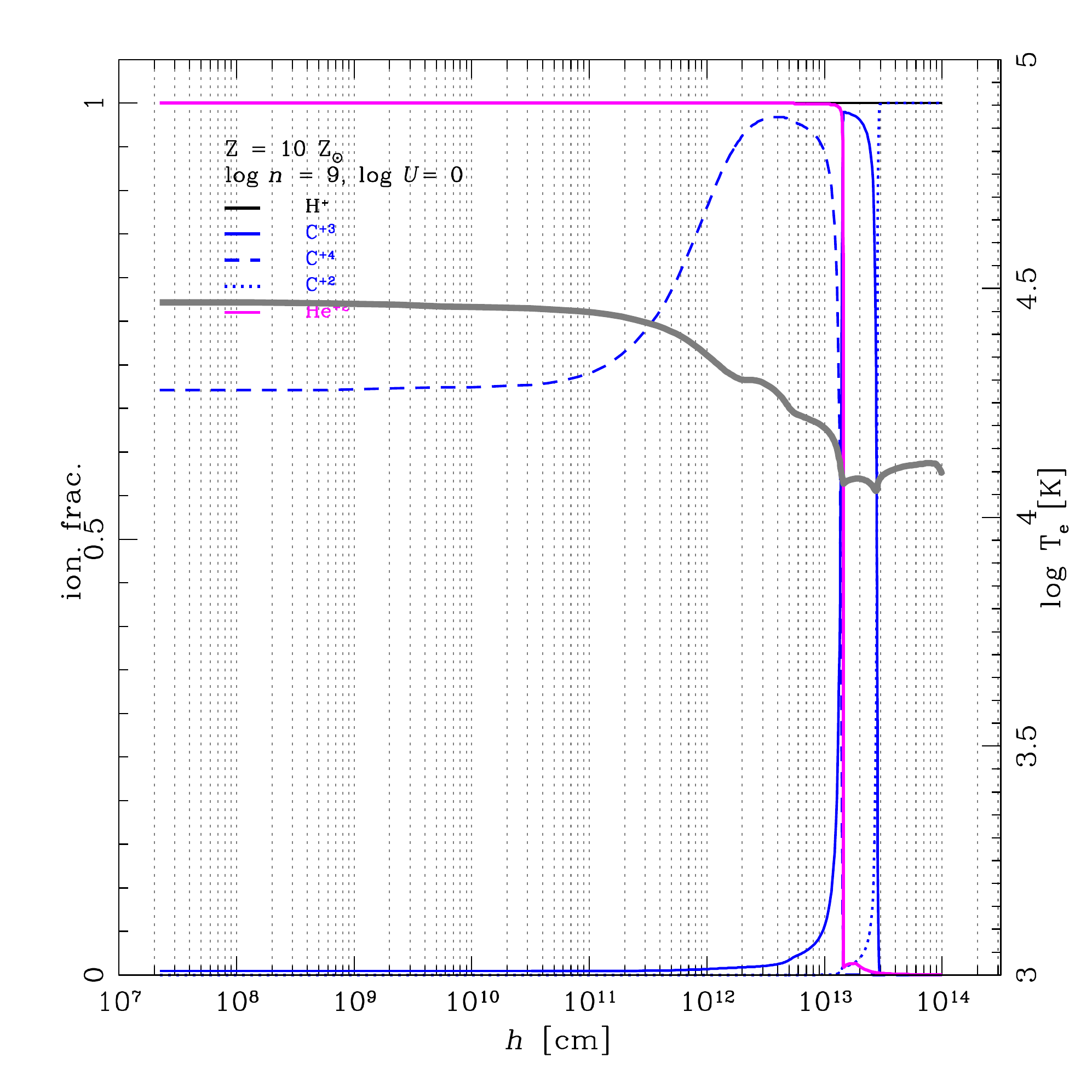}
\includegraphics[scale=0.38]{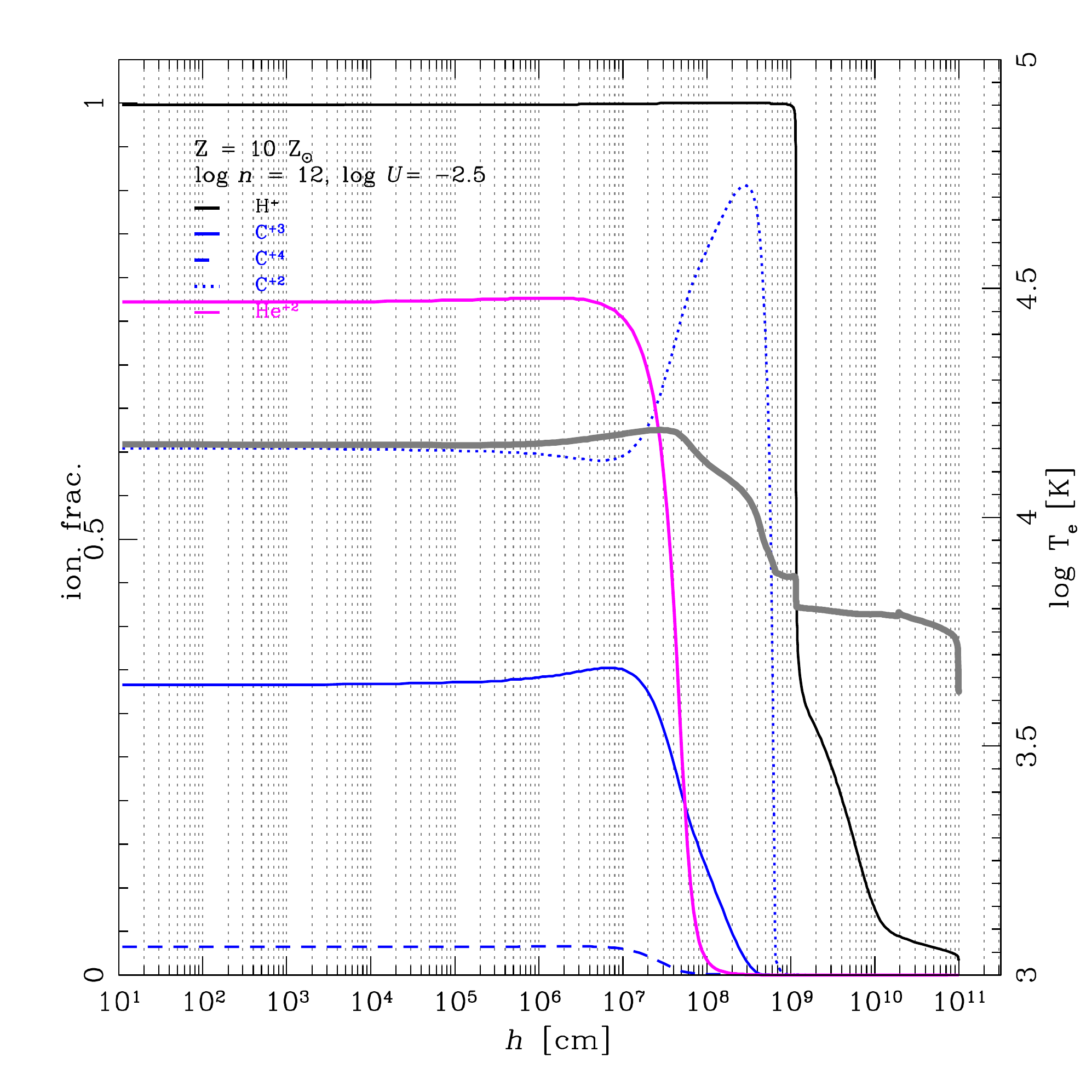}\\
\includegraphics[scale=0.38]{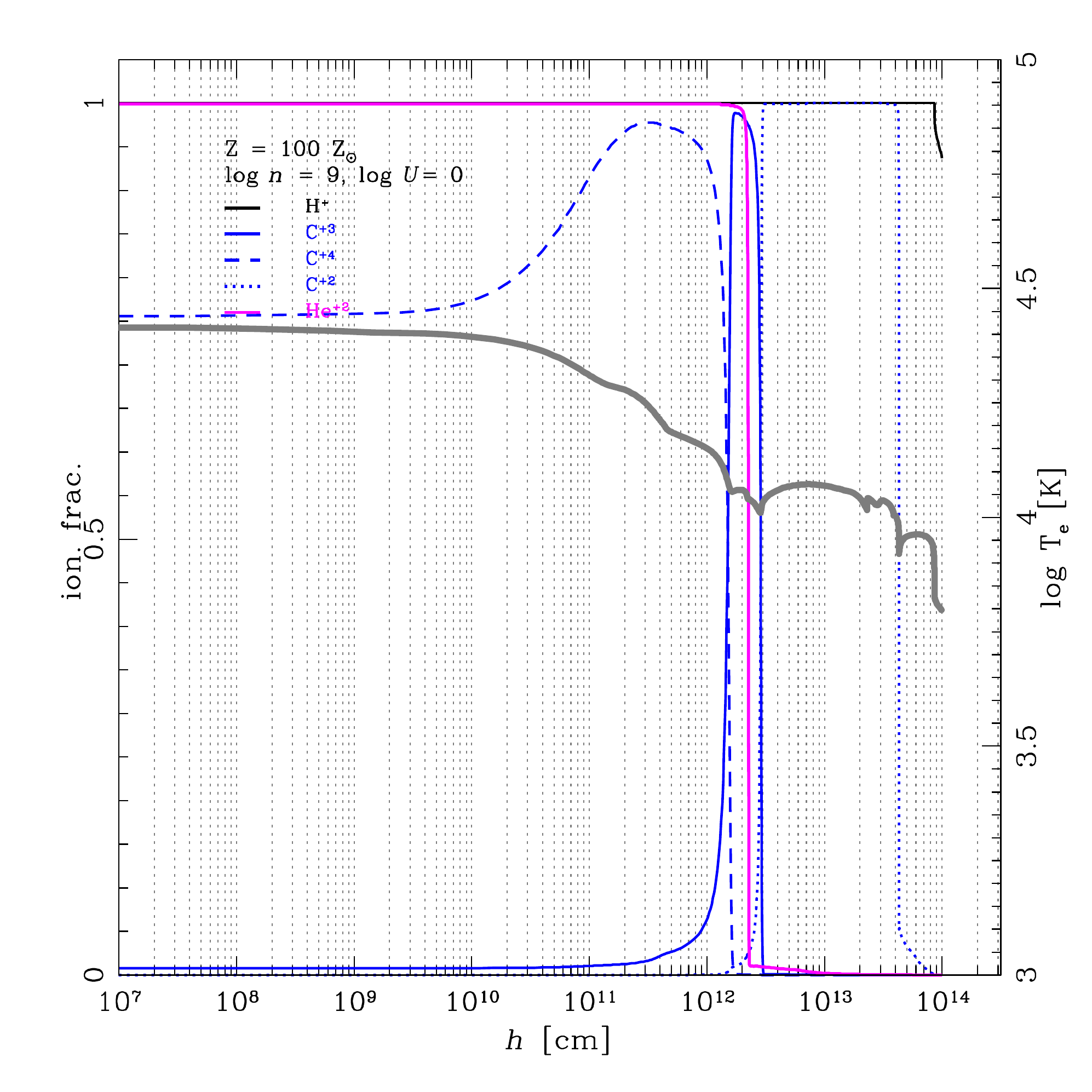}
\includegraphics[scale=0.38]{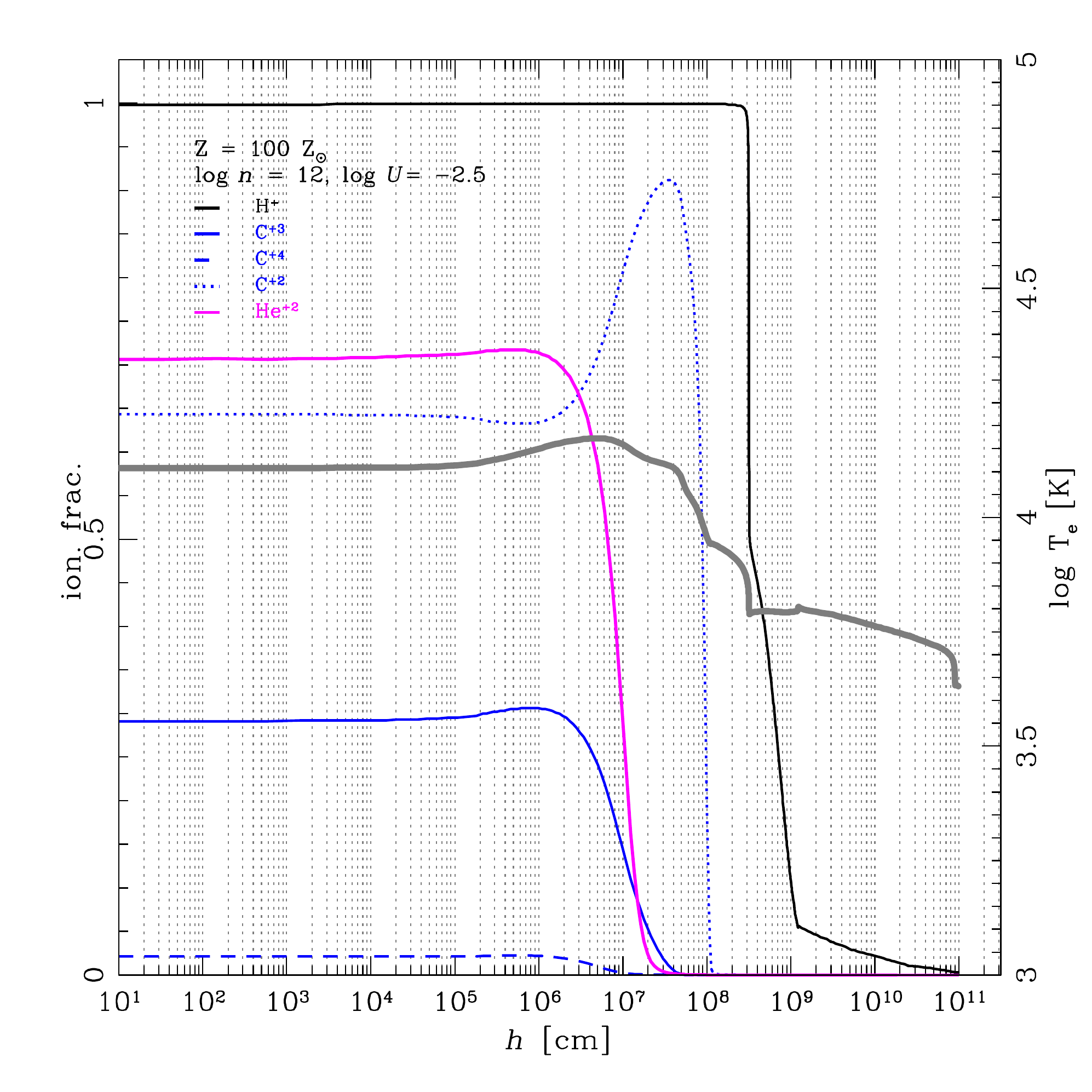}
 \caption{Ionization fraction and electron temperature (thick grey line) as a function of depth within the emitting gas slab, for physical parameters $U$\ and \nh\ fixed,     computed with {\tt CLOUDY 13.05}:  ($\log U$, $\log$ \nh) = (-1,9) (representative of BLUE and high ionization case, left) and ($\log U$, $\log$ \nh) = (-2.5,12) (representative of the low-ionization BLR, right), in order of increasing metallicity from top to bottom. 
 \label{fig:iontz}}
  \end{figure*}

\begin{figure}[ht]
\includegraphics[scale=0.45]{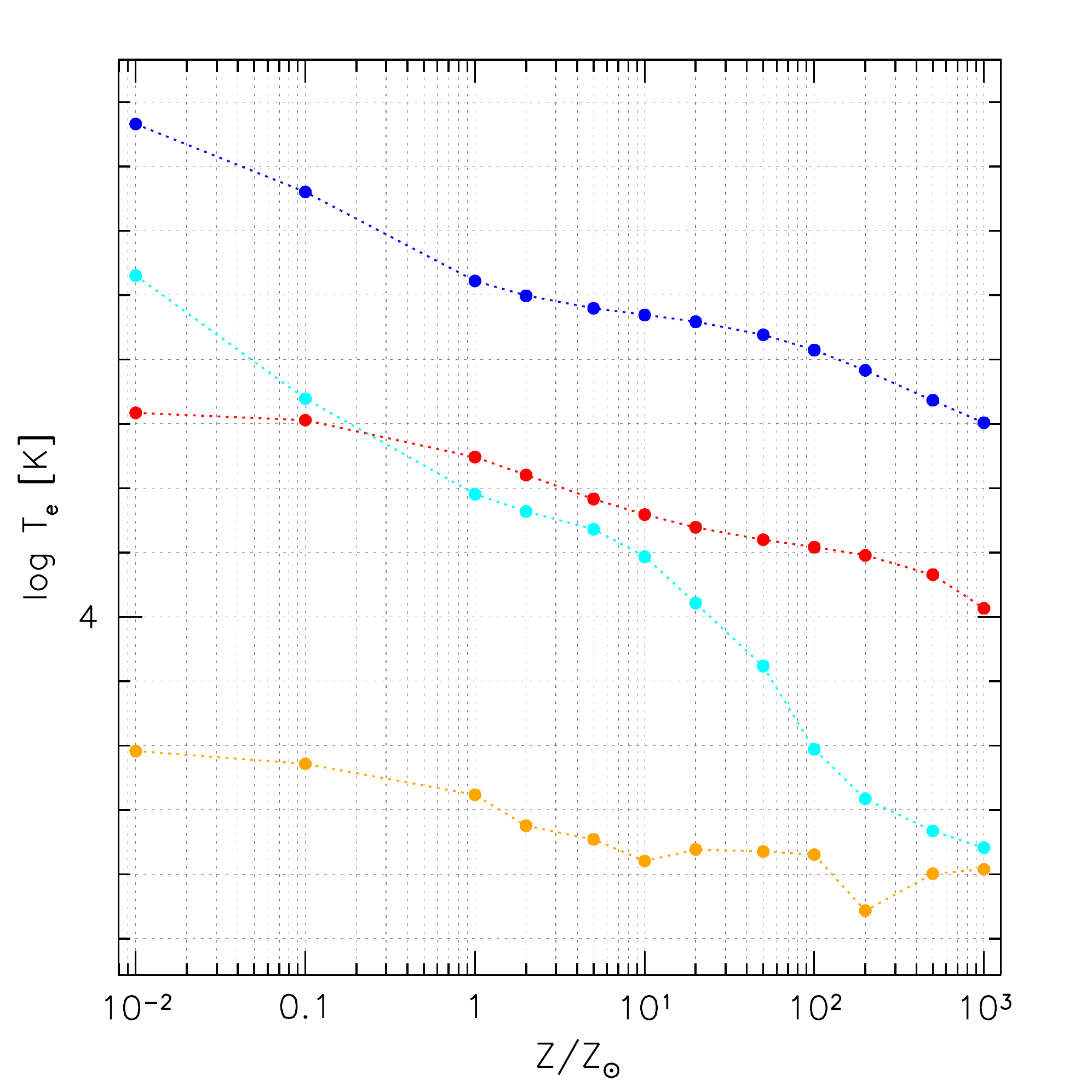}
  \caption{Electron temperature $T_\mathrm{e}$\ as a function of metallicity $Z$,  for physical parameters $U$\ and \nh\ fixed:  ($\log U$, $\log$ \nh) = (-1,9) (representative of BLUE and high ionization case, blue and cyan) and ($\log U$, $\log$ \nh) = (-2.5,12) (representative of the low-ionization BLR, red and orange),   from {\tt CLOUDY 13.05} computations. Blue and red refer to the first zone of the CLOUDY computation i.e., to the illuminated surface of the clouds; cyan and orange, to the side of the cloud farther from the continuum sources i.e., facing the observer.   \label{fig:tz}}
  \end{figure}

The ionization structure within the slab remains self similar over a wide metallicity range, with the same systematic differences between the high and low-ionization case (Fig. \ref{fig:iontz}), consistent with the assumption of a constant ionization parameter.  As expected, the electron temperature decreases with metallicity, and the transition between the fully and partially ionized zone (FIZ and PIZ) occurs at smaller depth. In addition, close to the illuminated side of the cloud the electron temperature remains almost constant;  the gas starts becoming colder before the transition from FIZ to PIZ. The depth at which $T_\mathrm{e}$\ starts decreasing is well-defined, and its value becomes lower with increasing $Z$\  (Fig. \ref{fig:iontz}). The effect is present for both the low- and high- ionization case, although it is more pronounced for the high-ionization. 
Fig. \ref{fig:tz} shows how an increase in metallicity is affecting the $T_\mathrm{e}$\  in the line emitting cloud. Fig. \ref{fig:tz} reports the behavior of $T_\mathrm{e}$\ at the illuminated face of the cloud  ($\tau \sim 0$) and at maximum $\tau$\ (corresponding to \nc = 10$^{23}$ cm$^{-2}$, the side facing the observer) for the high-ionization and low-ionization case. The $T_\mathrm{e}$\  monotonically decreases as a function of metallicity. The difference between the two cloud faces is almost constant for the low ionization case, with $\delta \log T_\mathrm{e} \approx 0.5$\ dex, while it increases for the high-ionization case, reaching $\delta \log T_\mathrm{e} \approx 0.75$\ dex at the highet $Z$\ value considered, 10$^3 Z_\odot$. 


%


\section{Results}
\label{sec:results}

 \subsection{Immediate Results}

The observational results of our analysis involve the measurements of the intensity of the line BC and BLUE component separately. The rest-frame spectra with the continuum placements, and the fits to the blends of the spectra are shown in Appendix \ref{app:spec}. Table \ref{tab:ciii} reports the measurement for the $\lambda$1900 blend. The columns list the SDSS identification code, the FWHM (in units of \kms) and equivalent width and  flux of \aliii\ (the sum of the doublet lines, in units of \AA\ and   {10$^{-14}$} erg s$^{-1}$ cm$^{-2}$, respectively), FWHM and flux of \ciii, and flux of \siiii\ (its FWHM is assumed equal to the one of the single \aliii\ lines.)   with Similarly, Table \ref{tab:civ} reports the parameter of the \civ\ blend:  equivalent width,  FWHM and flux of the \civ\ BC,  the flux of the \civ\ blueshifted component, as well as the fluxes of the BC and BLUE of \heii. FWHM values are reported but  especially values $\gtrsim 5000$ \kms\ should be considered as highly uncertain. There is the concrete possibility of an additional broadening ($\sim$ 10 \%\ of the observed FWHM) associated with non-virial motions for the \aliii\ line (del Olmo et al., in preparation). The fluxes of the BC and of BLUE  of  \siiv\ and \oiv\ are reported in Table \ref{tab:siiv}. Intensity ratios with uncertainties are reported in Table \ref{tab:lineratios}. The last row lists the median values of the ratios with their semi-interquartile ranges (SIQR). 

\subsubsection{Identification of xA sources and of ``intruders''}
\label{intruders}

\begin{figure}
\includegraphics[scale=0.37]{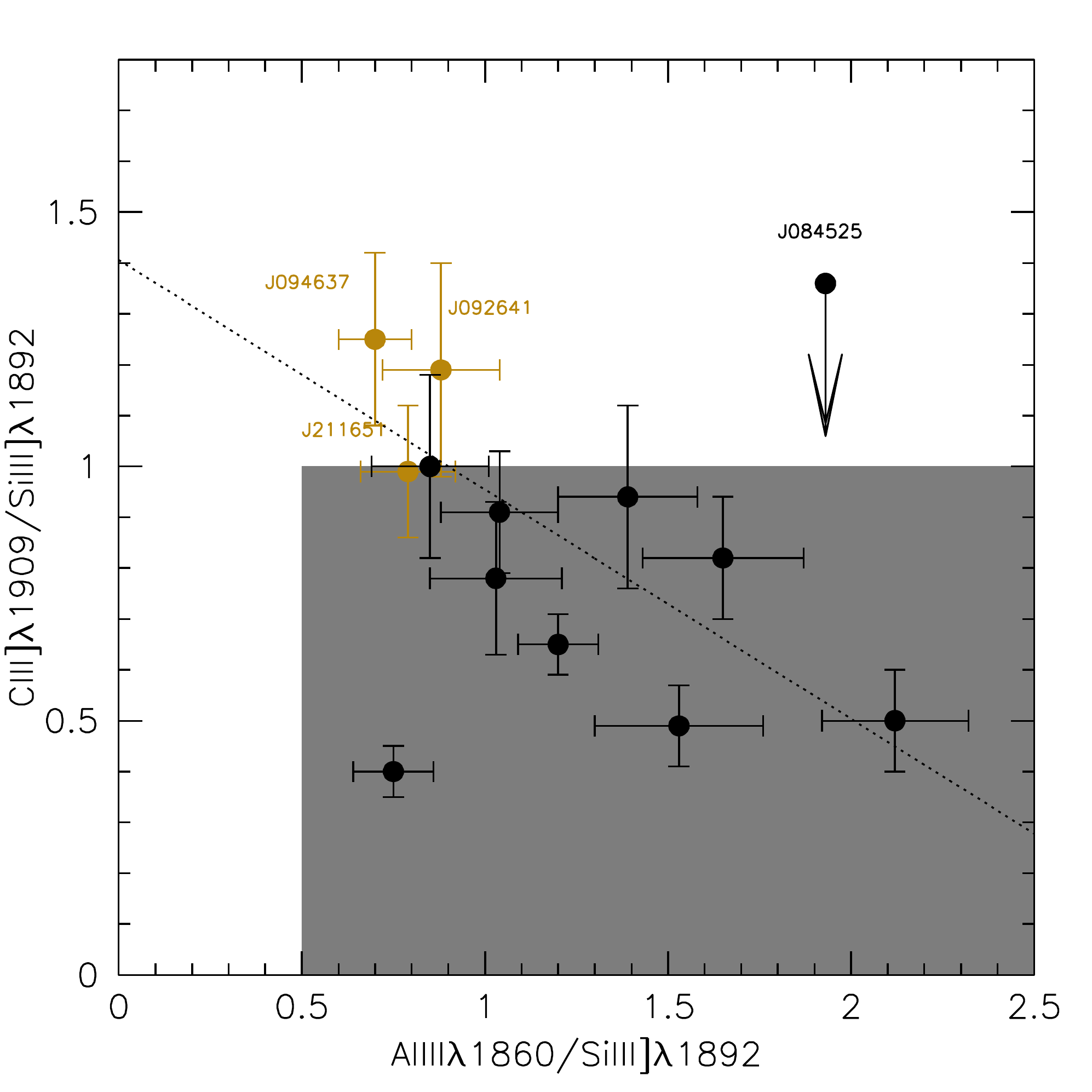}
\caption{Relation between intensity ratios AlIII$\lambda$1860/SiIII]$\lambda$1892 and
CIII]$\lambda$1909/SiIII]$\lambda$1892. The gray area corresponds to the parameter space occupied by the xA sources. Borderline sources are in orange color. 
\label{fig:lb}}
\end{figure}

Figure \ref{fig:lb} shows that the   majority of sources meet both UV selection criteria, and should be considered  xA quasars. The median value of the \aliii/\siiii\ (last row of Table \ref{tab:lineratios}) implies that the \aliii\ is strong relative to \siiii. Also \siiii\ is stronger than \ciii. Both selection criteria are satisfied by the median ratios.  Only one source (\object{SDSS J084525.84+072222.3}) shows \ciii/\siiii\ significantly larger than 1. This quasar is however confirmed as an xA by the very large \aliii/\siiii, by the blueshift of \civ, and by the prominent $\lambda$1400 blend comparable to the \civ\ emission. The lines in the spectrum of \object{SDSS J084525.84+072222.3} are broad, and any \ciii\ emission is heavily blended with \feiii\ emission. The \ciii\ value should be considered an upper limit. Three outlying/borderline data points (in orange) in Fig. \ref{fig:lb}\ have ratio \ciii/\siiii  $\sim 1$, and \aliii/\siiii\ consistent with the selection criteria within the uncertainties, but other criteria support their classification as xA.  The borderline sources will be further discussed in Section \ref{ind}.   In conclusion, all the 13 sources of the present sample save one should be considered bona-fide xA sources.  

It is intriguing that the intensity ratios \ciii/\siiii\ and \aliii/\siiii\ are apparently anti-correlated in Figure \ref{fig:lb}, if we exclude the two outlying points. Excluding the two outlying data point the Spearman rank correlation coefficient  is $\rho \approx 0.8$, which implies a $4\sigma$\ significance for a correlation, but the  correlation coefficient between the two ratios for the full sample is much lower. Given the small number of sources, a larger sample is needed to confirm the trend.   


\subsubsection{BC intensity ratios}

\begin{figure*}[ht]
  \centering
  \hspace{-2cm}
\includegraphics[width=21cm]{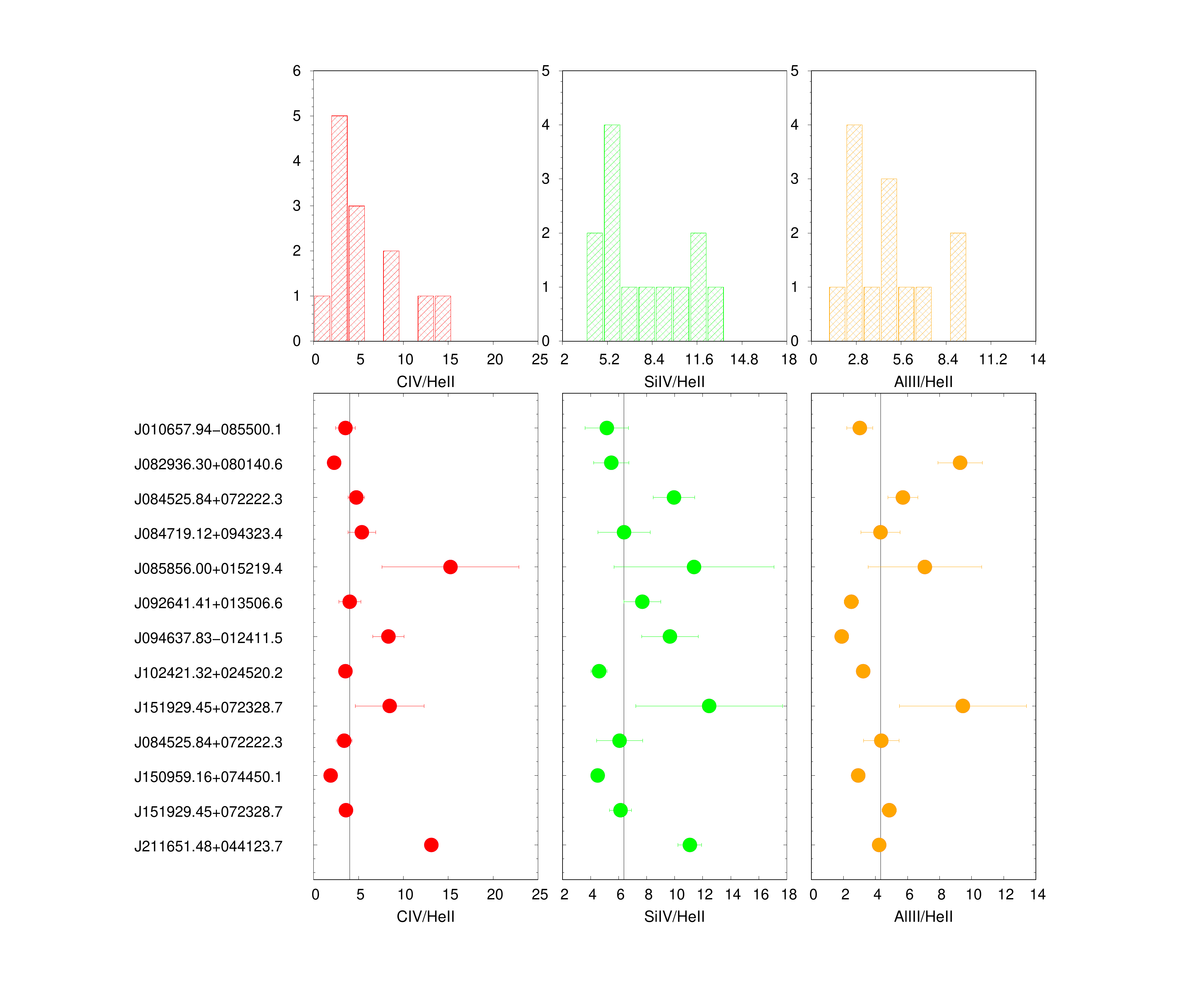}\vspace{-1cm}
\caption{{Distribution of  diagnostic intensity ratios based on the BC (top) for \civ/\heii\ (red), \siiv/\heii\ (green) and \aliii/\heii\ (orange). {The lower panels show results and associated uncertainties for individual sources with the same color-coding as histograms.} Vertical black line in lower panel represents median value of   sources measurements.}\label{fig:broad_histograms}}
\end{figure*}

Fig. \ref{fig:broad_histograms} shows the distributon of  diagnostic intensity ratios  \civ/\heii, \siiv/\heii, and \aliii/\heii\ for the BC. The lower panels of  Fig. \ref{fig:broad_histograms} shows the results for individual sources. 

The vertical lines identify the median values,  $\mu_{\frac{1}{2}}$(\civ/\heii)$ \approx 4.03$,  $\mu_{\frac{1}{2}}$(\aliii/\heii)$\approx$ 4.31, $\mu_{\frac{1}{2}}$(\siiv/\heii) $\approx$ 6.39. The higher value for \siiv/\heii\ than for \civ/\heii\ implies $\mu_{\frac{1}{2}}$(\civ/\siiv) $\approx$ 0.69, a value that is predicted by {\tt CLOUDY} for very low values of the ionization parameter (Appendix \ref{sec:iso}). The \civ/\aliii\ ratio is also constraining: the {\tt CLOUDY} simulations indicate  high $Z$\ and low ionization.   

The distribution of the data points is relatively well-behaved, with individual ratios showing small scatter around their median values. In the histogram, we see a tail made by a 3-5 objects suggesting systematically higher values. In particular,  at least two objects (\object{SDSS J102606.67+011459.0} and \object{SDSS J085856.00+015219.4}) show systematically higher ratios, with   \civ/\heii $\approx 10$, and \aliii/\heii $\approx 4$.  Both of them show extreme \civ\ blueshifts and \object{SDSS J102606.67+011459.0} shows the highest \aliii/\siiii\ ratio in the sample. 


Since the  three ratios are, for fixed physical conditions, proportional to metallicity, we expect an overall consistency in their behavior i.e., if one ratio is  higher than the median for one object, also the other intensity ratios should be also higher.  The lower diagrams are helpful to identify sources for which only one intensity ratio deviates significantly from the rest of the sample. A case in point is \object{SDSS J082936.30+080140.6} whose ratio \aliii/\heii $\approx 8$\ is one of the highest values, but whose \civ/\heii\ and \siiv/\heii\ are slightly below the median values. The fits of Appendix \ref{app:spec} show that this object is indeed extreme in \aliii\ emission. The \civ\ and $\lambda$1400 blends are dominated by the BLUE excess, and an estimate of the \civ\ and \siiv\ BC is very difficult, as it accounts for a small fraction of the line emission. The \heii\ emission is almost undetectable, especially in correspondence to the rest frame.   \object{SDSS J082936.30+080140.6}, along with other sources with high \aliii/\heii\ or \siiv/\heii\ ratios may indicate selective enhancement of Aluminium or Silicon (see also Sect. \ref{pollu}).

\subsubsection{BLUE intensity ratios}

\begin{figure*}[ht]
\centering
\hspace{-4cm}
\includegraphics[width=21cm]{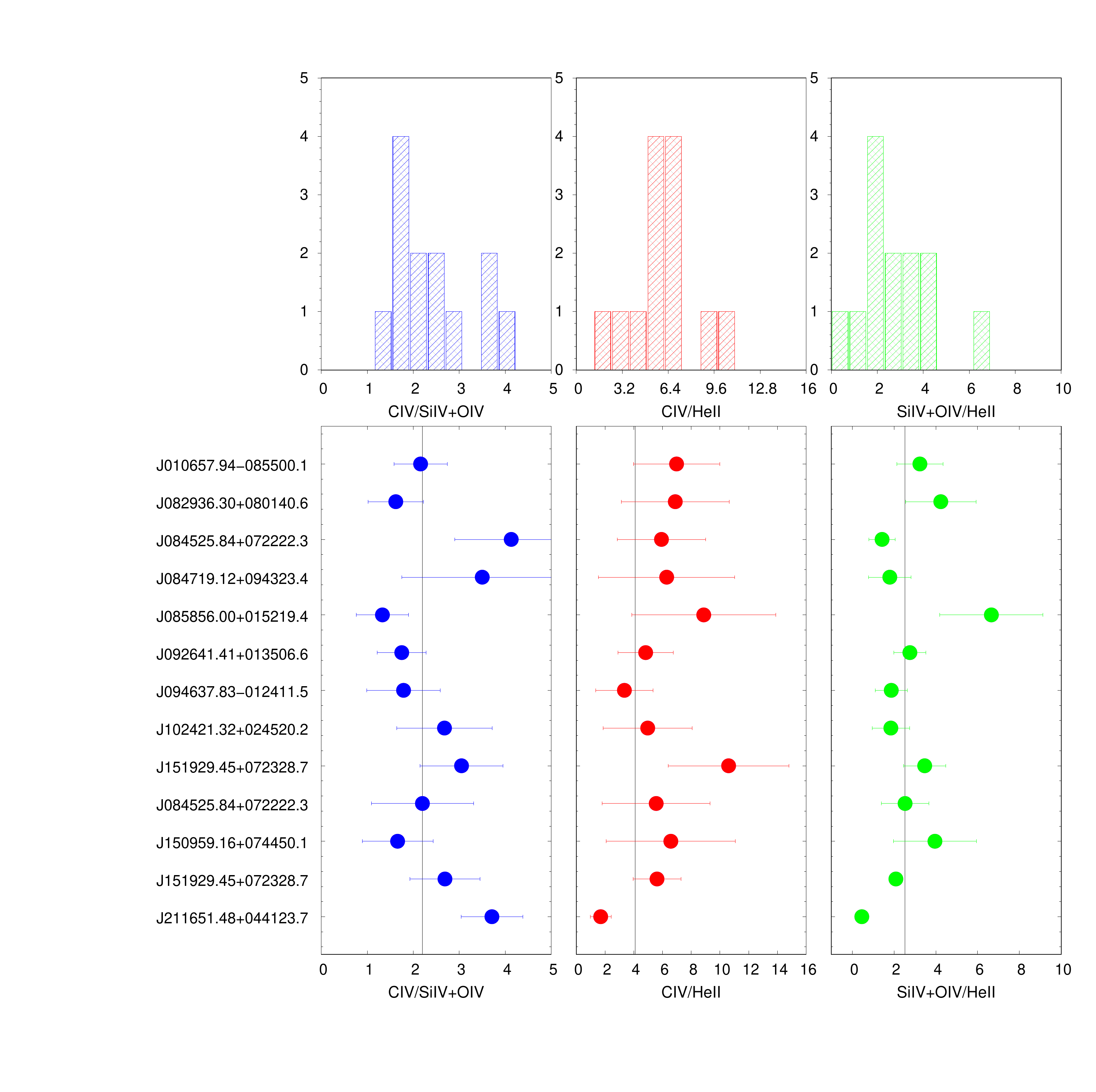}
\caption{Distribution of diagnostic intensity ratios based on the BLUE components (top) for \civ/\siiv+\oiv\ (blue), \civ/\heii\ (red) and \siiv+\oiv/\heii\ (green). {The lower panels show results and associated uncertainties for individual sources with the same color-coding as histograms, as in the previous Figure.} Vertical black line represents median value of ratios measurements.
\label{fig:blue_histograms}}
\end{figure*}

Similar considerations apply to the blue intensity ratios. We see systematic trends in Figure \ref{fig:blue_histograms} that imply consistency of the ratios for most sources, although the uncertainties are larger, especially for \civ/\heii. The ratio \civ/(\siiv + \oiv) values are systematically higher than for the BC, while the \civ/\heii\ is slightly higher (median BLUE 5.8 vs. median BC 4.38). The ratio (\siiv + \oiv)/\heii\ is much lower than for the BC (median BLUE 2.09 vs. median BC 6.27).  The difference might be in part explained by the difficulty of deblending \siiv\ from \oiv, and by the frequent occurrence of absorptions affecting the blue side of the blend. Both factors may conspire to depress BLUE. The lower diagrams of Fig. \ref{fig:blue_histograms} are again helpful to identify sources  for which  intensity ratios deviate significantly from the rest of the sample.
\object{SDSS  J102606.67+011459.0} shows a strong enhancement of \civ/\heii\ and \siiv+\oiv, confirming the trend seen in its BC.



\begin{deluxetable*}{ccccccc}
{\fontsize{6.2}{6}\selectfont\tabcolsep=2pt
\tablecaption{{Measurements in the 1900\AA\ blend region.}}
\tablehead{ \vspace{-0.3cm}
SDSS JCODE   & \colhead{\aliii} & \colhead{\aliii} & \colhead{\aliii} & \colhead{\ciii}  & \colhead{\ciii}  & \colhead{\siiii} \\ \vspace{-0.2cm} 
& $W$ \ & \colhead{FWHM} &  \colhead{Flux} &  \colhead{FWHM} &  \colhead{Flux} &   \colhead{Flux} \\
    (1) & (2) & (3) & (4) & (5) & (6) & (7)}
\startdata
 J010657.94-085500.1 &7.9  & 5560   & 5.41 $\pm$ 0.38 & 6050  & 2.88 $\pm$ 0.14 & 7.2 $\pm$ 0.93  \\
 J082936.30+080140.6 &10.2 & 5710   & 7.43 $\pm$ 0.42 & 5950  & 2.39 $\pm$ 0.21 & 4.85 $\pm$ 0.66 \\
 J084525.84+072222.3 &13.1 & 5510   & 6.04 $\pm$ 0.52 & 5570  & 4.25 $\pm$ 0.53 & 3.13 $\pm$ 0.08 \\
 J084719.12+094323.4 &9.9  & 5410   & 9.24 $\pm$ 0.24 & 5630  & 4.63 $\pm$ 0.36 & 5.61 $\pm$ 0.73 \\
 J085856.00+015219.4 &7.7  & 5520   & 4.57 $\pm$ 0.37 & 5660  & 3.98 $\pm$ 0.12 & 4.39 $\pm$ 0.58 \\
 J092641.41+013506.6 &8.0  & 5550   & 4.82 $\pm$ 0.3  & 5720  & 6.53 $\pm$ 0.23 & 5.48 $\pm$ 0.96 \\
 J094637.83-012411.5 &5.0  & 2730   & 3.54 $\pm$ 0.24 & 2090  & 6.38 $\pm$ 0.22 & 5.1 $\pm$ 0.67  \\
 J102421.32+024520.2 &10.1 & 5520   & 6.15 $\pm$ 0.31 & 6080  & 3.31 $\pm$ 0.2  & 5.1 $\pm$ 0.39  \\
 J102606.67+011459.0 &9.5  & 5590   & 7.7  $\pm$ 0.35  & 5470  & 1.83 $\pm$ 0.33& 3.64 $\pm$ 0.3  \\
 J114557.84+080029.0 &11.4 & 5520   & 8.74 $\pm$ 0.38 & 6060  & 5.91 $\pm$ 0.83 &  6.3 $\pm$ 0.81  \\
 J150959.16+074450.1 &11.8 & 5530   & 6.44 $\pm$ 0.53 & 6090  & 7.62 $\pm$ 0.25 & 7.58 $\pm$ 1.32 \\ 
 J151929.45+072328.7 &10.5 & 5320   & 7.37 $\pm$ 0.51 & 5310  & 5.6  $\pm$ 0.62 & 7.16 $\pm$ 1.12 \\
 J211651.48+044123.7 &6.1  & 5550   & 4.21 $\pm$ 0.4  & 5620  & 5.29 $\pm$ 0.2  & 5.35 $\pm$ 0.7  \\
 \enddata
\tablecomments{Columns are as follows: (1) SDSS name, (2) and (3) report the FWHM of \aliii\ and \ciii\ in  km s$^{-1}$; (4), (5) and (6) list the fluxes in units of {10$^{-14}$} erg s$^{-1}$cm$^{-2}$\ for \aliii, \ciii, and \siiii. }
\label{tab:ciii}}
\end{deluxetable*}

\begin{table*}[]
\centering
\caption{Measurements in the \civ\ spectral region}
\label{tab:civ}
\resizebox{0.65\textwidth}{!}{%
\hspace*{-2cm}\begin{tabular}{lcccccc}
\hline\hline\noalign{\vskip 0.1cm}
\noalign{\vskip 0.1cm}
\multicolumn{1}{c}{SDSS JCODE} & \civ\ & \civ\ BC & \civ\ BC & \civ\ BLUE & \heiiuv\ BC & \heiiuv\ BLUE \\
\multicolumn{1}{c}{} & \textit{W} & FWHM & Flux & Flux & Flux & Flux \\
\multicolumn{1}{c}{(1)} & (2) & (3) & (4) & (5) & (6) & (7) \\
\hline \noalign{\vskip 0.1cm}
J010657.94-085500.1&          18.6 & 5530     & 6.35 $\pm$ 1.04  & 13.46 $\pm$ 0.13 & 1.79 $\pm$ 0.46 & 1.93 $\pm$ 0.65 \\
J082936.30+080140.6&          15.5 & 3710 $\pm$ 670 & 1.83 $\pm$ 0.4   & 11.86 $\pm$ 0.03 & 0.8 $\pm$ 0.11  & 1.72 $\pm$ 0.69 \\
J084525.84+072222.3&           16.8 &3760    & 5.04 $\pm$ 0.68  & 11.11 $\pm$ 0.21 & 1.06 $\pm$ 0.15 & 1.87 $\pm$ 0.81 \\
J084719.12+094323.4&          17.6 & 5520     & 11.53 $\pm$ 0.53 & 13.24 $\pm$ 0.07 & 2.14 $\pm$ 0.61 & 2.1 $\pm$ 1.19  \\
J085856.00+015219.4&           22.8 & 5460     & 9.84 $\pm$ 0.82  & 11.39 $\pm$ 0.09 & 0.65 $\pm$ 0.32 & 1.28 $\pm$ 0.48 \\
J092641.41+013506.6&          25.5 & 5550   & 7.79 $\pm$ 1.98  & 7.08 $\pm$ 0.08  & 1.94 $\pm$ 0.33 & 1.47 $\pm$ 0.39 \\
J094637.83-012411.5&          23.6 & 3670    & 15.63 $\pm$ 0.6  & 5.89 $\pm$ 0.09  & 1.87 $\pm$ 0.39 & 1.76 $\pm$ 0.7  \\
J102421.32+024520.2&          20.1 & 5640    & 6.74 $\pm$ 0.45  & 10.9 $\pm$ 0.08  & 1.9 $\pm$ 0.19  & 2.19 $\pm$ 1.07 \\
J102606.67+011459.0&          17.3 & 3700 $\pm$ 650 & 6.92 $\pm$ 1.22  & 12.23 $\pm$ 0.05 & 0.82 $\pm$ 0.34 & 1.15 $\pm$ 0.32 \\
J114557.84+080029.0&           18.4 & 3500 $\pm$ 700 & 6.84 $\pm$ 0.3   & 12.57 $\pm$ 0.03 & 2.01 $\pm$ 0.5  & 2.26 $\pm$ 1.02 \\
J150959.16+074450.1&           16.8 & 3530 $\pm$ 690 & 4.19 $\pm$ 0.84  & 9.73 $\pm$ 0.22  & 2.21 $\pm$ 0.14 & 1.48 $\pm$ 0.74\\     
J151929.45+072328.7&          19.4 & 3470 $\pm$ 590 & 5.47 $\pm$ 0.41  & 12.04 $\pm$ 0.16 & 1.52 $\pm$ 0.09 & 2.14 $\pm$ 0.24 \\
J211651.48+044123.7&          19.1 & 4750  & 13.04 $\pm$ 0.68 & 3.01 $\pm$ 0.24  & 0.99 $\pm$ 0.03 & 1.77 $\pm$ 0.69\\ \hline  
\end{tabular}%
}
\label{tab:civ}
{
\begin{minipage}{0.65\textwidth}
{\sc Note.} Columns are as follows: (1) SDSS name, (2) rest-frame equivalent width of the total \civ\ emission i.e., \civ\ BLUE+BC, in \AA; (3)  the FWHM of the \civ\ line in  km s$^{-1}$; Cols. (4) and (5) list fluxes of the \civ\ BC and BLUE line; Cols. (6) and (7) report fluxes of the BC and BLUE components for the \heii\ line. All fluxed are in units of {10$^{-14}$} erg s$^{-1}$cm$^{-2}$.
\end{minipage}    

}
\end{table*}

\begin{table*}[]
\small
\centering
\caption{Measurements in the $\lambda1400$ region. }
\label{tab:siiv}
\resizebox{0.5\textwidth}{!}{%
\hspace*{-2cm}\begin{tabular}{lccc}
\hline\hline\noalign{\vskip 0.1cm}
\noalign{\vskip 0.1cm}
\multicolumn{1}{c}{SDSS JCODE} & \siiv+\oiv\ BC & \siiv+\oiv\ BC & \siiv+\oiv\ BLUE \\
 & FWHM & Flux & Flux \\
\multicolumn{1}{c}{(1)} & (2) & (3) & (4) \\
\hline \noalign{\vskip 0.1cm}
J010657.94-085500.1 & 5070  & 9.23 $\pm$ 1.4   & 6.24 $\pm$ 0.24 \\
J082936.30+080140.6 & 5560  & 4.39 $\pm$ 0.81  & 7.3 $\pm$ 0.06  \\
J084525.84+072222.3 & 5060  & 10.52 $\pm$ 0.58 & 2.69 $\pm$ 0.24 \\
J084719.12+094323.4 & 5550  & 13.69 $\pm$ 0.86 & 3.79 $\pm$ 0.19 \\
J085856.00+015219.4 & 6960  & 7.34 $\pm$ 0.69  & 8.55 $\pm$ 0.05 \\
J092641.41+013506.6 & 5540  & 14.89 $\pm$ 0.45 & 4.05 $\pm$ 0.32 \\
J094637.83-012411.5 & 4030  & 18.09 $\pm$ 0.7  & 3.29 $\pm$ 0.31 \\
J102421.32+024520.2 & 5530  & 8.77 $\pm$ 0.64  & 4.06 $\pm$ 0.09 \\
J102606.67+011459.0 & 5300  & 10.16 $\pm$ 0.57 & 4.01 $\pm$ 0.33 \\
J114557.84+080029.0 & 3760 & 12.19 $\pm$ 1.22 & 5.71 $\pm$ 0.08 \\
J150959.16+074450.1 & 3650 & 9.97 $\pm$ 0.75  & 5.86 $\pm$ 0.12 \\
J151929.45+072328.7 & 3670 & 9.32 $\pm$ 1.05  & 4.48 $\pm$ 0.31 \\
J211651.48+044123.7 & 4770 & 11.02 $\pm$ 0.77 & 0.81 $\pm$ 0.03 \\ \hline\\
\end{tabular}%
}
{
\begin{minipage}{0.55\textwidth}
{\sc Note.} Columns are as follows: (1) SDSS name, (2)  the FWHM of the \siiv\ line in  km s$^{-1}$. (3) and (4) list fluxes of the broad components and the blue component line in units of  {10$^{-14}$} ergs$^{-1}$cm$^{-2}$.
\end{minipage}
}
\end{table*}


\begin{table*}[]
\centering
\caption{Intensity ratios for the BC and BLUE line components}
\label{tab:lineratios}
\resizebox{\textwidth}{!}{%
\hspace*{-2cm}\begin{tabular}{lcccccccccc}
\hline\hline\noalign{\vskip 0.1cm}
\noalign{\vskip 0.1cm}
\multicolumn{1}{c}{SDSS JCODE} & \aliii/\siiii & \ciii/\siiii & \civ/\siiv & \multicolumn{1}{c}{\civ/\heii} & \multicolumn{1}{c}{\siiv/\heii} & \multicolumn{1}{c}{\civ/\aliii} & \multicolumn{1}{c}{\aliii/\heii} & \multicolumn{1}{c}{\civ/\heii} & \multicolumn{1}{c}{\civ/\siiv+ \oiv} & \multicolumn{1}{c}{\siiv+ \oiv/ \heii} \\
 & (BC) & (BC) & (BC) & \multicolumn{1}{c}{(BC)} & \multicolumn{1}{c}{(BC)} & \multicolumn{1}{c}{(BC)} & \multicolumn{1}{c}{(BC)} & \multicolumn{1}{c}{(BLUE)} & \multicolumn{1}{c}{(BLUE)} & \multicolumn{1}{c}{(BLUE)} \\
\multicolumn{1}{c}{(1)} & (2) & (3) & (4) & \multicolumn{1}{c}{(5)} & \multicolumn{1}{c}{(6)} & \multicolumn{1}{c}{(7)} & \multicolumn{1}{c}{(8)} & \multicolumn{1}{c}{(9)} & \multicolumn{1}{c}{(10)} & \multicolumn{1}{c}{(11)} \\
\hline \noalign{\vskip 0.1cm}
J010657.94-085500.1 & 0.75 $\pm$ 0.11 & 0.4 $\pm$ 0.05  & 0.69 $\pm$ 0.15 & 3.55 $\pm$ 1.09  & 5.16 $\pm$ 1.55  & 1.17 $\pm$ 0.21 & 3.02 $\pm$ 0.81 & 6.98 $\pm$ 3    & 2.16 $\pm$ 0.58  & 3.24 $\pm$ 1.1  \\
J082936.30+080140.6 & 1.53 $\pm$ 0.23 & 0.49 $\pm$ 0.08 & 0.42 $\pm$ 0.12 & 2.29 $\pm$ 0.6   & 5.48 $\pm$ 1.26  & 0.25 $\pm$ 0.06 & 9.28 $\pm$ 1.39 & 6.89 $\pm$ 3.75 & 1.62 $\pm$ 0.6   & 4.24 $\pm$ 1.69 \\
J084525.84+072222.3 & 1.93 $\pm$ 0.17 & 1.36 $\pm$ 0.17 & 0.48 $\pm$ 0.07 & 4.76 $\pm$ 0.92  & 9.95 $\pm$ 1.48  & 0.83 $\pm$ 0.13 & 5.71 $\pm$ 0.93 & 5.93 $\pm$ 3.07 & 4.13 $\pm$ 1.23  & 1.43 $\pm$ 0.63 \\
J084719.12+094323.4 & 1.65 $\pm$ 0.22 & 0.82 $\pm$ 0.12 & 0.84 $\pm$ 0.07 & 5.38 $\pm$ 1.55  & 6.39 $\pm$ 1.86  & 1.25 $\pm$ 0.07 & 4.31 $\pm$ 1.23 & 6.29 $\pm$ 4.74 & 3.5 $\pm$ 1.75   & 1.8 $\pm$ 1.02  \\
J085856.00+015219.4 & 1.04 $\pm$ 0.16 & 0.91 $\pm$ 0.12 & 1.34 $\pm$ 0.17 & 15.25 $\pm$ 7.63 & 11.38 $\pm$ 5.71 & 2.15 $\pm$ 0.25 & 7.08 $\pm$ 3.54 & 8.87 $\pm$ 5.01 & 1.33 $\pm$ 0.57  & 6.66 $\pm$ 2.47 \\
J092641.41+013506.6 & 0.88 $\pm$ 0.16 & 1.19 $\pm$ 0.21 & 0.52 $\pm$ 0.13 & 4.03 $\pm$ 1.23  & 7.69 $\pm$ 1.31  & 1.62 $\pm$ 0.42 & 2.49 $\pm$ 0.45 & 4.83 $\pm$ 1.92 & 1.75 $\pm$ 0.53  & 2.76 $\pm$ 0.77 \\
J094637.83-012411.5 & 0.7 $\pm$ 0.1   & 1.25 $\pm$ 0.17 & 0.86 $\pm$ 0.05 & 8.34 $\pm$ 1.74  & 9.66 $\pm$ 2.02  & 4.41 $\pm$ 0.35 & 1.89 $\pm$ 0.41 & 3.35 $\pm$ 1.99 & 1.79 $\pm$ 0.8   & 1.87 $\pm$ 0.77 \\
J102421.32+024520.2 & 1.2 $\pm$ 0.11  & 0.65 $\pm$ 0.06 & 0.77 $\pm$ 0.08 & 3.55 $\pm$ 0.43  & 4.61 $\pm$ 0.58  & 1.1 $\pm$ 0.09  & 3.23 $\pm$ 0.37 & 4.97 $\pm$ 3.09 & 2.68 $\pm$ 1.04  & 1.85 $\pm$ 0.9  \\
J102606.67+011459.0 & 2.12 $\pm$ 0.2  & 0.5 $\pm$ 0.1   & 0.68 $\pm$ 0.13 & 8.48 $\pm$ 3.83  & 12.46 $\pm$ 5.23 & 0.9 $\pm$ 0.16  & 9.45 $\pm$ 3.96 & 10.6 $\pm$ 4.21 & 3.05 $\pm$ 0.9   & 3.47 $\pm$ 1.01 \\
J114557.84+080029.0 & 1.39 $\pm$ 0.19 & 0.94 $\pm$ 0.18 & 0.56 $\pm$ 0.06 & 3.41 $\pm$ 0.87  & 6.07 $\pm$ 1.64  & 0.78 $\pm$ 0.05 & 4.36 $\pm$ 1.11 & 5.56 $\pm$ 3.76 & 2.2 $\pm$ 1.11   & 2.53 $\pm$ 1.14 \\
J150959.16+074450.1 & 0.85 $\pm$ 0.16 & 1 $\pm$ 0.18    & 0.42 $\pm$ 0.09 & 1.9 $\pm$ 0.4    & 4.51 $\pm$ 0.44  & 0.65 $\pm$ 0.14 & 2.92 $\pm$ 0.3  & 6.58 $\pm$ 4.49 & 1.66 $\pm$ 0.77  & 3.96 $\pm$ 1.98 \\
J151929.45+072328.7 & 1.03 $\pm$ 0.18 & 0.78 $\pm$ 0.15 & 0.59 $\pm$ 0.08 & 3.61 $\pm$ 0.35  & 6.14 $\pm$ 0.79  & 0.74 $\pm$ 0.08 & 4.86 $\pm$ 0.45 & 5.62 $\pm$ 1.66 & 2.69 $\pm$ 0.76  & 2.09 $\pm$ 0.28 \\
J211651.48+044123.7 & 0.79 $\pm$ 0.13 & 0.99 $\pm$ 0.13 & 1.18 $\pm$ 0.1  & 13.12 $\pm$ 0.78 & 11.08 $\pm$ 0.84 & 3.1 $\pm$ 0.34  & 4.23 $\pm$ 0.42 & 1.7 $\pm$ 0.73  & 3.71 $\pm$ 0.67  & 0.46 $\pm$ 0.18 \\ \hline
Median              & 1.04 $\pm$ 0.34 & 0.91 $\pm$ 0.17 & 0.68 $\pm$ 0.16 & 4.03  $\pm$ 2.395  & 6.39   $\pm$ 2.23 & 1.10 $\pm$ 0.42 & 4.31 $\pm$ 1.35 & 5.93 $\pm$  0.96 & 2.2 $\pm$ 0.65  & 2.53 $\pm$ 0.81 \\ \hline
\end{tabular}%
}
\end{table*}

\subsubsection{Correlation between diagnostic ratios}

\begin{figure}
     \centering
     \includegraphics[scale=0.25]{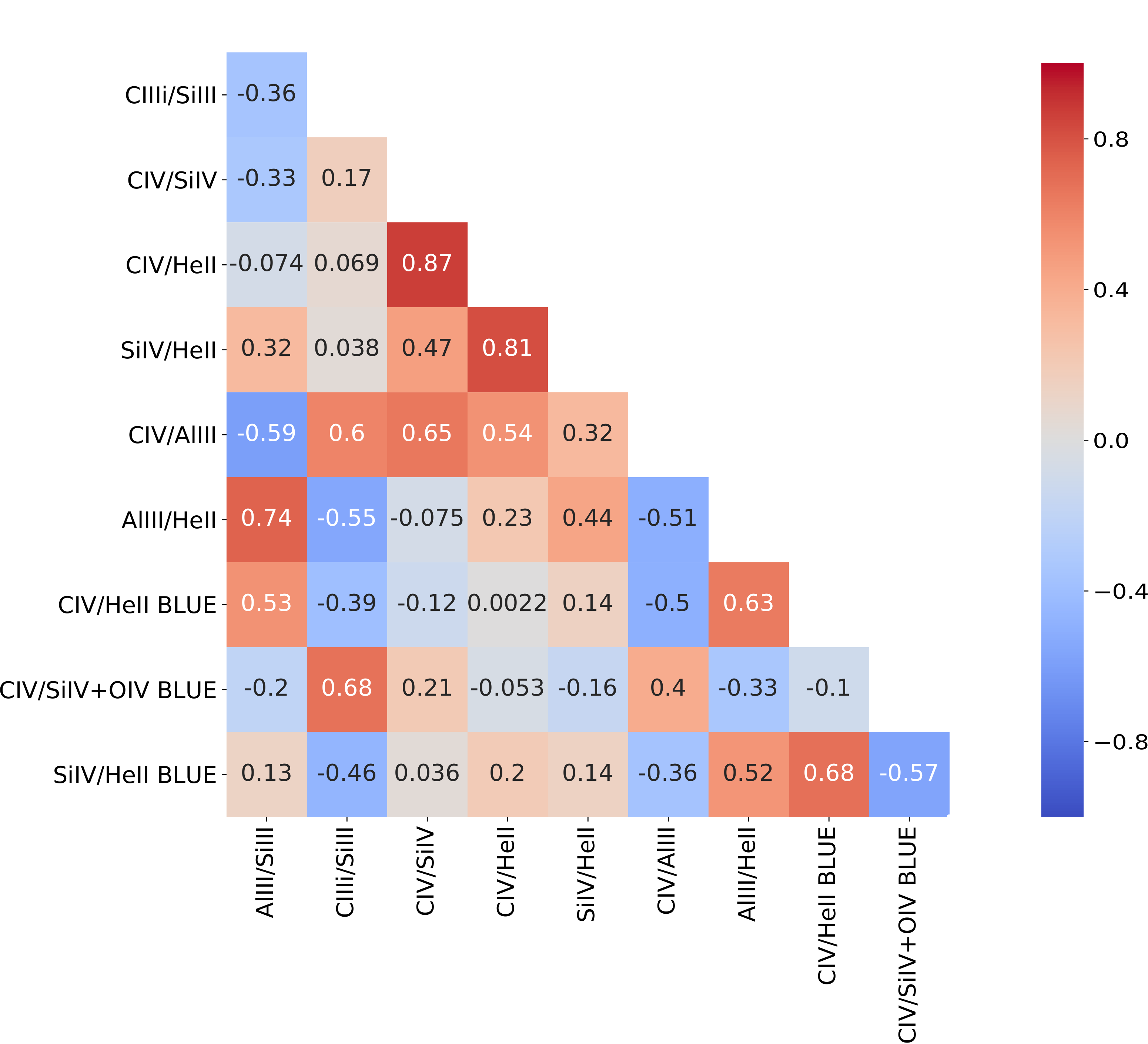}
     \caption{{The correlation matrix between diagnostic ratios in BC and BLUE. The numbers in each square show the Spearman rank correlation coefficient. Red colors indicate a positive correlation, blue colors indicate a negative correlation..}}
     \label{fig:cor-ratios}
 \end{figure}

 Fig. \ref{fig:cor-ratios} shows a matrix of  correlation coefficients for all diagnostic ratios which we considered in this work. The 2$\sigma$ confidence level of significance for the Spearman's rank correlation coefficient for 13 objects is achieved at $\rho \approx 0.54$. The highest degree of correlation is found between the ratios \civ/\heii\ and \civ/\siiv (0.87), and between \civ/\heii\ and \siiv/\heii\ (0.81). A milder degree of correlation is found between \aliii/\heii\  and \civ/\heii\ (0.23) and \siiv/\heii\ (0.44). These results imply that \siiv\ and \civ\ are likely affected in a related way by a single parameter.  The main parameter is expected to be $T_\mathrm{e}$, and hence $Z$\ (Sect. \ref{basic}). The \aliii\ (normalized to the \heii\ flux) line shows much lower values of the correlation coefficient.  The \aliii\ line has a different dependence on $U$, \nh\ and optical depth variations. The prominence of \ciii\ with respect to \siiii\ decreases with \siiv/\heii\ BLUE, \civ/\heii, \siiv/\heii\ BLUE, and increases with \civ/\siiv+\oiv. Apparently the \ciii/\siiii\ ratio is strongly affected by an increase in metallicity and more in general by ratios that are indicative of ``extremeness" in our sample. For BLUE, the two main independent $Z$ estimators are   correlated ($\rho \approx 0.68$).
 

%

\subsection{Analysis of Z distributions: global inferences on sample}

\subsubsection{Fixed  ($U$,\nh)}

\begin{figure*}[ht]
  \centering
\includegraphics[width=17cm]{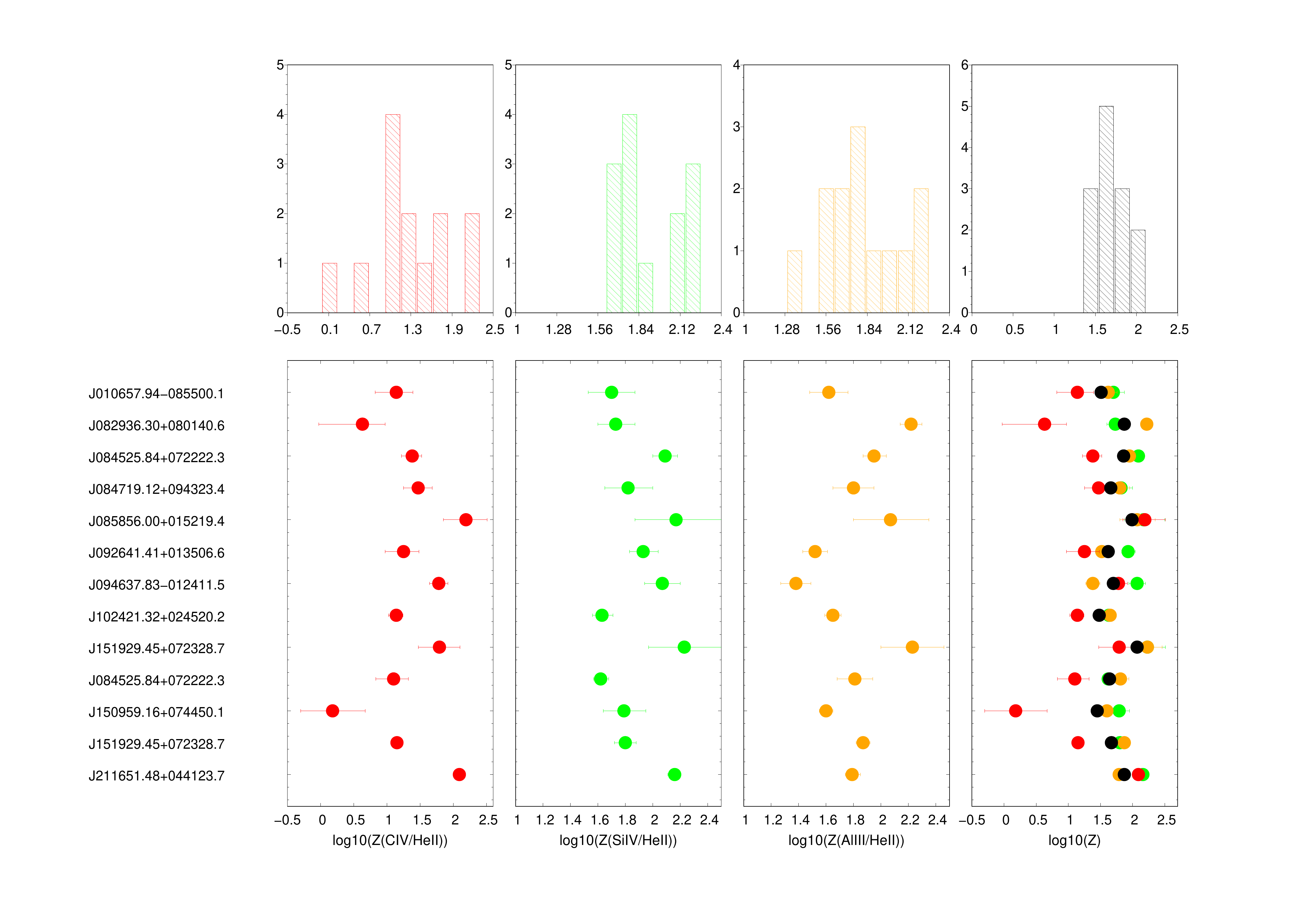}
\caption{ {(Upper panel) Distribution of metallicity measurements for broad component obtained from ratios \civ/\heii\ (red), \siiv/\heii\ (green), \aliii/\heii\ (orange) and the mean metallicity obtained from all ratios (black). (Lower panel) Results and associated uncertainties for individual sources with the same color-coding as histograms. The last panel contains all metallicity measurements and the mean of them for each object.}\label{fig:bc_histograms_z}}
\end{figure*}

\begin{figure*}[ht]
  \centering
\includegraphics[width=17cm]{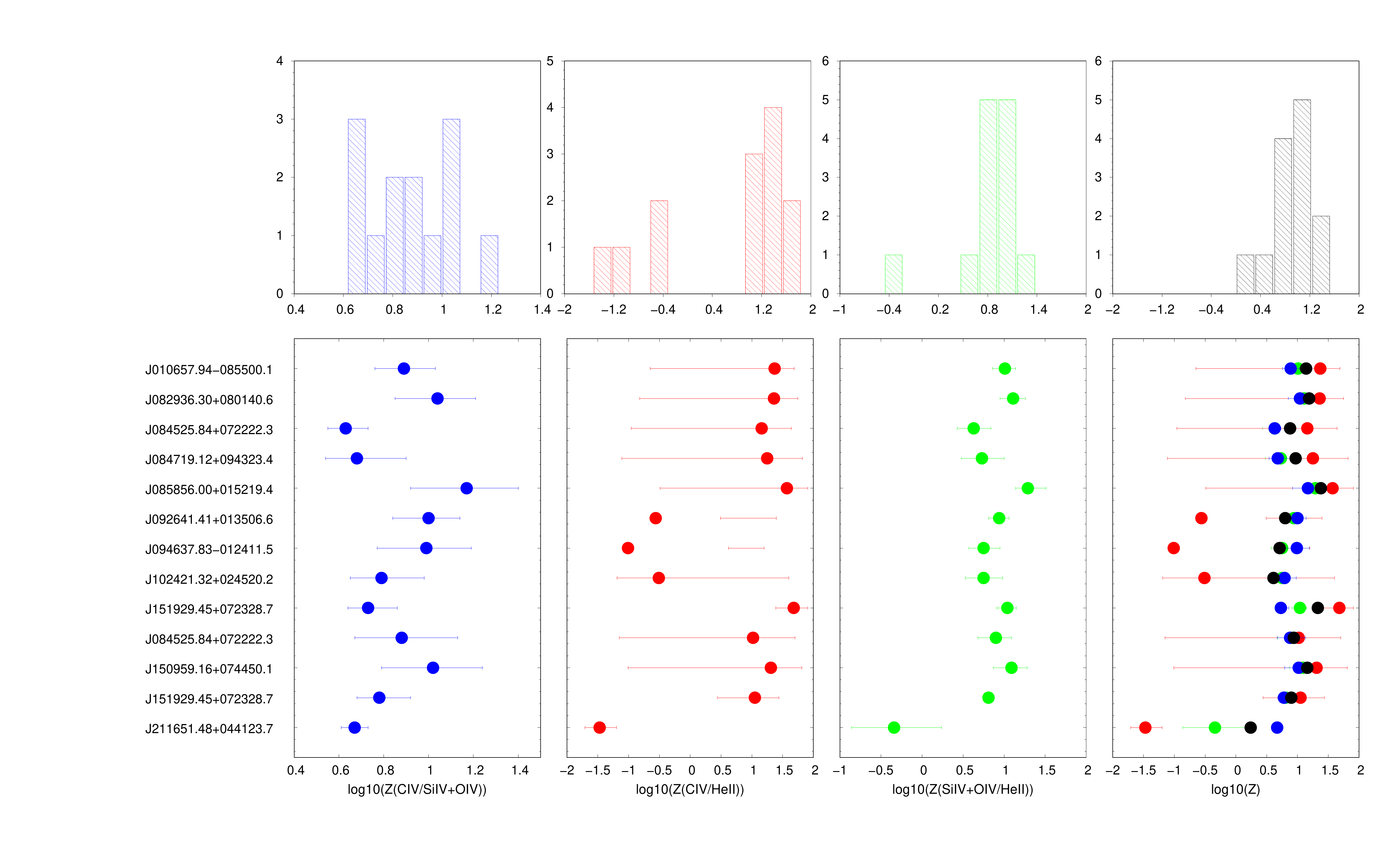}
\caption{{(Upper panel) Distribution of metallicity measurements for blue component obtained from ratios: \civ/\siiv+\oiv\ (blue), \civ/\heii\ (red), \siiv+\oiv/\heii\ (green) and the mean metallicity obtained from all ratios (black). (Lower panel) Results and associated uncertainties for individual sources with the same color-coding as histograms. The last panel contains all metallicity measurements and the mean of them for each object.} \label{fig:blue_histograms_z}}
\end{figure*}

We propagated the diagnostic intensity ratios measured on the BC and BLUE components with their lower and upper uncertainties following the relation between ratios and $Z$\ in the Figures \ref{fig:civheiifixed}, for the fixed physical conditions assumed in the low- and high-ionization region. The results are reported in Table \ref{tab:bc_z} and Table \ref{tab:blue_z} for the BC and for the blueshifted component, respectively. The last row reports the median values of the individual sources $Z$\ estimates  with the sample  SIQR.  The distributions are shown in Figs. \ref{fig:bc_histograms_z} and \ref{fig:blue_histograms_z}, along with a graphical presentation of each source  and its associated uncertainties. 

Table \ref{tab:bc_z} and Table \ref{tab:blue_z} permit   to quantify the systematic differences that are apparent in   Figs. \ref{fig:bc_histograms_z} and \ref{fig:blue_histograms_z}.  
The agreement between the various estimators is in good on average (the medians scatter around $\log Z  \approx 1$ by less than 0.2 dex).  However, there are systematic differences between the $Z$\ obtained from the various diagnostic ratios. \siiv\ and \aliii\ over \heii apparently overestimate  the $Z$ \ by a factor 2 with respect to \civ/\heii. The out-of-scale values of \civ/\siiv\ and \civ/\aliii\ may suggest that metallicity scaling according to solar proportion may not be strictly correct (Sect. \ref{sec:segregation}).  In the case of BLUE, several estimates from\ \civ/\heii\ strongly deviate from the ones obtained with the other ratios, due to the non-monotonic behavior of the relation between $Z$\ and \civ/\heii, right in the range of metallicity that is expected. Fig. \ref{fig:isoph1} shows that the non-monotonic behavior as a function of $Z$\ occurs for $-1 \lesssim \log U \lesssim 0$, assuming $\log $ \nh = 9. 

The median values of all three ratios consistently suggest high metallicity with a firm lower limit $Z \approx 5$, and in the range $10 Z_{\odot} \lesssim Z \lesssim 100 Z_{\odot}$, with typical values between 20 and 50 $Z_\odot$.  There is apparently a systematic difference between BC and BLUE, in the sense that $Z$\ derived from the BC is systematically higher than $Z$ from blue. The difference is small in the case of \civ/\heii\ but is significant in the case of (\siiv + \oiv)/\heii, where $Z$\ from BLUE are a factor of 10 lower. We have stressed earlier that there are often absorptions affecting the BLUE of \siiv + \oiv/\heii. Absorptions and the blending with C{\sc ii}$\lambda$1332 and \siiv\ BC lines make it difficult to properly define the continuum underlying the $\lambda$1400 blend at negative radial velocities. We think that the \siiv + \oiv\ BLUE intensity estimate  is more of a lower limit. Another explanation might be related to the assumption of a constant density and $U$\ for all sources. While there are observational constraints supporting this condition for the BC \citep{pandaetal18,pandaetal19,pandaetal19a}, there are no strong clues to the BLUE properties, save a high ionization degree.

\subsection{$Z$ for individual sources for fixed $U$, \nh}
\label{ind}

Table \ref{tab:ind_z} reports the   $Z$\ estimates for the BC, BLUE, and a combination of BC and BLUE for each individual object. The values reported are the median values of the individual objects' estimates from the different ratios. 
Here the $Z$ value for each object is computed by vetting the ratios according to concordance. {  If the discordance is not due on   physical ground, but rather to instrumental problems (for example, contamination by absorption lines, non linear dependence on $Z$\ of some ratios), a proper strategy is to use estimators such as the median that eliminate discordant values even for small sample sizes ($n \ge3$). Measuring medians and SIQR is an efficient way to deal with the measurements of large samples of objects.} All estimates $\log Z \lesssim 0$ were excluded, as either the product of heavy absorptions (\siiv+\oiv/\heii) or of difficulties in relating the ratio (\civ/\heii) to $Z$;  apart from J211651.48+044123.7, the upper uncertainty of the negative estimates is so large that $Z$\ is actually unconstrained. The difference between BLUE and BC is even more evident: the median (last row) indicates a factor $\approx 6$\ difference between BLUE and BC. The BC suggests a median $Z \approx 60 Z_\odot$, while the BLUE   $Z \approx 10 Z_\odot$. The assumption that the wind and disk component have the same $Z$ in each object is a reasonable one, with the caveats mentioned in Sect. \ref{sec:segregation}. Therefore the two estimates, for BLUE and BC could be considered two independent estimators of $Z$. If the two estimates are combined for each individual object, $10 Z_\odot \lesssim Z \lesssim 100 Z_\odot$, with a median value of $Z \approx 20 Z_\odot$.  

 There is a  good  agreement between the $Z$\ median estimates from the BC and BLUE of \civ, $\log Z \approx 1.27$ vs $\log Z \approx 1.13$, respectively (Tables \ref{tab:bc_z} and \ref{tab:blue_z}). Ignoring \siiv+\oiv\ and \aliii, the   $Z \approx 20 Z_\odot$ value derived for the \civ\ BC  is not affected by a possible enhancement of [Si/C] and [Al/C] with respect to the solar values. If the Carbon abundance is used as a reference, the BC  $Z$ estimate from \aliii\ and \siiv\ could point toward a selective enhancement of Si and Al with respect to C. 

The disagreement between BLUE and BC $Z$\ estimates rests on the blueshifted component of \siiv + \oiv. The disagreement between the $Z$\ estimates from ratios involving \siiv+\oiv\ BC and BLUE  might be explained if one considers that the measurement of the \siiv+\oiv\ BLUE is most problematic and the \siiv+\oiv\  intensity might be systematically underestimated. 


%
%


\begin{deluxetable}{lccccl}
\tabletypesize{\scriptsize}
\tabcolsep=2pt
\tablecaption{Metallicity ($\log Z$) of the BC assuming fixed $U$, \nh\ } 
 \tablehead{ 
\colhead{SDSS JCODE} & \colhead{\civ/\heii} & \colhead{\siiv/\heii}  & \colhead{\aliii/\heii}  &  
}
\startdata
J010657.94-085500.1     &		1.16	$_{		-0.33	}^{+	0.26	}$&	1.85	$_{	-0.20	}^{+	0.20	}$&	1.7		$_{	-0.15	}^{+	0.16	}$\\
J085856.00+015219.4    &		2.22	$_{		-0.35	}^{+	0.36	}$&	2.41	 $_{	-0.36	}^{+	0.59	}$&	2.2		$_{	-0.29	}^{+	0.31	}$\\
J082936.30+080140.6    &		0.63	$_{		-0.65	}^{+	0.36	}$&	1.89	 $_{	-0.15	}^{+	0.16	}$&	2.36		$_{	-0.09	}^{+	0.10	}$\\
J084525.84+072222.3    &		1.41	$_{		-0.16	}^{+	0.14	}$&	2.31	 $_{	-0.11	}^{+	0.11	}$&	2.07		$_{	-0.09	}^{+	0.10	}$\\
J084719.12+094323.4    &		1.50	$_{		-0.23	}^{+	0.20	}$&	1.99 	 $_{	-0.19	}^{+	0.21	}$&	1.91		$_{	-0.17	}^{+	0.17	}$\\
J092641.41+013506.6    &		1.27	$_{		-0.29	}^{+	0.24	}$&	2.12	 $_{	-0.12	}^{+	0.12	}$&	1.59		$_{	-0.10	}^{+	0.11	}$\\
J094637.83-012411.5     &		1.80	$_{		-0.14	}^{+	0.15	}$&	2.29	 $_{	-0.15	}^{+	0.16	}$&	1.44		$_{	-0.13	}^{+	0.12	}$\\
J102421.32+024520.2    &		1.16	$_{		-0.12	}^{+	0.11	}$&	1.77	 $_{	-0.08	}^{+	0.09	}$&	1.74		$_{	-0.06	}^{+	0.07	}$\\
J102606.67+011459.0    &		1.82	$_{		-0.32	}^{+	0.30	}$&	2.49	 $_{	-0.32	}^{+	0.51	}$&	2.38		$_{	-0.26	}^{+	0.27	}$\\
J114557.84+080029.0     &		1.12	$_{		-0.27	}^{+	0.22	}$&	1.96	 $_{	-0.18	}^{+	0.19	}$&	1.92		$_{	-0.15	}^{+	0.15	}$\\
J150959.16+074450.1    &		0.19	$_{		-0.47	}^{+	0.49	}$&	1.76	 $_{	-0.07	}^{+	0.06	}$&	1.69		$_{	-0.06	}^{+	0.05	}$\\
J151929.45+072328.7    &		1.18	$_{		-0.09	}^{+	0.08	}$&	1.97	 $_{	-0.09	}^{+	0.08	}$&	1.98		$_{	-0.05	}^{+	0.05	}$\\
J211651.48+044123.7   &			2.11	$_{		-0.04	}^{+	0.04	}$&	2.39	 $_{	-0.06	}^{+	0.06	}$&	1.9		$_{	-0.06	}^{+	0.06	}$\\ 
 \hline
Median & 1.27 $\pm$  0.32 & 1.99$\pm$ 0.21 & 1.91$\pm$ 0.18\\
 \hline 
\enddata
\tablecomments{Columns: (1) SDSS identification, (2), (3) and (4) metallicity values for \civ/\heii\, \siiv+\oiv/\heii\ and \aliii/\heii\ with uncertainties.}
\label{tab:bc_z}
\end{deluxetable}

\begin{deluxetable}{lcccl}
\tabletypesize{\scriptsize}
\tabcolsep=2pt
\tablecaption{{Metallicity ($\log Z$) of BLUE assuming fixed $U$, \nh\ }  \label{tab:blue_z}
}
\tablehead{ 
\colhead{SDSS JCODE}          & \colhead{\siiv+\oiv/\heii} & \colhead{\civ/\siiv+\oiv}  & \colhead{\civ/\heii}  &  
}
\startdata
J010657.94-085500.1      	&	1.00	$_{		-0.16	}^{+	0.12	}$&	0.88	$_{	-0.12	}^{+	0.14	}$&		1.34	$_{	-1.99	}^{+	0.34	}$\\
 J085856.00+015219.4     	&	1.26	$_{		-0.13	}^{+	0.20	}$&	1.12	$_{	-0.21	}^{+	0.23	}$&		1.55	$_{	-2.05	}^{+	0.35	}$\\
 J082936.30+080140.6     	&	1.10	$_{		-0.16	}^{+	0.14	}$&	1.03	$_{	-0.19	}^{+	0.16	}$&		1.33	$_{	-2.14	}^{+	0.41	}$\\
 J084525.84+072222.3     	&	0.62	$_{		-0.19	}^{+	0.20	}$&	0.63	$_{	-0.09	}^{+	0.09	}$&		1.13	$_{	-2.07	}^{+	0.50	}$\\
 J084719.12+094323.4     	&	0.72	$_{		-0.24	}^{+	0.27	}$&	0.68	$_{	-0.15	}^{+	0.21	}$&		1.22	$_{	-2.32	}^{+	0.59	}$\\
 J092641.41+013506.6     	&	0.93	$_{		-0.14	}^{+	0.11	}$&	0.99	$_{	-0.16	}^{+	0.14	}$&		-0.56	$_{	-0.46	}^{+	1.93	}$\\
 J094637.83-012411.5     	&	0.74	$_{		-0.18	}^{+	0.20	}$&	0.98	$_{	-0.22	}^{+	0.20	}$&		-0.99	$_{	-1.06	}^{+	2.16	}$\\
 J102421.32+024520.2     	&	0.73	$_{		-0.20	}^{+	0.24	}$&	0.78	$_{	-0.14	}^{+	0.19	}$&		-0.52	$_{	-0.65	}^{+	2.10	}$\\
 J102606.67+011459.0     	&	1.02	$_{		-0.12	}^{+	0.11	}$&	0.73	$_{	-0.10	}^{+	0.12	}$&		1.67	$_{	-0.31	}^{+	0.24	}$\\
 J114557.84+080029.0     	&	0.89	$_{		-0.22	}^{+	0.19	}$&	0.87	$_{	-0.20	}^{+	0.25	}$&		0.99	$_{	-2.12	}^{+	0.70	}$\\
 J150959.16+074450.1     	&	1.07	$_{		-0.21	}^{+	0.18	}$&	1.01	$_{	-0.23	}^{+	0.22	}$&		1.28	$_{	-2.28	}^{+	0.52	}$\\
 J151929.45+072328.7     	&	0.79	$_{		-0.06	}^{+	0.07	}$&	0.78	$_{	-0.11	}^{+	0.13	}$&		1.02	$_{	-0.59	}^{+	0.40	}$\\
 J211651.48+044123.7     	&	-0.57	$_{		-0.43	}^{+	0.81	}$&	0.66	$_{	-0.05	}^{+	0.06	}$&		-1.46	$_{	-0.23	}^{+	0.27	}$\\ \hline
Medians &  
 0.89 $\pm$ 0.15 &
 0.87$\pm$ 0.13 &
 1.13$\pm$  0.92\\
 \enddata
\tablecomments{Columns: (1) SDSS identification, (2), (3) and (4) metallicity values for \siiv+\oiv/\heii\, \civ/\siiv+\oiv\ and \civ/\heii\ with uncertainties.}
\end{deluxetable}

\begin{deluxetable}{lcccl}
\tabletypesize{\scriptsize}
\tabcolsep=2pt
\tablecaption{{Metallicity ($\log Z$) of individual quasars assuming fixed $U$, \nh\ }  \label{tab:ind_z}
}
\tablehead{ 
\colhead{SDSS JCODE}          & \colhead{BC} & \colhead{BLUE}  & \colhead{Combined}      
}
\startdata
J010657.94-085500.1 & 1.70 $\pm$  0.34 & 1.00 $\pm$   0.23 & 1.34 $\pm$   0.35  \\
J082936.30+080140.6 & 1.89 $\pm$  0.86 & 1.10 $\pm$   0.15 & 1.33 $\pm$   0.43  \\
J084525.84+072222.3 & 2.07 $\pm$  0.45 & 0.63 $\pm$   0.25 & 1.41 $\pm$   0.72 \\
J084719.12+094323.4 & 1.91 $\pm$  0.25 & 0.72 $\pm$   0.27 & 1.50   $\pm$	0.60 \\
J085856.00+015219.4 & 2.22 $\pm$  0.11 & 1.26 $\pm$   0.21 & 2.20 $\pm$   0.48  \\
J092641.41+013506.6 & 1.59 $\pm$  0.42 & 0.93 $\pm$   0.77 & 1.27 $\pm$   0.33  \\
J094637.83-012411.5 & 1.80 $\pm$  0.42 & 0.74 $\pm$   0.99 & 1.44 $\pm$   0.53 \\
J102421.32+024520.2 & 1.74 $\pm$  0.30 & 0.73 $\pm$   0.65 & 1.16 $\pm$   0.51 \\
J102606.67+011459.0 & 2.38 $\pm$  0.33 & 1.02  $\pm$  0.47 & 1.82 $\pm$   0.68 \\
J114557.84+080029.0 & 1.92 $\pm$  0.42 & 0.89  $\pm$  0.06 & 1.12 $\pm$   0.52 \\
J150959.16+074450.1 & 1.69 $\pm$  0.78 & 1.07  $\pm$  0.14 & 1.28 $\pm$   0.34  \\
J151929.45+072328.7 & 1.97 $\pm$  0.40 & 0.79  $\pm$  0.12 & 1.18 $\pm$   0.59  \\
J211651.48+044123.7\tablenotemark{a} & 2.11 $\pm$  0.24 & 0.66 $\pm$   \ldots & 2.11 $\pm$   0.34  \\\hline
 Median &             1.91  $\pm$ 0.16  &   0.89 $\pm$ 0.12 & 1.34 $\pm$ 0.11     \\  \hline
 \enddata
 \tablenotetext{a}{\siiv\ and \civ\ affected by absorptions on the blue wings.}
\tablecomments{Columns: (1) SDSS identification, (2), (3) and (4) metallicity medians for BC, BLUE and the combination of the two, with uncertainties. Column (5) yields the number of ratios used for the BLUE estimates. No uncertainty is reported for BLUE of \object{SDSS J211651.48+044123.7} since only one ratio was used. }
\end{deluxetable}

\subsection{Estimates of $Z$ relaxing the constraints on $U$\ and \nh}

  We computed the $\chi^2$\ in the following form, to identify the value of the metallicity for median values of the diagnostic ratios and for the diagnostic ratios of individual objects relaxing the assumption of fixed density and ionization parameters. For each object $k$, and for each component $c$, we can write:
\begin{eqnarray}
\chi^2_\mathrm{kc}(n_\mathrm{H},U,Z) & = & \sum_\mathrm{{i}} w_\mathrm{ci} \left(\frac{R_\mathrm{kci}- R_\mathrm{kci,mod}(n_\mathrm{H},U,Z)}{\delta R_\mathrm{kci}}\right)^2    \label{eq:chi}
\end{eqnarray}
where the summation is done over the available diagnostic ratios, and the $\chi^2$ is computed with respect to the results of the CLOUDY simulations as a function of $U$, \nh, and $Z$\ (subscript `mod'). Weights $w_\mathrm{ci} = 1$ were assigned to \civ/\heii, \siiv/\heii, and \aliii/\heii; $w_\mathrm{ci} = 0$ or 0.5 were assigned to \civ/\aliii\ and and \civ/\siiv. For BLUE, the three diagnostic ratios were all assigned  $w_\mathrm{ci} = 1$.  The $Z$ estimates for the BC are  based on the three  ratios involving \heii\ normalization.  

To gain a global, bird's eye view of the $Z$\ dependence on the physical parameters,  Fig. \ref{fig:mediancr}    shows the 3D space  $U$, \nh, $Z$. Each point in this space corresponds to an element of the grid of {\tt CLOUDY} the parameter space and is consistent with the minimum $\chi^2$\ within the uncertainties at 1$\sigma$ confidence level.  The case shown in the panels of  Fig. \ref{fig:mediancr}   is the one with the median values of the sample objects.  

The distribution of the data points   is constrained in a relatively narrow range of $U$, \nh, $Z$, at very high density, low ionization, and high metallicity. Within the limit in  $U$, \nh, the distribution of $Z$\ is flat and thin, around $Z \sim 50 - 100 Z_\odot$. This implies that, for a change of the $U$\ and \nh\ within the limits allowed by the data, the estimate of $Z$\ is stable and independent on $U$\ and \nh. Table \ref{tab:ind_z} reports the individual $Z$\ estimates and the SIQR for the sources in the sample (the last row is the median).  


The allowed parameter space volume for BLUE is by far less constrained. The right panel of Fig. \ref{fig:mediancr} shows the parameter space  for the $Z$\ estimates from the 3 BLUE intensity ratios. The condition on the $\chi^2$\ distribution is the same, as used for the BC, namely that the data points all satisfy the condition $\chi^2  \le \chi^2_\mathrm{min}+1$. A similar shape is obtained if we consider  the condition that all the 3 ratios agree with the ones predicted by the model within 1$\sigma$.  The spread in ionization and density is very large, although the concentration of data points is higher in the case of low \nh\ ($\log $ \nh $\sim$ 8-9 [cm$^{-3}$]) and high ionization ($\log U \sim 0$).  At any rate the spread of the data points indicate that solutions at low ionization and high density are also possible. The results for individual sources tend to disfavor this scenario for the wide majority of the objects, but the properties of the gas emitting the BLUE  component  are less constrained than the ones of the gas emitting the BC. What is missing for BLUE is especially a firm diagnostic of density that in the case of BC is provided mainly by the ratio \aliii/\heii. Results on $Z$\ are however as stable as for the BC, even if the dispersion is larger, and suggest values in the range $10 \lesssim Z \lesssim 50 Z_\odot$.

Summing up, all meaningful estimators converge toward high $Z$ values, definitely super-solar, with $Z \gtrsim 10 Z_\odot$. Ratios \civ/\siiv\ significantly less than $< 1$ are   predicted in the parameter space. \siiv/\heii\ seems to give the largest estimates of $Z$. Also the high \aliii/\civ\ requires   high values of $Z$. A conclusion has to be tentative, considering the possible systematic errors affecting the estimates of the \civ\ and \siiv\ intensities: for \civ, the BC in the most extreme cases is often buried under an overwhelming BLUE; a fit is not providing  a reliable estimate of the BC (by far the fainter component) but provides a reliable BLUE intensity; for \siiv\ we may overestimate the intensity due to  ``cancellation" of the BLUE by absorptions. This said, the present data are consistent with the possibility of a selective enhancement of Al and Si, as already considered by \citet{negreteetal12}. The issue will be briefly discussed in Sect. \ref{Discussion}. 

At any rate, the absence of correlation between BLUE and BC parameters (Fig. \ref{fig:cor-ratios}), the difference in the diagnostic ratios and differences in inferred $Z$, as well as the results for individual sources described below justify the approach followed in the paper to maintain a separation between BLUE and BC.  The meaning of possible systematic differences between the BC and BLUE are further discussed in \S \ref{Discussion}.


\subsubsection{Individual sources}

\paragraph{BC} The best \nh, $U$, and $Z$ for each object have been obtained by minimizing the $\chi^2$ as defined in Eq. \ref{eq:chi}, and they are reported in Table \ref{tab:zun}. The $\chi^2_\mathrm{min}$\ values are listed in the second Column of Tab. \ref{tab:zun}.   At the side of each value  there is the uncertainty range for each parameter defined from volume in the parameter space satisfying the condition $\chi^2 \approx \chi^2_\mathrm{min} + 1$.\footnote{ This approach follows a standard procedure \citep[][p. 209]{bevingtonrobinson03} for the determination of the confidence intervals, and we see in the matrix $\chi^2(Z, U,$\nh) a well-defined global minimum around which the $\chi^2$\ increases systematically. }  In other words, the choice of the best physical conditions was obtained by minimizing the sum of the deviations between the model predictions and the observer diagnostic ratios.  The last two rows list the minimum $\chi^2$\ values for the median (with the SIQR of the sample from Table \ref{tab:lineratios}) and for the median of the values  reported for individual sources in Table \ref{tab:zun}.  The obtained values of $Z$ cover the range $5 \lesssim Z \lesssim 100$, with 9 out of 13 sources with $50 \lesssim Z \lesssim 100 Z_\odot$, and medians of intensity ratios yielding $Z \sim 50 Z_\odot$. There is some spread in the ionization parameter values, $-4 \lesssim U \lesssim -1$, but in most cases $U$\ indicates low or very low ionization level. The   Hydrogen density is very high: in only a few cases $\log$\nh $\sim 10 - 11$, and in several cases \nh\ reaches 10$^{14}$\ cm$^{-3}$. The median values from the ratios are   $\log U =-2.25$, $\log $ \nh =13.75 (with a range 12.5 -- 14), therefore validating the original assumption of   $\log U =-2.5$, $\log $ \nh = 12 for fixed physical condition.   The results for individual sources confirm the scenario of Fig. \ref{fig:mediancr} for the wide majority of the sample sources. The higher \nh\ values are consistent with recent inferences for the low-ionization BLR derived from \citet{templeetal20}, based on the FeIII UV emission which is especially prominent in the UV spectra of xA quasars \citep{martinez-aldamaetal18a}. It is interesting to note that two of the borderline objects (\aliii/\siiii $\approx 0.5$, \ciii/\siiii $\approx 1$)\ show higher values of the ionization parameter ($\log U \approx -1. - -1.5$). 

 Large ($\gg 1$)\ values of $\chi^2_\mathrm{min}$\ are associated with cases in which the BC components of \aliii\ and/or of \siiv\ are strong with respect to the BC of \civ, and are further increased by small uncertainty ranges (which are more likely to occur if a line is strong). Intensity ratios \civ/\aliii $\lesssim 1$ and \civ/\siiv $\lesssim 1$ are   reproduced by photoionization simulations in conditions of very low ionization. The ratio \civ/\aliii\ tend to decrease with increasing $Z$, although the trend is steep in the case of $\log$ \nh $\sim $ 9, and more shallow at the higher densities appropriate for the BC emission (Appendix \ref{sec:iso}).  However, high values of the \aliii\ and \siiv+\oiv\ over \heiiuv\ ratios induced by overabundances could bias the $U$ and  lower  its values.


\paragraph{BLUE} The inferences are less clear from BLUE (Table \ref{tab:zun_blue}, organized like Table \ref{tab:zun}). In most cases, the permitted volume in  the 3D  parameter space for individual sources covers a broad range in $U$\ and \nh\ as for the median (Fig. \ref{fig:mediancr}).  Fig. \ref{fig:mediancr} shows that there is a strip of $\chi^2$\ values statistically consistent with the minimum $\chi^2$\ that crosses the full domain of the parameter space. Along this strip of permitted values \nh\ and $U$\ are linearly dependent, with $\log U \approx -0.5 \cdot n_\mathrm{H} +4 $\ for the median composite ratios.  In most sources the $U$\ value implies a  high degree of ionization, $-1 \lesssim \log U \lesssim  0$, but in three cases  (for example \object{SDSS J150959.16+074450.1}) there is apparently a low-ionization solution with $U$\ comparable to that of the low-ionization BLR. The median are $\log$\nh $\sim 7.75$, $\log$U $\sim$-0.5, close to the values that we assumed for the fixed ($U$, \nh) approach. The results on metallicity suggest in most cases $Z \gtrsim 20 Z_\odot$. However, within 1 SIQR from the minimum $\chi^2$, $Z$\ values up to 30 are also possible.  

 $Z$ values from BLUE are systematically lower than the those from BC. The medians differ by a factor of 2. However, a Welch t-test \citep{welch47} fails to detect a significant difference between the average values of the metallicity for the two components: $t \approx 0.86$ for 5 degrees of freedom (computed using the Welch-Satterthwaite equation) implies a significance of just 80\%. Three cases in which the disagreement is large, more than a factor 5, namely J084525.84+072222.3, J084719.12+094323.4, J102606.67+011459.0  are apparently not strongly affected by absorption lines, but the constraints from Table \ref{tab:zun_blue} and Table \ref{tab:zun} are poor, implying that also for BLUE the $Z$\ could be much higher. Therefore, we cannot substantiate any claim of a systematic difference between BLUE and BC $Z$\ estimates.   

\paragraph{The $Z$, \ $U$, \nh\ parameter space occupation of xA quasars} In summary, the low-ionization BLR of xA sources seems to be consistently  characterized by  low ionization, { extremely high density and very high metallicity, under the assumption that $Z$\ scales with the solar chemical composition.  Diagnostics on BLUE is less constraining, and measurements are more difficult. The 0-order results are however consistent again with high metallicity $Z \gtrsim 5 Z_\odot$.  

 The 3D distribution in Fig. \ref{fig:mediancr} indicates that, although there might be a large range of uncertainty in the $U$ and \nh\ especially for BLUE, the $Z$ values tend to remain constrained within a narrow strip around a well-defined $Z$, parallel to the $U$, \nh\ plane. In other words, $Z$\ estimates should be stable, as they  are not strongly dependent on the assumed physical parameters. 

Comparing the individual $Z$\ estimates for fixed and free \nh\ and $U$\ (Cols. 2 of Table \ref{tab:bc_z} and Table  \ref{tab:zun}) for the BC, the agreement is good, with a median difference of 0.22 and a SIQR of 0.15, with the fixed \nh\ and $U$\ being therefore a factor $\approx 1.65$\ higher than the one derived assuming a free  \nh\ and $U$. Two sources (\object{J151929.45+072328.7} and \object{J114557.84+080029.0}) show a large disagreement, in the sense that the $Z$ values leaving $U$\ and \nh\ free are much lower. These $Z$ estimates are however highly uncertain, with a shallow $\chi^2$\ distribution around the minimum especially for \object{J151929.45+072328.7}. For this object the maximum metallicity covered by the simulations $Z = 1000 Z_\odot$ is still consistent within the uncertainties.

\begin{deluxetable*}{cccccccc}
\tabletypesize{\scriptsize}
\tabcolsep=2pt
\tablecaption{{$Z$, $U$, \nh\ of individual sources and median derived from the BC}  \label{tab:zun}
}
\tablehead{ 
\colhead{SDSS JCODE}      &\colhead{$\chi^2_\mathrm{min}$  }  & \colhead{$\log Z$\ [$Z_\odot$]} & \colhead{$\delta \log Z$\ [$Z_\odot$]} & \colhead{$\log U$} & \colhead{$\delta \log U$}  & \colhead{$\log $ \nh} & \colhead{$\delta \log $ \nh}  \\ \vspace{-0.4cm}
(1) & (2) & (3) & (4)  & (5) & (6) & (7)\\ 
}
\startdata
   J010657.94-085500.1 & 2.6894      & 1.7  &    0.7 -- 1.7  &   -2.00   & -2.75  --   -1.25   &  13.50   & 12.00 -- 14.00   \\  
   J082936.30+080140.6 & 6.6414	 & 1.3  &    0.7 -- 2.7  &   -4.00   & -4.00  --   -3.50 & 14.00   & 12.75 -- 14.00 \\
   J084525.84+072222.3 & 13.700	 & 1.7  &    1.0 -- 3.0  &   -1.75   & -3.75  --   -0.25 & 14.00   & 12.00  -- 14.00  \\
   J084719.12+094323.4 & 1.7004   & 1.7  &    1.0 -- 2.0  &   -2.25   & -2.50  --   -1.50 & 13.75   & 12.50 -- 14.00   \\  
   J085856.00+015219.4 & 0.0012   & 2.0  &    1.7 -- 2.7  &   -2.25   & -3.00  --   -1.75 & 12.50   & 12.00 -- 14.00 \\
   J092641.41+013506.6 & 1.7982   & 1.7  &    1.7 -- 2.0  &   -1.50	 & -1.75  --   -1.00 & 14.00   & 13.50 -- 14.00  \\ 
   J094637.83-012411.5 & 0.3253   & 1.7  &    1.7 -- 2.3  &   -1.00   & -1.25  --   -0.75 & 12.25   & 11.75 -- 14.00 \\
   J102421.32+024520.2 & 15.819   & 1.7  &    1.3 -- 2.0  &   -2.25   & -3.00  --   -1.50 & 13.50   & 12.75 -- 14.00  \\
   J102606.67+011459.0 & 1.5020	 & 2.0  &    1.7 -- 2.3  &   -1.75   & -2.50  --   -0.75 & 14.00   & 12.50  -- 14.00 \\  
   J114557.84+080029.0 & 6.6174	 & 0.7  &    0.7 -- 2.0  &   -3.75   & -3.75  --   -1.25 & 14.00   & 12.25 -- 14.00  \\
   J150959.16+074450.1 & 45.287   & 1.3  &    0.5 -- 2.0  &   -1.75   & -0.50  --   -3.50 & 13.75   & 12.25 -- 14.00   \\
   J151929.45+072328.7 & 29.220	 & 0.7  &    0.3 -- 3.0  &   -3.75   & -4.00  --   -2.75 & 14.00   & 11.00  -- 14.00  \\
 	J211651.48+044123.7 & 1.4476   & 2.0  &    1.7 -- 2.0  &   -2.00   & -2.00  --   -0.75 & 12.25   & 10.75 -- 12.25  \\	  
 \hline                                                 
$\mu_\frac{1}{2}$(Ratios)&  1.9914 & 1.7  &    0.7  --   2.0 &  -2.25   & -2.75  --  -1.25 &  13.75   &  12.50  -- 14.00  \\
$\mu_\frac{1}{2}$(Objects)&  \ldots & 1.7  &    1.5  --   1.9 &  -2.00   & -2.25  --  -1.75 &  13.75   &  13.50  -- 14.00 \\       
 \enddata
\end{deluxetable*}

\begin{deluxetable*}{cccccccc}
\tabletypesize{\scriptsize}
\tabcolsep=2pt
\tablecaption{{$Z$, $U$, \nh\ of individual sources and median derived from BLUE}  \label{tab:zun_blue}
}
\tablehead{ 
\colhead{SDSS JCODE}  &\colhead{$\chi^2_\mathrm{min}$  }        & \colhead{$\log Z$\ [$Z_\odot$]} & \colhead{$\delta \log Z$\ [$Z_\odot$]} & \colhead{$\log U$} & \colhead{$\delta \log U$}  & \colhead{$\log $ \nh} & \colhead{$\delta \log $ \nh} \\ \vspace{-0.4cm}
(1) & (2) & (3) & (4)  & (5) & (6) & (7)\\ 
}
\startdata
  J010657.94$-$085500.1 & 0.00985    & 1.70 &  0.70  --  1.7     &  0.75 &  -1.5 --   0.75  &       7.75   &  7.50 -- 9.75   \\   
  J082936.30+080140.6   & 0.01506   & 1.30 &  1.00  --  1.70     & -0.25 &  -2.0 --  -0.00  &       8.00  &   7.75 -- 12.25  \\ 
  J084525.84+072222.3   & 0.00143   & 1.0 &  0.00  --  2.30     & -1.50 &  -2.50 -- -1.25  &       7.25  &   7.00 -- 11.25  \\   
  J084719.12+094323.4   & 0.00097   & 1.00 &  0.00  --  3.00     & -2.24 &  -2.50 -- 	0.75  &       8.75  & 7.50 -- 11.50   \\ 
  J085856.00+015219.4   & 0.00305   & 1.70 &  1.0  --  2.00       &  0.50 &  -1.75 -- 	0.50  &       8.25  & 7.00 -- 11.75  \\
  J102606.67+011459.0   & 0.00111   & 0.70 &  0.70  --  3.00       & -1.00 &  -2.50 --  -0.5  &       9.25  &   8.25 -- 11.25  \\
  J114557.84+080029.0   & 0.00164   & 1.30 &  0.30  --  1.70     & -2.50 &  -2.50 --  0.25	&      11.50  &   7.25 -- 12.50   \\
  J150959.16+074450.1   & 0.00288   & 1.30 &  0.30  --  2.00       & -2.00 &  -2.75 --  0.75  &      11.75  &   7.00 -- 13.75  \\  
  J151929.45+072328.7   & 0.00032   & 0.30 &  0.30  --  1.30     & -0.50 &  -0.75 -- -0.25  &      10.00  &   7.75 -- 10.25  \\   
 \hline
$\mu_\frac{1}{2}$(Ratios)&    0.07336&     1.30 &  1.0  --  1.30     & -0.50 &  -0.50 -- -0.50  &       7.75  &   7.75 -- 8.00   \\
$\mu_\frac{1}{2}$(Objects)  & \ldots   & 1.30 &  1.15  --  1.45   & -1.00 &  -0.125 -- -1.875 &      8.75  &   7.75 -- 9.75   \\
\hline 
\enddata
\end{deluxetable*}

\begin{figure*}[ht]
  \centering
\includegraphics[width=8cm]{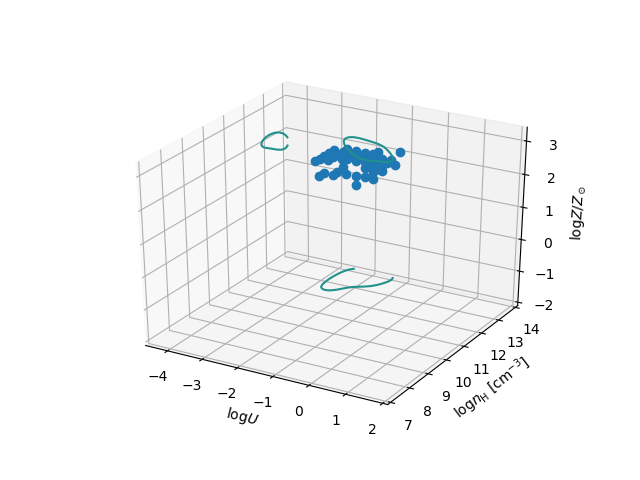}
\includegraphics[width=8cm]{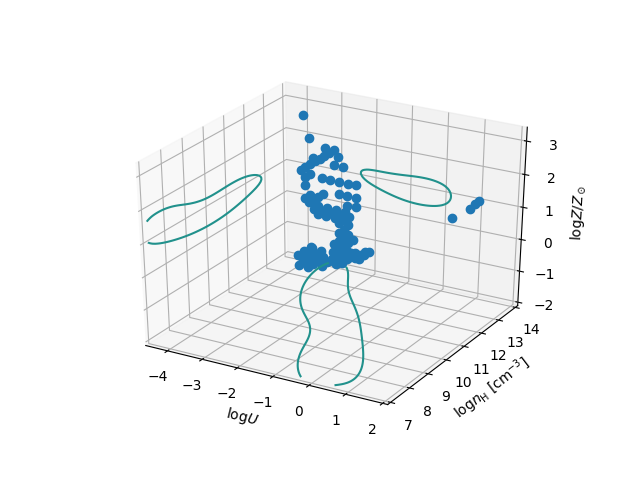}
\caption{ The parameter space \nh, $U$, $Z$. Left: data points in 3D space are elements in the grid of the parameter space selected for not being different from $\chi^2_\mathrm{min}$\ by more than $\delta \chi^2 \approx 1$, computed from the emission line ratios measured for the BC and referring to the median values in Table \ref{tab:lineratios}.  Right: same for the BLUE component, also including the condition that that the data points are in agreement with the three diagnostic ratios  within the SIQR of  the median estimate from Table \ref{tab:lineratios}. The individual contour line was smoothed with a Gaussian kernel.        \label{fig:mediancr}}
\end{figure*}



\section{Discussion}
\label{Discussion}

\subsection{A method to estimate $Z$}

The determination of the metal content of the broad line emitting region of xA quasars  was made possible by the following procedure:
\begin{enumerate}
    \item the estimation of an accurate redshift. Even if all lines are affected by significant blueshifts which reduce the values of measured redshift, in the absence of information from the \hb\ spectral range the \aliii\ and the $\lambda$1900 blend can be used as proxies of proper redshift estimators. The blueshifts are the smallest in the intermediate ionization lines at $\lambda$1900\ (del Olmo et al., in preparation). 
    \item the separation of the BC and BLUE, for \civ\ and the $\lambda$1400 blend. The line width of the individual components of the \aliii\ doublet can be used as a template BC. The component BLUE is defined as the excess emission on the blue side of the BC.
    \item a first estimate of metallicity can be obtained from the assumption that the low-ionization BLR associated with the BC and wind/outflow component associated with BLUE can be described by similar physical conditions in different objects. Several diagnostic ratios can be associated with the intensity ratios predicted by an array of photoionization simulations, namely 
    \begin{itemize}
        \item for the BC: \aliii/\heii, \civ/\heii, \siiv/\heii, assuming ($\log U$, $\log$\nh) = (-2.5,12) or  ($\log U$, $\log$\nh) = (-2.5,13)
     \item for the BLUE:   \civ/\heii, \siiv+\oiv/\heii, \civ/\siiv+\oiv\  assuming  ($\log U$, $\log$\nh) = (0,9).
    \end{itemize}
   \item Estimates can be refined for individual sources relaxing the constant ($\log U$, $\log$\nh) assumptions. Tight constraints can be obtained for the BC. The BLUE is more problematic, because of both observational difficulties and the absence of unambiguous diagnostics. 
\end{enumerate}

 Our method relies on ratios involving \heii\ that have not been much considered in previous literature.  In addition, we have considered fixed SED, turbulence (equal to 0), and column density in the simulations (\nc = 10$^{23}$)\ as fixed. The role of turbulence is further discussed in Sect. \ref{turb}, and is found to be not relevant unlike in the case of \feii\ emission in the optical spectral range, where effects of self- and \lya-fluorescence are important,  \citep[e.g.,][]{verneretal99,pandaetal18}, while the \nc\ effect is most likely negligible. 

Extension of the method to the full Population A is a likely possibility, since we do not expect a very strong effect of the SED on the metallicity estimate, as long as the SED has a prominent big blue bump, as it seems to be case for Population A. The role of SED is likely important if the method has to be extended to sources of Pop. B along the main sequence.  At least two SED cases should be considered, if the aim is to apply the method presented in this paper to a large sample of quasars.

Intensity ratios involving \heii\ are difficult to measure in the xA spectra, but may be more accessible for Population B spectra. \citet{ferlandetal20} have shown significant differences in the SED as a function of \lledd, with a much flatter SED at low \lledd. The extension to Pop. B would therefore require a new dedicated array of simulations.


\phantom{
Metallicity increases with \mbh\ \citep{matsuokaetal11}, and ratios involving \nv\ are apparently correlated with \lledd, suggesting a delay of $\sim 10^{8}$\ for the onset of nuclear activity after star forming events \ldots
Difficulties in moddeling of \siiv+\oiv\ are generally known in the literature (cite ???). \siiv\ is probably overestimated, while \oiv\ is underestimated (why???). Moreover, the \siiv+\oiv\ blend is strongly 
affected by absorption (show in which figures we can see it???). 
The BC emission in \siiv+\oiv\ blend is dominated by \siiv\ due to collisional deexcitation of the intercombination \oiv\ (cite ???).
The BLUE component emission comes from an inextricable contribution of both \siiv+\oiv. 
Table 5. Measured Quantities on Composite Spectra - Pop. A Average fluc values lya and nv - Sulentic et al.: Spectra of Quasars at $\approx$ 2.3. We must preliminarily point out that great care should be
exerted in analyzing data based on the Nv 1240/Civ 1549 and
Nv 1240/\heii\ ratios because: (1) the intensity of the
Nv 1240 line is very difficult to estimate unambiguously unless
a reliable model of the \lya\ wings are build as done...
L. E. Simon and F. Hamann: Metallicity and FIR luminosity of high-z QSOs - Figure 3.Gaussian fits for \lya,andNVin the left panels and CIVin theright panels for each normalized composite spectrum.
As shown in Figure 8, low-z AGNs with
log[L/L Edd ] $<  -1.5$ (that is not covered in the high-z sample)
show systematically lower N V /C IV flux ratios, leading to
the more evident dependence on L/L Edd . FROM SHIN 12:
R ESULTS OF S PEARMAN ’ S RANK - ORDER CORRELATION TEST - N V /C IV AND N V /He II. FROM SHIN 12: In Figure 5 (bottom panel) we compare the N V  1240 flux
and the sum of the Si IV  1397 and O IV ] 1402 fluxes, which
are used as a numerator of metallicity indicators. Again a
clear positive correlation is present between them, with a
somewhat larger scatter than that shown in the comparison
between the C IV and He II fluxes. This larger scatter is partly
caused by the fact that the N V flux depends on Z BLR as well
as the relative abundance of N V (see, e.g., Matsuoka et al.
2011b; Araki et al. 2012), while the sum of the Si IV and O IV ]
fluxes mainly depends on Z BL
}


\subsection{Accretion parameters of sample sources}
\label{acc}



The bolometric luminosity has been computed assuming a flat $\Lambda$CDM cosmological model with $\Omega_{\Lambda}$ = 0.7, $\Omega_{m}$ = 0.3, and H$_0$ = 70 \kms{} Mpc$^{-1}$. Following \cite{marzianisulentic14} we decided to use  \aliii\ as virial broadening estimator for computing the \mbh. Our estimates adopt two different scaling laws: (1) the scaling laws of \citet{vestergaardpeterson06} for \civ\ and a second, unpublished one based on \aliii\ (del Olmo et al., in preparation).  Eddington ratios have been obtained using  the Eddington luminosity  $L_\mathrm{Edd} \approx 1.3 \times 10^{38} (M_\mathrm{BH}/M_{\odot})$ erg s$^{-1}$. The luminosity range of the sample is very limited, less than a factor 3, $46.8 \lesssim \log L \lesssim 47.3$, in line with the requirement of similar redshift and high flux values. Correspondingly, the \mbh\ and the Eddington ratio are constrained in the range $8.8 \lesssim \log $\mbh$ \lesssim 9.5$\ and $-0.55 \lesssim \log$ \lledd $\lesssim 0.18$, respectively.  The \mbh\ sample dispersion is   small, with $\log$ \mbh $\sim$ 9.4 $\pm 0.2$ [$M_\odot$]. The scatter in \mbh\ and \lledd\ is reduced to $\approx$ 0.1 dex if we exclude one object with the lowest \mbh\ and highest \lledd. Applying a small correction (10\%) to the FWHM to account for an excess broadening in \aliii\ due to non-virial motions will decrease the \mbh\ by 0.1 dex (as found by \citealt{negreteetal18} for \hb), and increase \lledd\ correspondingly. If this correction is applied the median \lledd\ is $\approx$0.6.   Using the \civ\ BC FWHM as a virial broadening estimator further decreases the \mbh\ median estimate by 0.1 dex.  The accretion  parameters are consistent with extreme quasars of Population A at high mass and luminosity; they are mainly at the low \lledd\  end of Sample 3 (based on \mbh\ estimates from \aliii) of \citet{marzianisulentic14}. The  small dispersion in physical properties of the present sample ($0.2$ dex) focuses the analysis on properties that may differ for fixed  accretion parameters, and fixed ratio of radiation and gravitation forces, perhaps related to different enrichment histories.


\subsubsection{{Correlation between diagnostic ratios and AGN physical properties}}

 
 \begin{figure}
     \centering
     \hspace{-1cm}
     \includegraphics[scale=0.3]{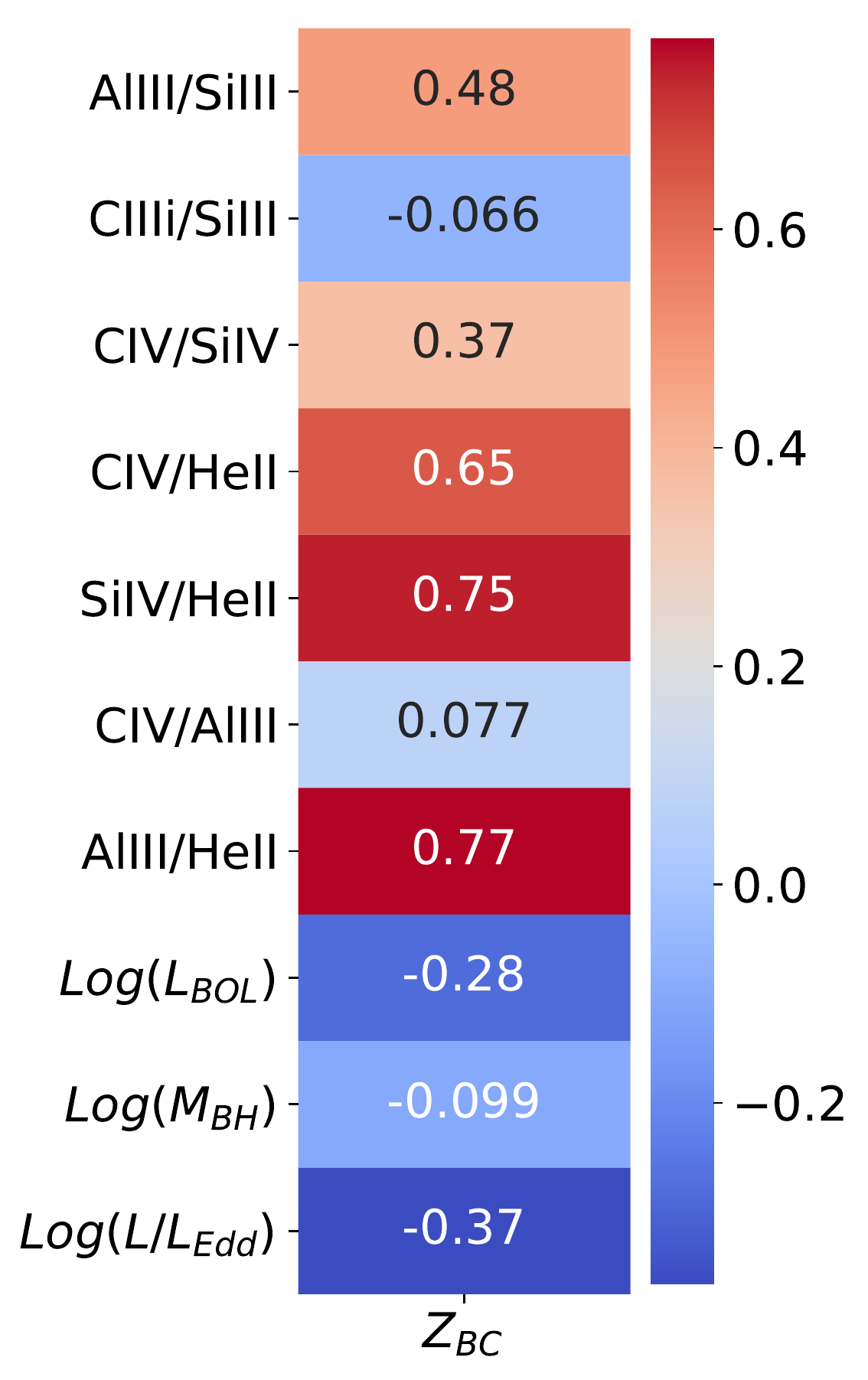}
     \includegraphics[scale=0.3]{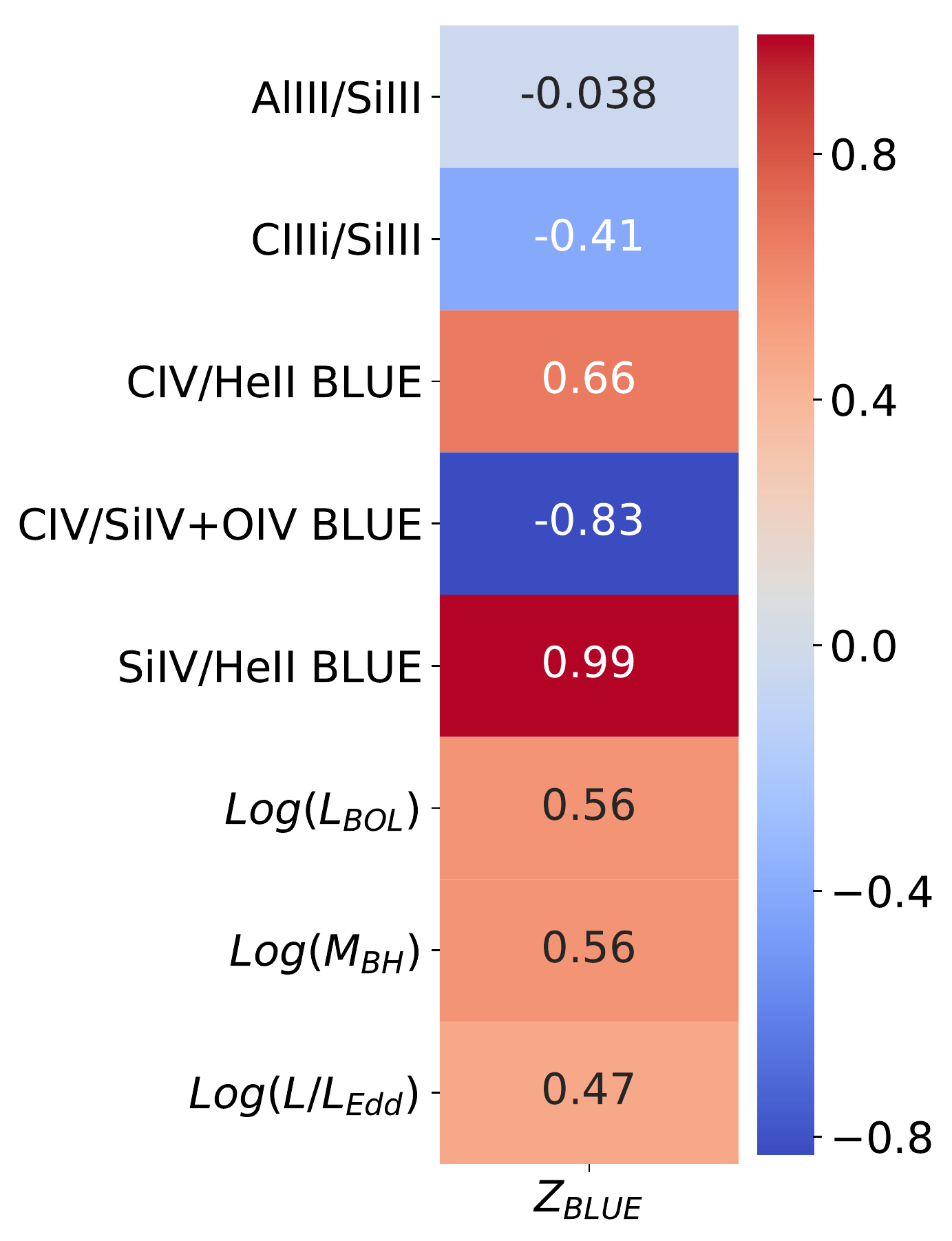}
     \caption{{{Left panel: the correlation matrix between $Z$\ computed  for the  BC, and BC diagnostic ratios assuming fixed \nh\ and $U$ along with log of bolometric luminosity, log of black hole mass and log of Eddington ratio. Right panel: Same, but for the BLUE component. The numbers in each square show the Spearman rank correlation coefficient. The color hue is proportional to the correlation, from dark blue (strong negative correlation) to red (strong positive correlation).}}}
     \label{fig:cor-ratios-phys-metallicites}
 \end{figure}

Considering the small dispersion in \mbh, \lledd\ and bolometric luminosity, it is hardly surprising that none of the ratios utilized in this paper is significantly correlated with the accretion parameter. The highest degree of correlation is seen between \lledd\ and \civ/\aliii, but still below the minimum $\rho$\ needed for a statistically significant correlation.  

In Figure \ref{fig:cor-ratios-phys-metallicites} we present the correlation between metallicity and diagnostic ratios along with log of bolometric luminosity, log of black hole mass and log of Eddington ratio for BC and BLUE. The strongest correlation between Z$_{BC}$ and intensity ratios are with \siiv/\heii\ (0.75) and \aliii/\heii\ (0.77). For Z$_\mathrm{BLUE}$, \siiv/\heii\ (BLUE components) correlates strongly (0.99). Z$_\mathrm{BLUE}$ correlates with physical parameters, whereas Z$_{BC}$ rather anti-correlates with them but not at a statistically significant level.  Considering the limited range in luminosity and \mbh, and small sample size, these trends should be confirmed. 

The metallicity values we derive are very high  among  quasars analyzed with similar techniques \citep[e.g., ][]{nagaoetal06,shinetal13,sulenticetal14}: as mentioned, typical values for high-$z$\ quasars are around $5 Z_\odot$. This value could be taken as a reference over a broad range of redshift, and also for the  sample considered in the present paper, as there  is no evidence of metallicity evolution in the BLR up to $z \approx 7.5$\ \citep[e.g., ][]{nagaoetal06,juarezetal09,xuetal12,onoueetal20}. This is in line with the results of \citet{negreteetal12} who found very similar intensity ratios for the prototypical NLSy1 and xA source I Zw 1, of relatively low luminosity at low $z$, and a luminous xA object at redshift $z \approx 3.23$. Even if these authors did not derive $Z$\ from their data, the I Zw 1 intensity ratios reported in their paper indicate very high metallicity consistent with the values derived for the present sample. 

More than inferences on the global enhancement of $Z$\ in the host galaxies, the absence of evolution  points toward a circumnuclear source of metal enrichment, ultimately associated with a Starburst \citep[e.g.,][]{collinzahn99,xuetal12}. 

 A detailed comparison with previous work on the dependence of $Z$ on accretion parameters is hampered by two difficulties. (1) Before comparing the intensity ratios of this paper, we should consider that other authors do not distinguish between BLUE and BC when computing the ratios. This has the unfortunate implications that in some cases such as \aliii/\civ, the ratio is taken between lines emitted predominantly in different regions (virialized and wind), presumably in very different physical conditions. Not distinguishing between BC and BLUE yields \civ/\aliii $\sim 10 \gg 1 $.  (2) Methods of \mbh\ estimate differ. For example \citet{matsuokaetal11} use the \citet{vestergaardpeterson06} scaling laws without any correction to the line width of \civ. This might easily imply overestimates of the \mbh\ by factor 5 -- 10 \citep{sulenticetal07}.  The analysis by \citet{shemmeretal04} instead used \hb\ from optical and IR observations to compute \mbh\ and to examine the dependence of metallicity on accretion parameters. These authors found the strongest dependence on Eddington ratio (with respect to luminosity and mass) over 6 orders of magnitude in luminosity, suggesting that luminosity and black hole mass are  less relevant (as also found, for example, by \citealt{shinetal13}).  


\subsection{A posteriori analysis of \nv\ strength \label{nv}}
\label{nv}

{As it was stressed in several works (e.g. \citealt{wangetal12,sulenticetal14}, the intensity of the \nv\ line is difficult to estimate due to blending with \lya\ and strongly affected by absorption. We model \lya\ and \nv\ using the same criteria as in \siiv\ modelling. However, in this work, we give  only a qualitative judgement of \nv\ strength for our sample, because of large uncertainties due to effect mentioned above. For sources in the highest metallicity range obtained from ratios from BC, the \nv\ broad component intensity is slightly higher or comparable to \lya\ broad component. Blue components dominate both lines. We notice also significantly higher intensity of blue component in comparison to broad one in \siiv\ and \civ\ blends. An example of  source of this type is shown in the upper half Fig. \ref{fig:obj_3816_ly}. On the contrary, sources with the lowest metallicities obtained from BC intensity ratios  show the \lya\ BC intensity higher than  in \nv\ and the BC is stronger than BLUE.  The same behaviour of strong broad component we see in the \siiv\ and \civ\ ranges. An example of sources of this type is shown in the lower panels of Fig. \ref{fig:obj_4077_ly}. \citet{shinetal13} compared \siiv+\oiv\ and \nv\ fluxes and found strong, significant correlation between them ($\rho$ = 0.75).  The \nv\ over \heii\ or \hb\ should be a strong tracer of $Z$, as it is sensitive to secondary $Z$\ production and hence proportional to $Z^2$ \citep{hamannferland99}. Therefore, we conclude that the \nv\ emission is extremely strong, and consistent with very high metal content. A much more thorough investigation of the quasar absorption / emission system is needed to include \nv\ as a $Z$\ estimator. This is deferred to further work. 
}

   \begin{figure*}
     \centering
     \includegraphics[scale=0.5]{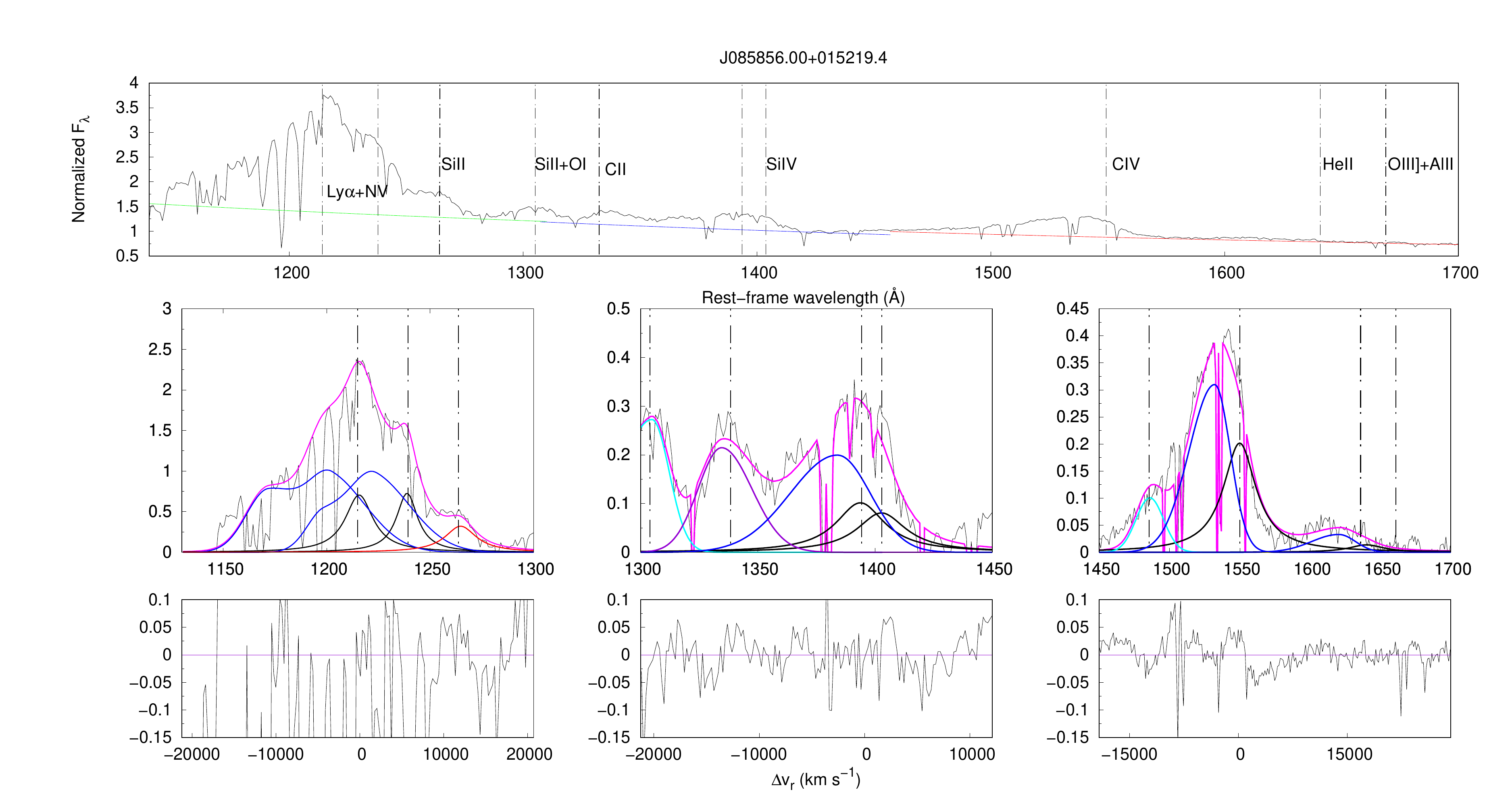}
       \includegraphics[scale=0.5]{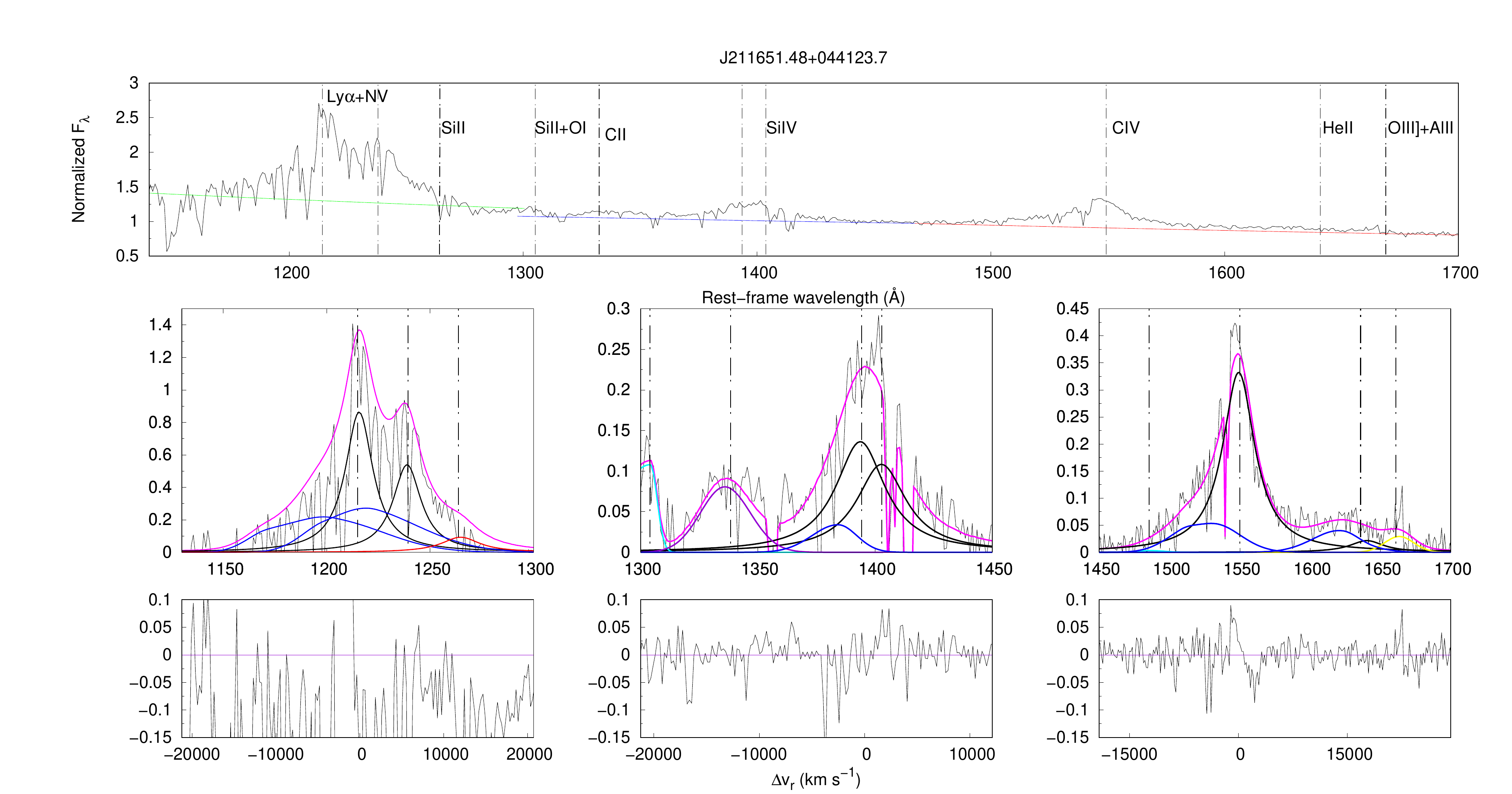}
   
     \caption{Analysis of sources showing the \lya\ + \nv\ blend. Top: calibrated rest-frame spectrum of \object{SDSS J085856.00+015219.4}. before continuum subtraction. 
      Global or local continuum are specified by a continuous coloured line, while the black line mark rest-framed data. Dot-dashed vertical lines correspond to the rest-frame wavelength of each emission line. Bottom: multicomponent fits after continuum subtraction for the L$\alpha \lambda$1216, SiIV$\lambda$1397 and CIV$\lambda$1549 spectral ranges. The continuous black line marks the broad component at rest-frame associated to Ly$\alpha\lambda$1216, NV$\lambda$1240, SiIV$\lambda$1397 and CIV$\lambda$1549. The blue one marks the blueshifted component associated to each emission. 
      The magenta line correspond to the fit to the whole spectrum. 
      In the left bottom panel, the red line corresponds to SiII$\lambda$1265.
 In the middle bottom panel, the light blue line marks the contribution of OI + SiII$\lambda$1304 blend, while the violet line corresponds to the CII$\lambda$1335 emission line. 
 In the CIV$\lambda$1549 region, NIV]$\lambda$1486 is represented by a light blue line, while the gold one corresponds to the OIII]$\lambda$1663 + AlII$\lambda$1670 blend.  Lower panels correspond to the residuals, in radial velocity units km s$^{-1}$ and in \AA.
 Bottom: Same as in the previous panels, for SDSS J211651.48+044123.7.}
     \label{fig:obj_3816_ly}     \label{fig:obj_4077_ly}
 \end{figure*}

\subsection{Role of column density}

The column density assumed in the present paper is $\log$\nc =$23$ [cm$^{-2}$]. With this value the emitting clouds in the low-ionization conditions remain optically thick to the Lyman continuum for most of the geometrical depth of the cloud. Even if the value  $\log$\nc =$23$\ may appear as a lower limit for the low-ionization BLR, as higher values are required to explain low-ionization emission such as \caii\ and \feii\ \citep{pandaetal20a,pandaetal20}, the emission of the intermediate and high-ionized region is confined within the fully ionized part of the line emitting gas whose extension is already  much less than the geometrical depth of the gas slab for $\log$\nc =$23$. Therefore, we expect no or negligible  effect from an increase in the column density for the low ionization part of the BLR.

For BLUE, the situation  is radically different, and we have no actual strong constraints on column density. Most emission may come from a clumpy outflow \citep[][and references therein]{matthews16}, and therefore assuming a constant \nc\ may not be appropriate. Considering the poor constrain that we are able to obtain, we leave the issue to an eventual investigation.

\subsection{Role of turbulence}
\label{turb}

\begin{figure}
\includegraphics[scale=0.45]{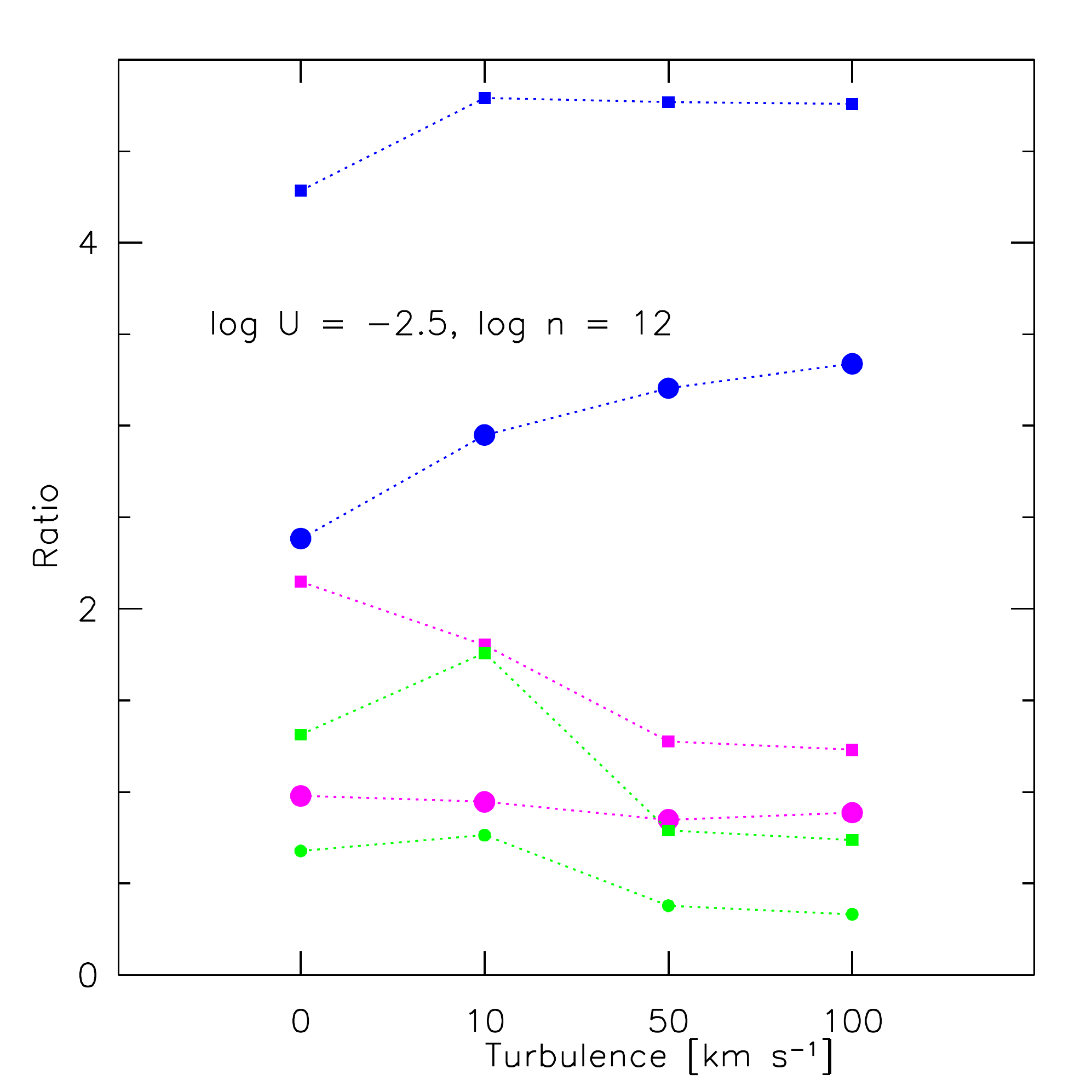}\\
\includegraphics[scale=0.45]{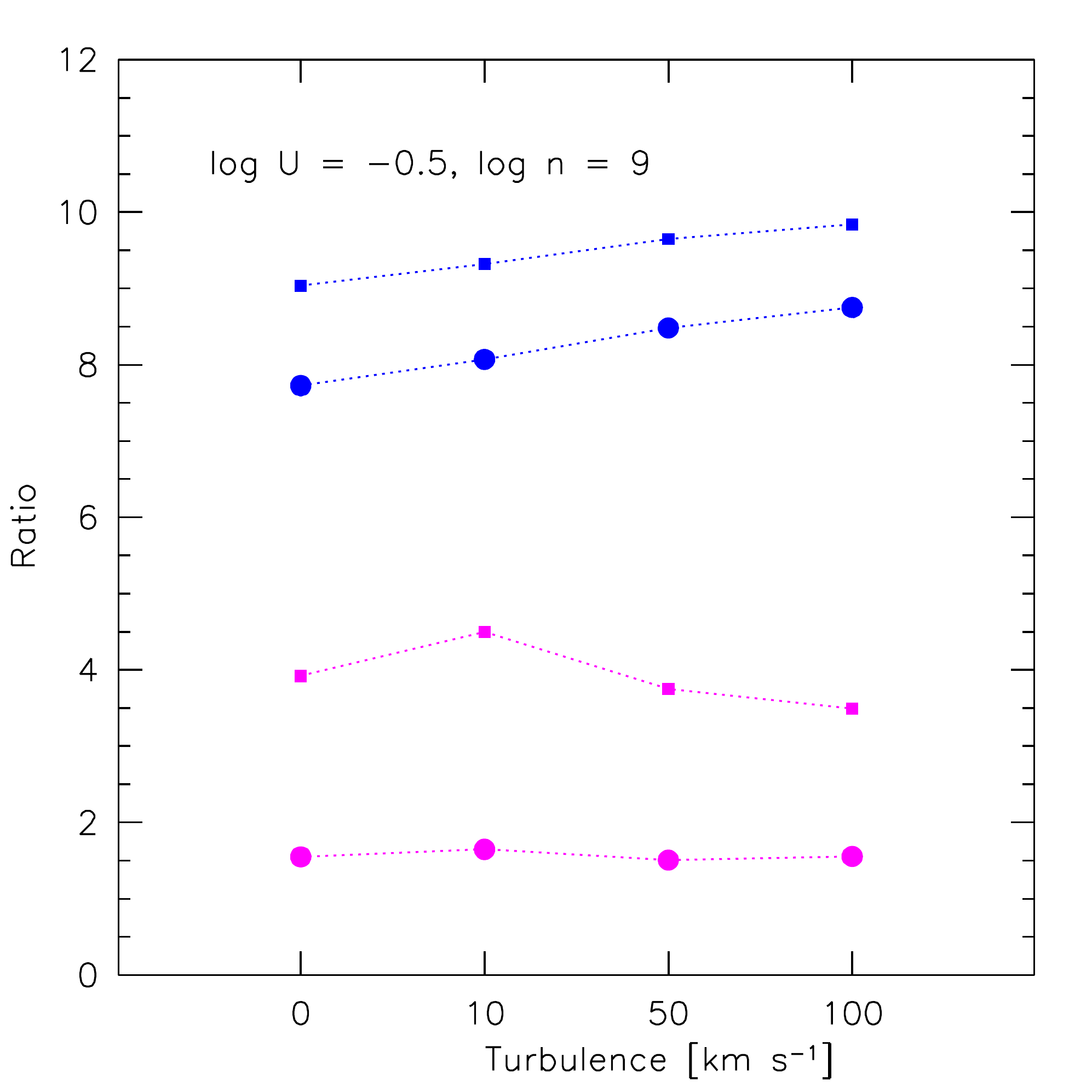}
\caption{Effects of turbulence on diagnostic line ratios, for \civ/\heii\ (blue) and \siiv/\heii (magenta), considering 5 (circles) and 20 $Z\odot$\ (squares),  according to {\tt CLOUDY} 13.05 computations. The top panel assumes the low ionization conditions appropriate for   BC emission, the bottom one for  BLUE. In the top panel the green lines trace the same trends for the FeII blend at $\lambda 4570$ over H$\beta$ ratio i.e., \rfe.  \label{fig:turb}}
\end{figure}

The results presented in this work refer to the case in which there is no significant micro turbulence included in the {\tt CLOUDY} computations. Fig. \ref{fig:turb} shows that at low ionization the effect is relatively modest, and that in the high-ionization case appropriate for BLUE  the effect is very modest.  Less obvious is the behavior at low-ionization for \rfe: it shows an increase for $t = 10$ km s$^{-1}$, but then it has a surprising drop at larger value of the micro-turbulence.  While the increase can be explained by an increase of the transitions for which fluorescence is possible, the decrease is not of obvious interpretation. It has been however confirmed by the independent set of simulation of \citet{pandaetal18,pandaetal19}.

\subsection{Metal segregation?}
\label{sec:segregation}

Metals are expected to be preferentially accelerated by resonance scattering \citep[e.g.,][]{proga07,risalitielvis10}. In principle, for a sufficiently large photon flux, the acceleration of metals by radiation pressure might become  larger than the Coulomb friction, therefore causing a decoupling of the metals with respect to their parent plasma \citep{baskinlaor12}. This possibility has been explored in the context of the BALs, and broad absorption and emission components are expected to be related \citep{elvis00,xuetal20}.  The ionization parameter values are however several orders of magnitudes higher than the ones derived for the BLUE emission component.  In addition our $Z$ estimates for the BLUE suggest, if anything, values lower or equal than  for the BC, whose $Z$\ might be related more to the original chemical composition of the gas in the accretion material. However, we ascribe the systematic differences between BC and BLUE as uncertainties in the method and measurement, so that $Z$ from BLUE and BC should be considered intrinsically equal.   

Considering that the most metal rich stars, galaxies, and molecular clouds  in the Universe do not exceed $Z \approx 5 Z_\odot$  \citep{maiolinomannucci19}, circumnuclear star formation is needed for the chemical enrichment of the BLR gas \citep[e.g.,][]{collinzahn99,collinzahn99a,wangetal11a,wangetal12a}. Star formation may occur in the self gravitating, outer part of the disk. An alternative possibility is that a massive star could be formed inside the disk by accretion of disk gas \citep{cantielloetal20}.

\subsection{Abundance pollution?}
\label{pollu}

An implication of the scenarios involving circumnuclear or even nuclear star formation  is that there could be an alteration of the relative abundance of elements with respect to the standard solar composition  \citep{andersgrevesse89,grevessesauval98}. Support for this hypothesis is provided by the extreme \civ/\siiv\ and \civ/\aliii\ that may hint to a selective enhancement  of Al with respect to C. As suggested by \citet{negreteetal12}, core-collapse supernov\ae\ with very massive progenitors could be at the origin of a selective enhancement. Supernovae with progenitors of masses between 15 and 40 $M_\odot$\ have selective enhancement in their yields of Al and Si by factors of $\approx$ 100 and 10 relative to Hydrogen with respect to solar \citep{chieffilimongi13}. Since Carbon is also increased by  a factor $\sim$10 with respect to solar, the [Al/C] is expected to be a factor $\sim 10$\ the solar value in supernova ejecta. The case for Silicon is less clear, as the enhancement is of the same order of magnitude of the one expected for Carbon.  Pollution of gas by supernov\ae\ may therefore lead to an estimate of the $Z$ higher than the actual one, if  solar relative abundances are assumed. This possibility will be explored in an eventual work (Garnica et al., in preparation). 


\subsection{Implications for quasar structure evolution}
\label{quasar_evolution}

Metallicity and the outflow prominence of quasars were found to be highly correlated \citep{wangetal12,shinetal17}. The implication of these results is that xA sources, which show the highest blueshifts \citep{sulenticetal17,vietrietal18,martinez-aldamaetal18,martinez-aldamaetal18a}, should also be the most metal rich. The xA sources should be at the top of the $Z$ outflow parameter correlation of \citet{wangetal12}, if $Z \gtrsim 10 Z_\odot$. 

There is evidence of a metallicity  correlation between BLR and NLR \citep{duetal14}, as expected if the outflows on spatial scales of kpc are originating in a disk wind. \citet{zamanovetal02} derived very small spatial scales   at low luminosity. This provides additional support to the idea that xA sources -- which at low-$z$ phenomenologically appear as \feii-strong NLSy1s, are relatively young sources. Their low \oiii\ equivalent width implied young age more than orientation effects \citep{risalitietal11,bisognietal17}. The $z\approx 2$\ quasars of the present sample are radiating at relatively high \lledd\ although there are no examples of the extremes of xA sources showing blueshifted emission in \aliii\ as prominent as the one of \civ\ \citep[e.g.,][]{martinez-aldamaetal17}. There is no evidence of heavy obscuration. They are certainly out of the obscured early evolution stage in which the accreting black hole is enveloped by gas and dust (see the sketch in \citealt{donofriomarziani18}). The $W$\ \civ\ distribution  covers the upper half of the one of \citet{martinez-aldamaetal18a}. There are no weak-lined quasars following \citet{diamond-stanicetal09}.  The xA sources of the present sample may have reached a sort of stable equilibrium between gravitation and radiation forces made perhaps possible by the development by an optically thick, geometrically thick accretion disk, and by its anisotropic radiation properties \citep[e.g., ][]{abramowiczetal88,szuszkiewiczetal96,sadowskietal14}.

The median value of the peak displacement of the BLUE component is around $\approx 3500$ \kms, and the centroid at half maximum is shifted by 5000 \kms.   The extreme blueshifts in the metal lines imply  outflows that may  not remain bound to the potential well of the black hole and of the inner bulge of the host galaxies \citep[e.g.,][and references therein]{marzianietal16}. The high metal content of the outflows, estimated by the present work to be in the range $10 - 50 Z_\odot$, implies that these sources are likely to be a major source of metal enrichment of the interstellar gas of the host galaxy and of the intergalactic medium.  Using a standard estimate for the mass outflow rate $\dot{M}$\ \citep{marzianietal16}, $\dot{M} \approx 15 L_\mathrm{CIV,45} v_{5000} r_\mathrm{1pc}^{-1} n_9^{-1}$ $M_\odot$ yr$^{-1}$, we obtain an outflow rate of $ \dot{M}  \approx
20 M_\odot$ yr$^{-1}$, assuming   median values for the sources of our sample: median outflow velocity from the peak of BLUE $\approx -3500$ \kms, a median luminosity of the \civ\ BLUE (corrected because of Galactic extinction) of 4.2$\cdot$ 10$^{44}$ erg s$^{-1}$, a median radius 5.9 $\cdot$ 10$^{17}$\ cm from the \citet{kaspietal07} radius-luminosity correlation for \civ, and $n_9 =1$. For a duty cycle of $\sim 10^8$ yr, the expelled mass of heavily enriched-gas  could be $\sim 10^9 M_\odot$.

\section{Conclusion}

The sources at the extreme end of Population A along the main sequence are defined by the prominence of their \feii\ emission and, precisely, by the selection criterion \rfe $\gtrsim 1$\ \citep{marzianisulentic14,duetal16}.
Their properties as a class are scarcely known. Even if there has been a long history of studies focused on \feii-strong sources since \citet{liparietal93,grahametal96}, their relevance to galactic and large scale structure evolution is being reconsidered anew with the help of the quasar main sequence. This paper adds to other aspects that were considered by previous investigations (for example, the very powerful outflows, the disjoint low- and high-ionization emitting regions, first suggested by \citealt{collinsouffrinetal88}), a quantitive analysis of the chemical composition of xA sources.  The main aspects of the present investigations can be summarized as follows:

\begin{itemize}
    \item We distinguish between two emission line components most likely origination from emitting in widely different physical conditions: a virialized low-ionization BLR, and a high-ionization region associated to a very strong blueshifted excess in the \civ\ emission line. This is the {\em conditio sine qua non} for meaningful $Z$\ estimates.  
       \item The physical conditions in the low and high regions were confirmed to be very different, with the low ionization ($U$, \nh) $\approx$(-2.75, 12.5 -- 14) and the high ionization ($U$, \nh) $\approx$(-0.5, 8). The high ionization region parameters are however poorly constrained. 
   \item Using intensity ratios between the strongest metal lines and  \heii\ emission at $\lambda 1640$\ we derive metallicity values in the range $10 \lesssim Z \lesssim 50 Z_\odot$, with most likely values around several tens of the time solar metallicity. Incidentally, we note that the low equivalent width is consistent the high $Z$\ of the emitting regions.
   \item We find evidence of overabundance of Al  with respect to C. This result points toward possible pollution of the broad line emitting gas chemical composition by supernova ejecta.  
  \end{itemize}

 xA quasars are perhaps the only quasars whose ejection are able to overcome the potential well of the black hole and of the host galaxy.  Applying the method to large samples of quasars would permit to constrain the metal enrichment processes on a galactic scale. 

The present analysis relied heavily on the \heii\ line which is of low equivalent width and with a flat, very broad profile. Therefore, a more precise analysis would require spectra of moderate dispersion but of higher S/N ratio. A large part of the scatter and/or systematic difference for various $Z$\ estimators is related to the difficulty to isolate faint and broad emission in relatively noisy spectra.
\vfill

\section*{ACKNOWLEDGEMENTS}

MS acknowledges the support of the Erasmus+ programme of the European Union and would like to express very great appreciation to Istituto Nazionale di Astrofisica (INAF) Osservatorio Astronomico
di Padova, University of Padova and Astronomical Observatory of the University of Warsaw for enabling to complete an internship. The project was partially supported by the Polish Funding Agency National Science Centre project 2017/26/\-A/ST9/\-00756 (MAESTRO  9) and MNiSW grant DIR/WK/2018/12.  PM  acknowledges  the Hypatia of Alexandria visiting grant SO-IAA (SEV-2017-0709) through the Center of Excellence Severo Ochoa, and is deeply indebted to Drs. J. Perea and A. del Olmo for the generous allocation of computing resources and for a stay at IAA. AdO acknowledges financial support from the Spanish grants MCI PID2019-106027GB-C41 and the State Agency for Research of the Spanish MCIU through the “Center of Excellence Severo Ochoa” award for the Instituto de Astrof\'{\i}sica de Andaluc\'{\i}a
(SEV-2017-0709).

Funding for the Sloan Digital Sky Survey (SDSS) has been provided by the Alfred P. Sloan Foundation, the Participating Institutions, the National Aeronautics and Space Administration, the National Science Foundation, the U.S. Department of Energy, the Japanese Monbukagakusho, and the Max Planck Society. The SDSS Web site is http://www.sdss.org/.

The SDSS is managed by the Astrophysical Research Consortium (ARC) for the Participating Institutions. The Participating Institutions are The University of Chicago, Fermilab, the Institute for Advanced Study, the Japan Participation Group, The Johns Hopkins University, Los Alamos National Laboratory, the Max-Planck-Institute for Astronomy (MPIA), the Max-Planck-Institute for Astrophysics (MPA), New Mexico State University, University of Pittsburgh, Princeton University, the United States Naval Observatory, and the University of Washington.
\vfill


\pagebreak\pagebreak
\bibliographystyle{apj} 
\bibliography{main}
\pagebreak

\begin{appendix}
\section{Rest-frame spectra and fits}
\label{app:spec}

The spectral analysis of the 13 objects of our sample is shown in the figures below. 

\begin{figure}[h]
     \centering
     \includegraphics[scale=0.37]{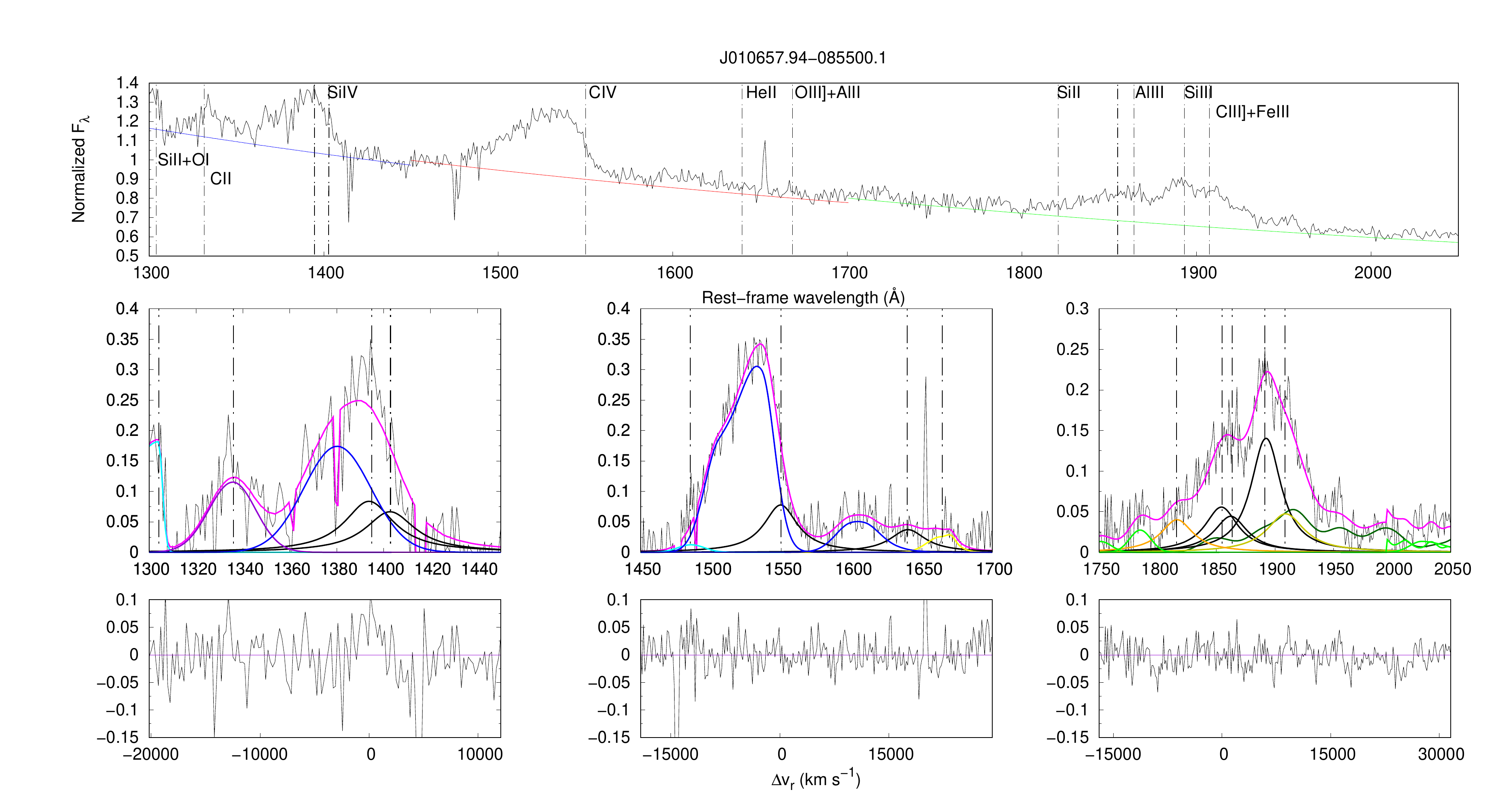}
     \caption{Top panels: calibrated rest-frame spectrum of SDSS J010657.94-085500.1 before continuum substraction. 
      Global or local continuum are specified by a continuous coloured line, while the black line mark rest-framed data. Dot-dashed vertical lines correspond to the rest-frame wavelength of each emission line. Bottom: multicomponent fits after continuum subtraction for the SiIV$\lambda$1397, CIV$\lambda$1549 and $\lambda$1900 blend spectral ranges. The continuous black line marks the broad component at rest-frame associated to SiIV$\lambda$1397,  CIV$\lambda$1549 and AlIII$\lambda$1860, the blue one marks the blueshifted component associated to each emission. 
      The magenta line correspond to the fit to the whole spectrum. 
 In the left bottom panel, the light blue line marks the contribution of OI + SII$\lambda$1304 blend, while the violet line corresponds to the CII$\lambda$1335 emission line. 
 In the CIV$\lambda$1549 region, NIV]$\lambda$1486 is represented by a light blue line,
while the gold one corresponds to the OIII]$\lambda$1663 + AlII$\lambda$1670 blend. In the $\lambda$1900 blend range, FeIII and FeII contributions are marked dark and light green lines respectively.  Violet lines mark the NIII]$\lambda$1750, the orange one corresponds to the SIII$\lambda$1816 and the gold one to the CIII]$\lambda$1909 line. Lower panels correspond to the residuals, in radial velocity units km s$^{-1}$ and in \AA.}
     \label{fig:obj_0658}
 \end{figure}
 
 \addtocounter{figure}{-1}

\begin{figure}
     \centering
     \includegraphics[scale=0.37]{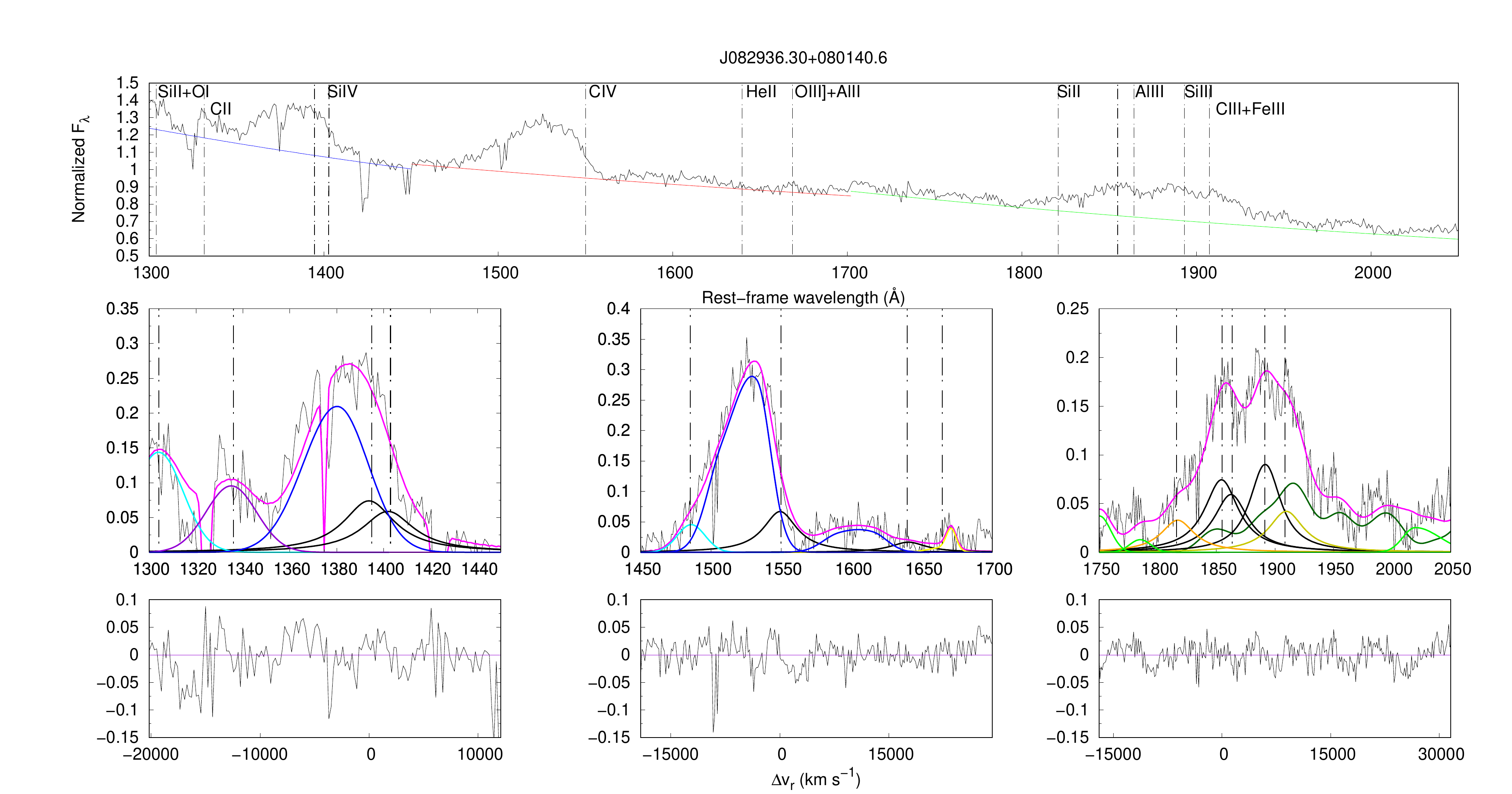}\\
     \vspace{-0.25cm}
     \includegraphics[scale=0.37]{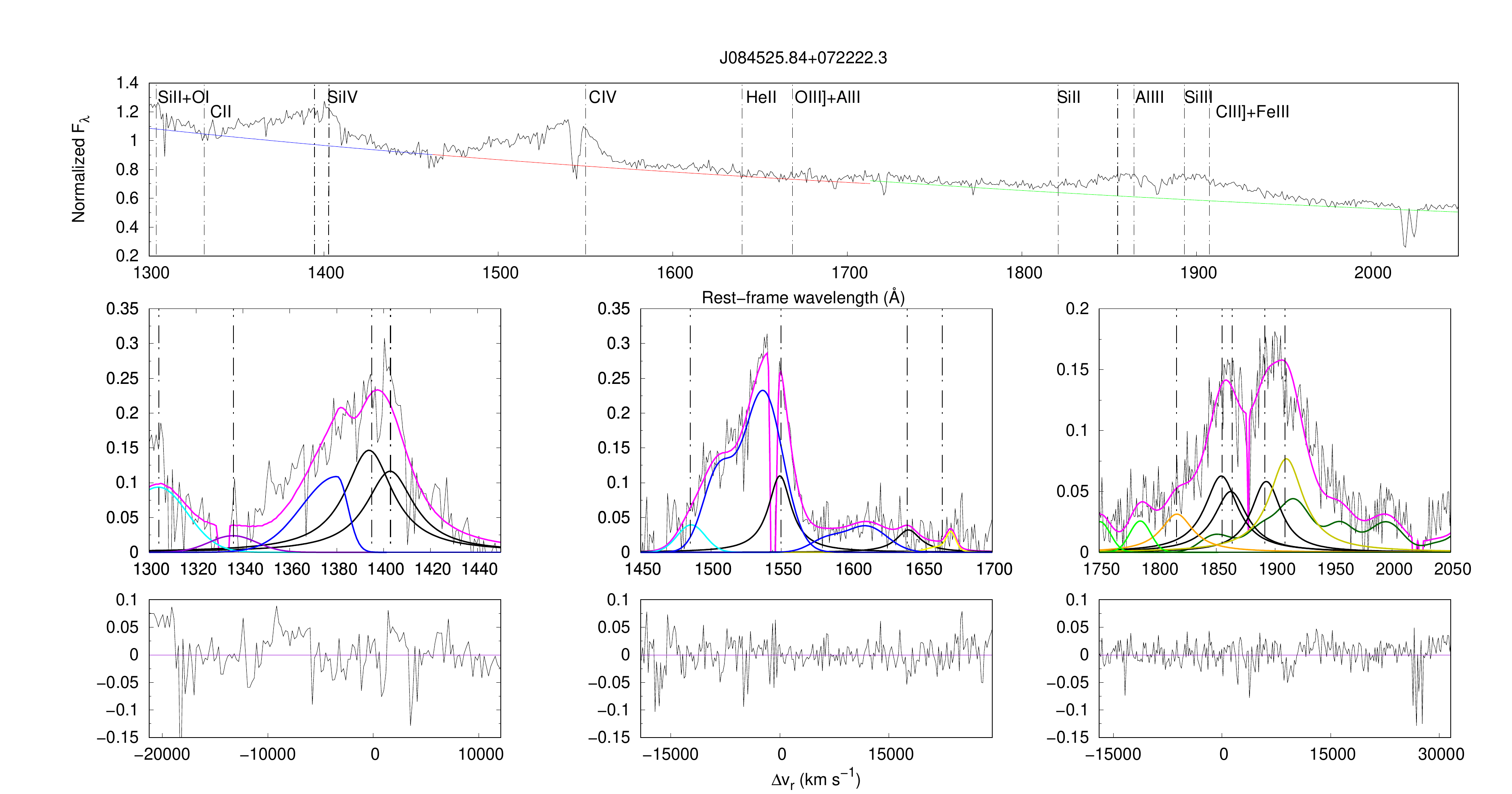}\\
     \vspace{-0.25cm}
     \includegraphics[scale=0.37]{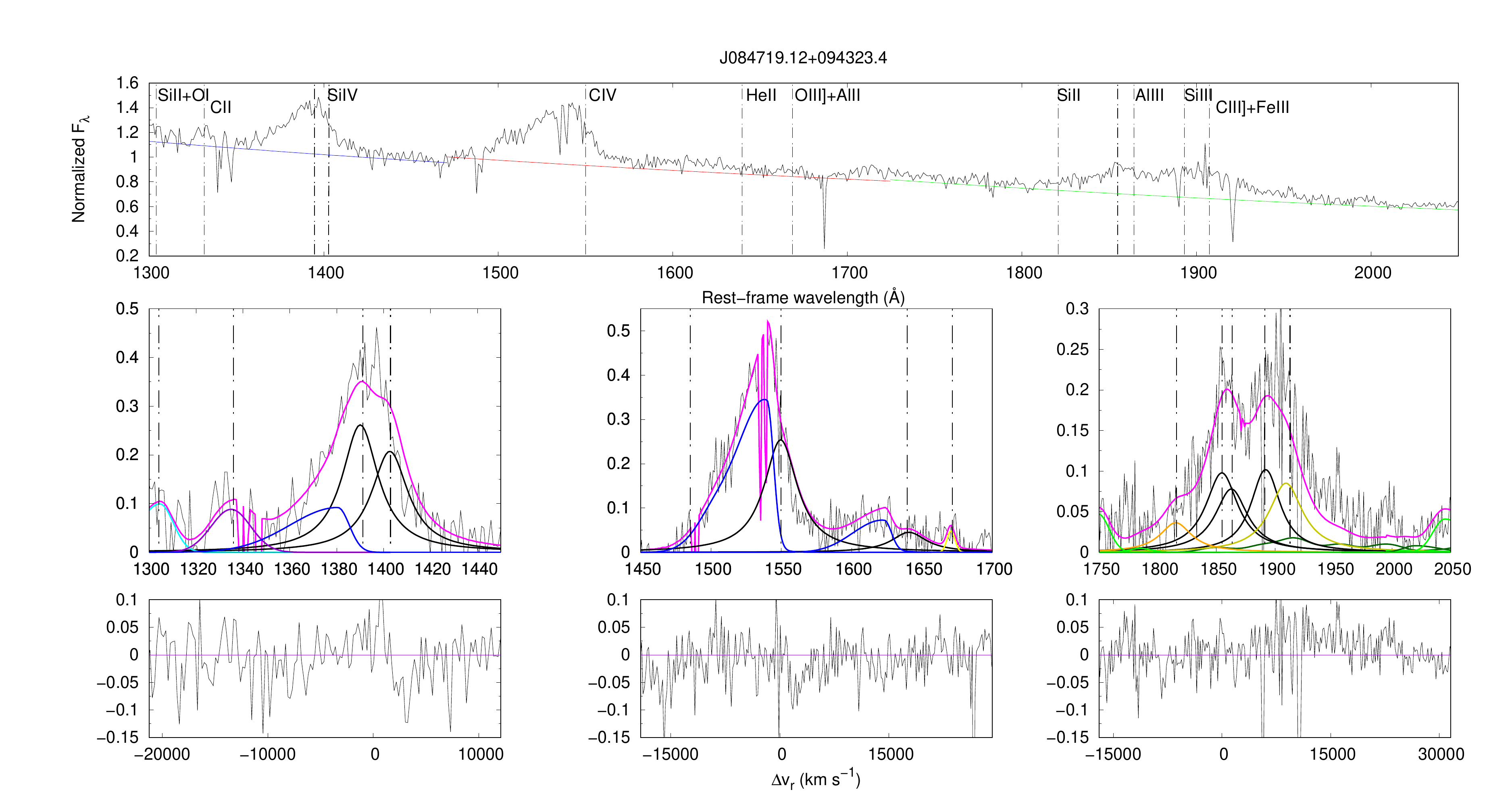}
     \caption{Same of the previous panel, for SDSS J082936.30+080140.6 and SDSS J084525.84+072222.3, and SDSS J084719.12+094323.4.}
     \label{fig:obj_4866}
 \end{figure} 
 \addtocounter{figure}{-1}
 

  \begin{figure}
     \centering
     \includegraphics[scale=0.37]{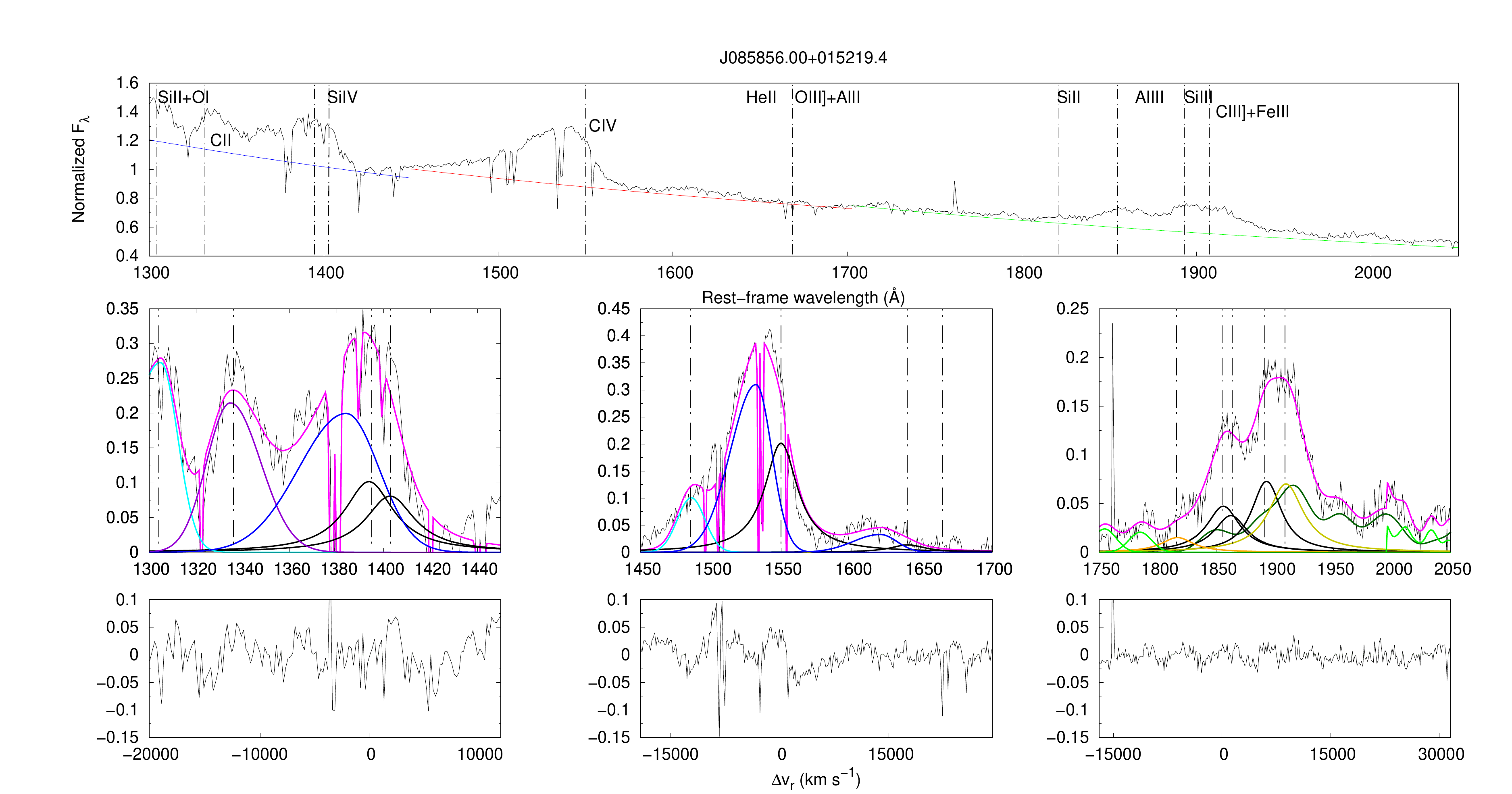}
      \vspace{-0.25cm}
     \includegraphics[scale=0.37]{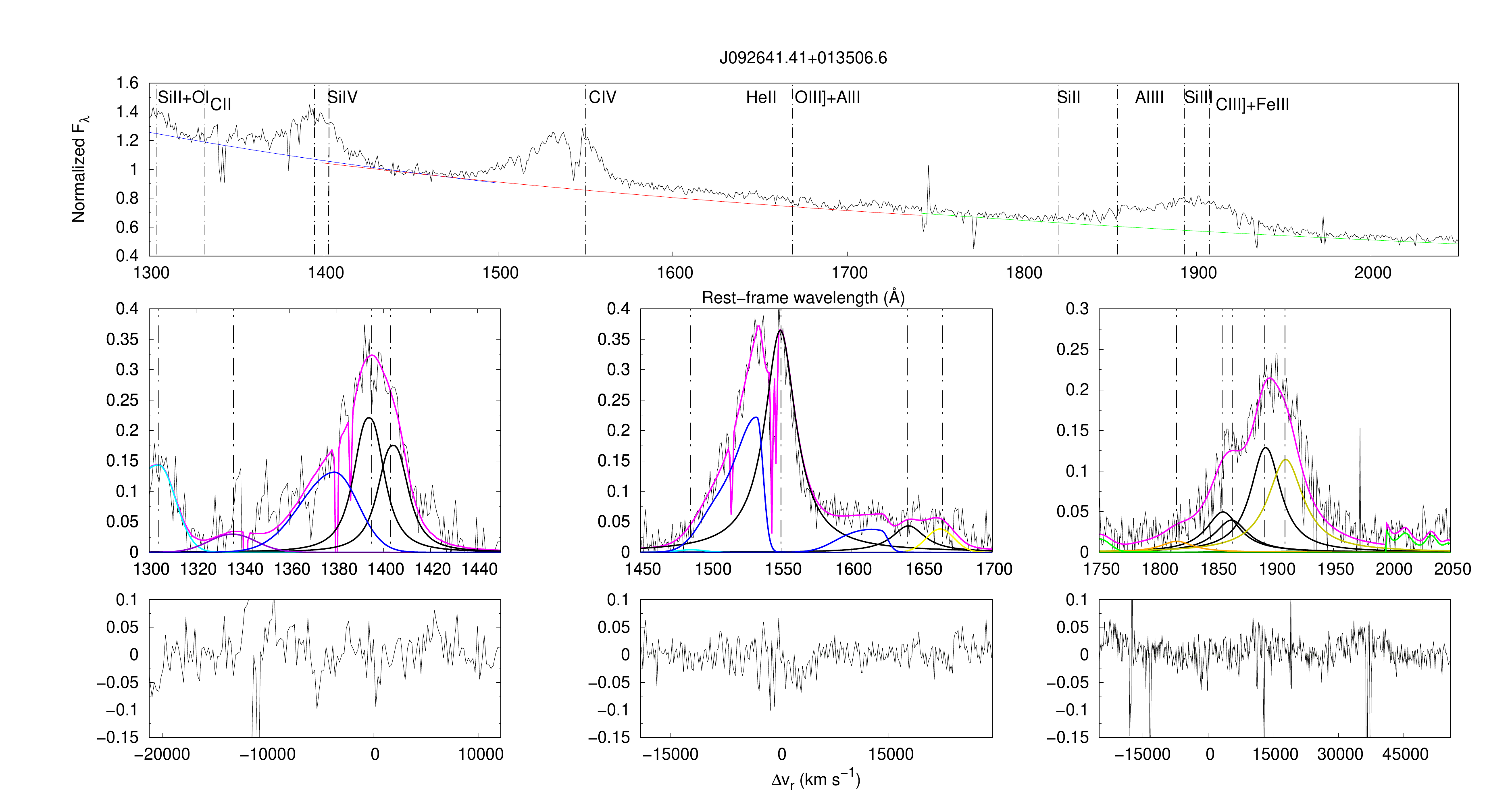}
        \vspace{-0.25cm}
     \includegraphics[scale=0.37]{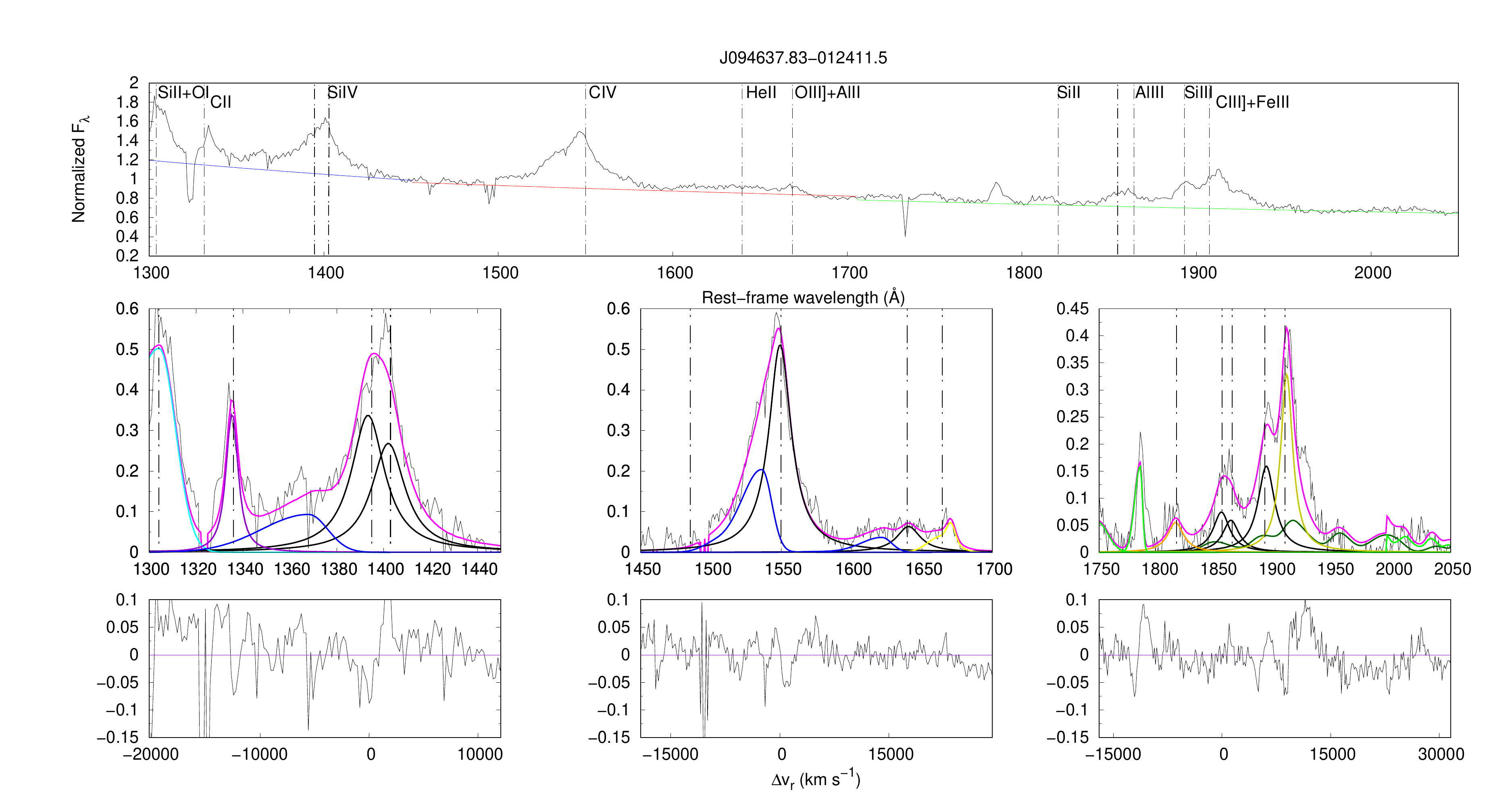}
     \caption{Same of the previous panel, for  SDSS J085856.00+015219.4, SDSS J092641.41+013506.6, and SDSS J094637.83-012411.5.}
     \label{fig:obj_3782}
 \end{figure}
\addtocounter{figure}{-1}

 
   \begin{figure}
     \centering
     \includegraphics[scale=0.37]{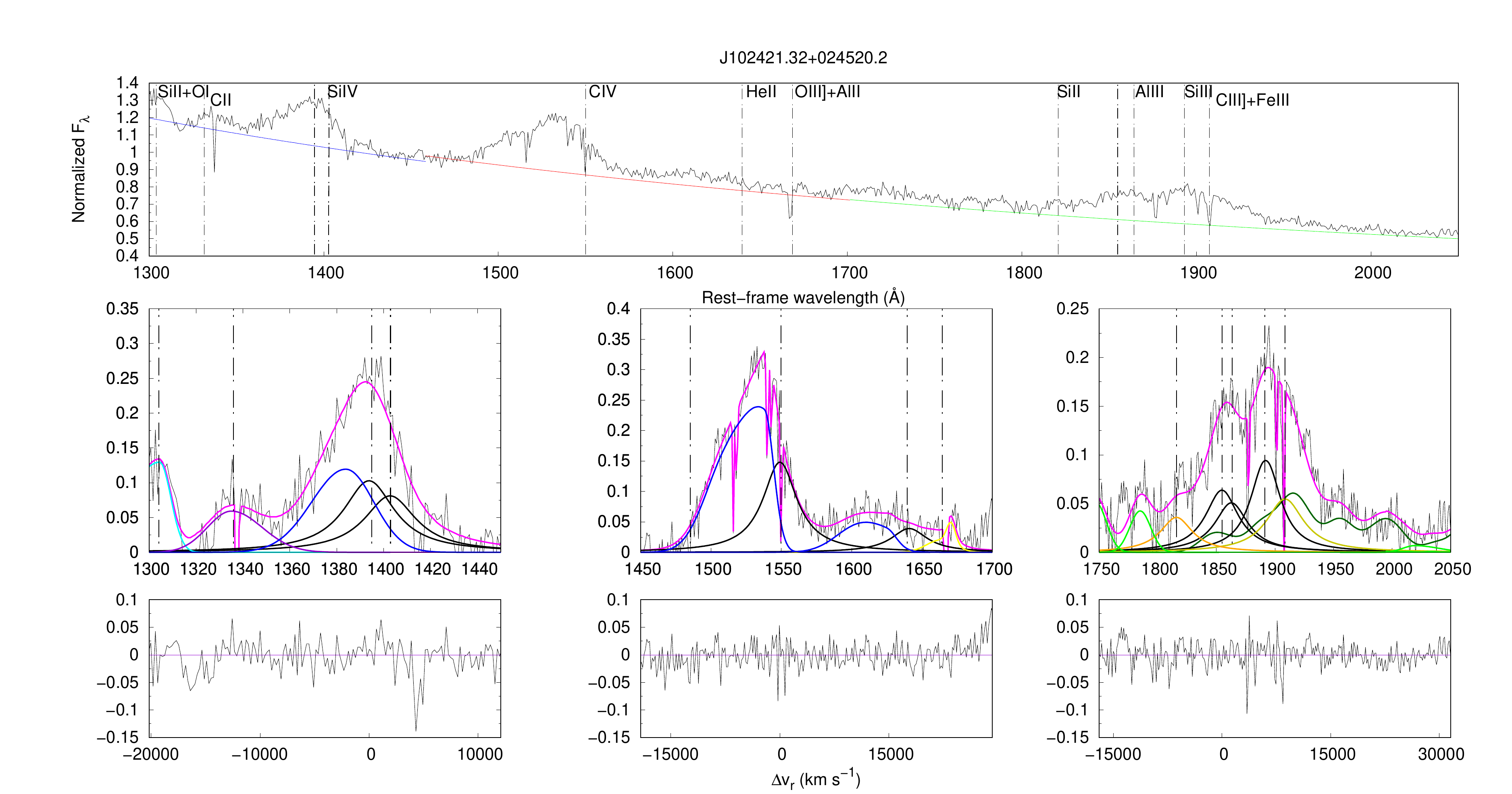} 
       \vspace{-0.25cm}
       \includegraphics[scale=0.37]{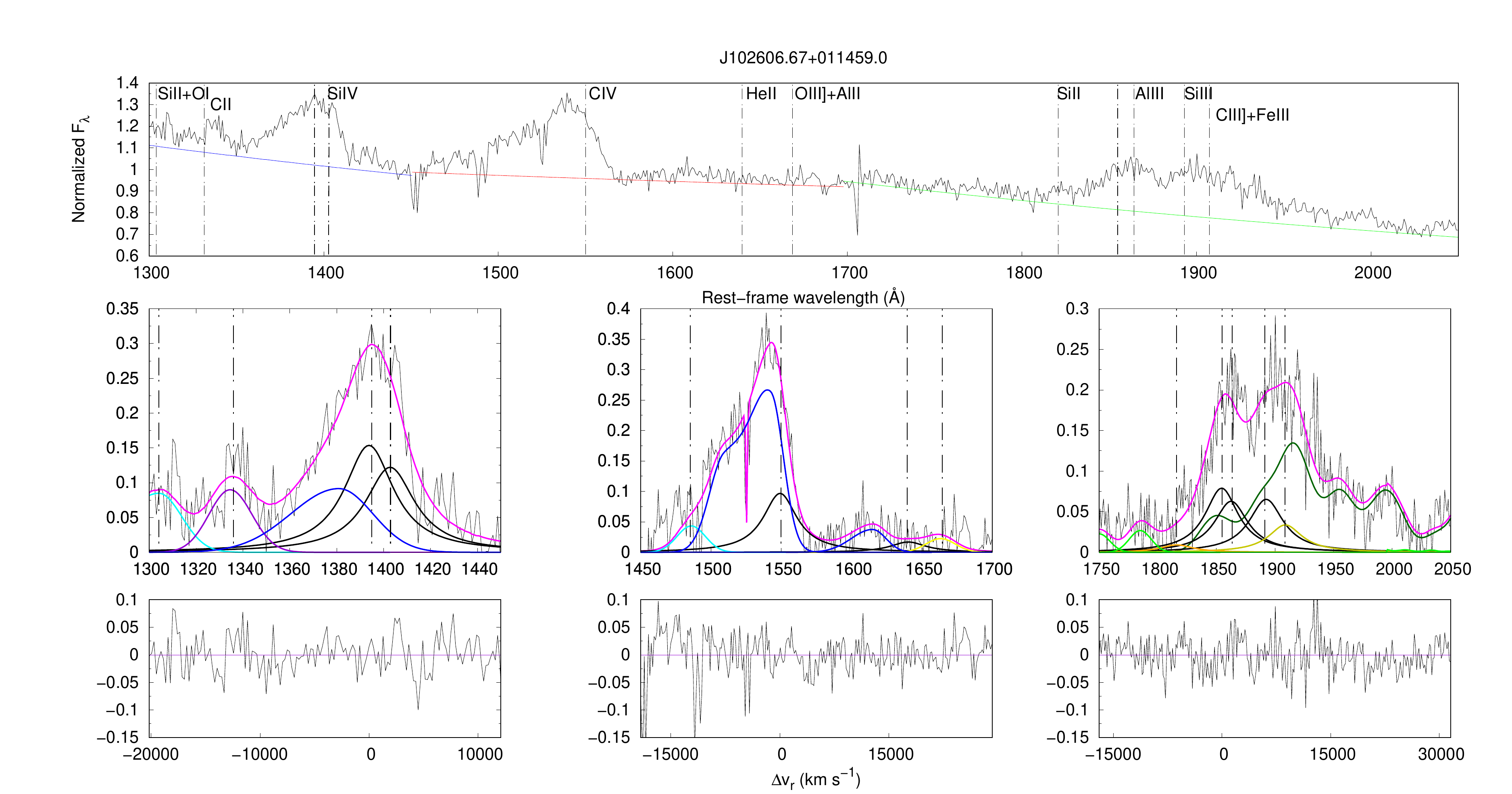}
           \vspace{-0.25cm}
          \includegraphics[scale=0.37]{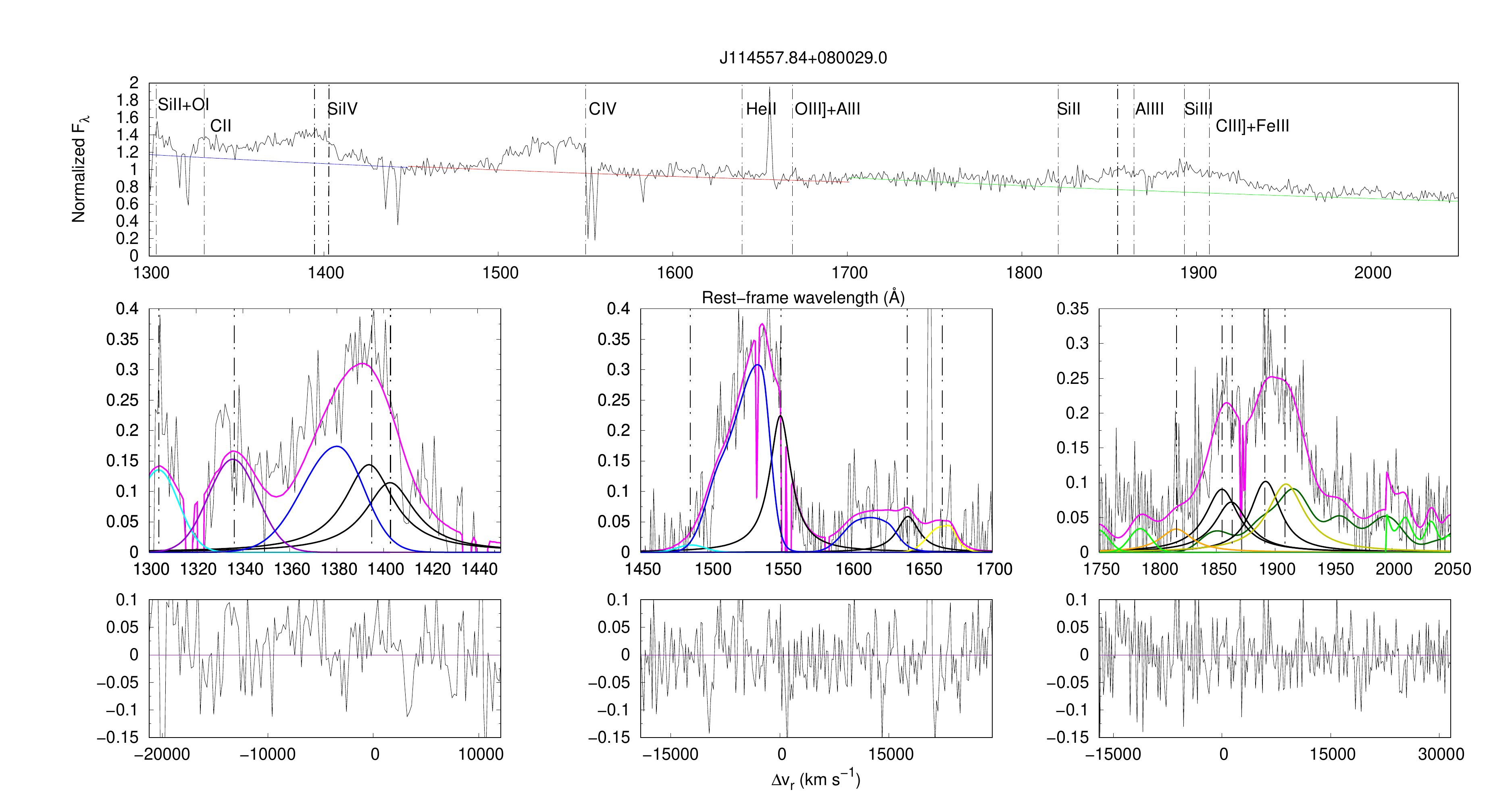}
       \caption{Same of the previous panel, for SDSS J102421.32+024520.2 SDSS J102606.67+011459.0 SDSS J114557.84+080029.0.}
     \label{fig:obj_4739}
 \end{figure}
 \addtocounter{figure}{-1}
 
    \begin{figure}
     \centering
     \includegraphics[scale=0.37]{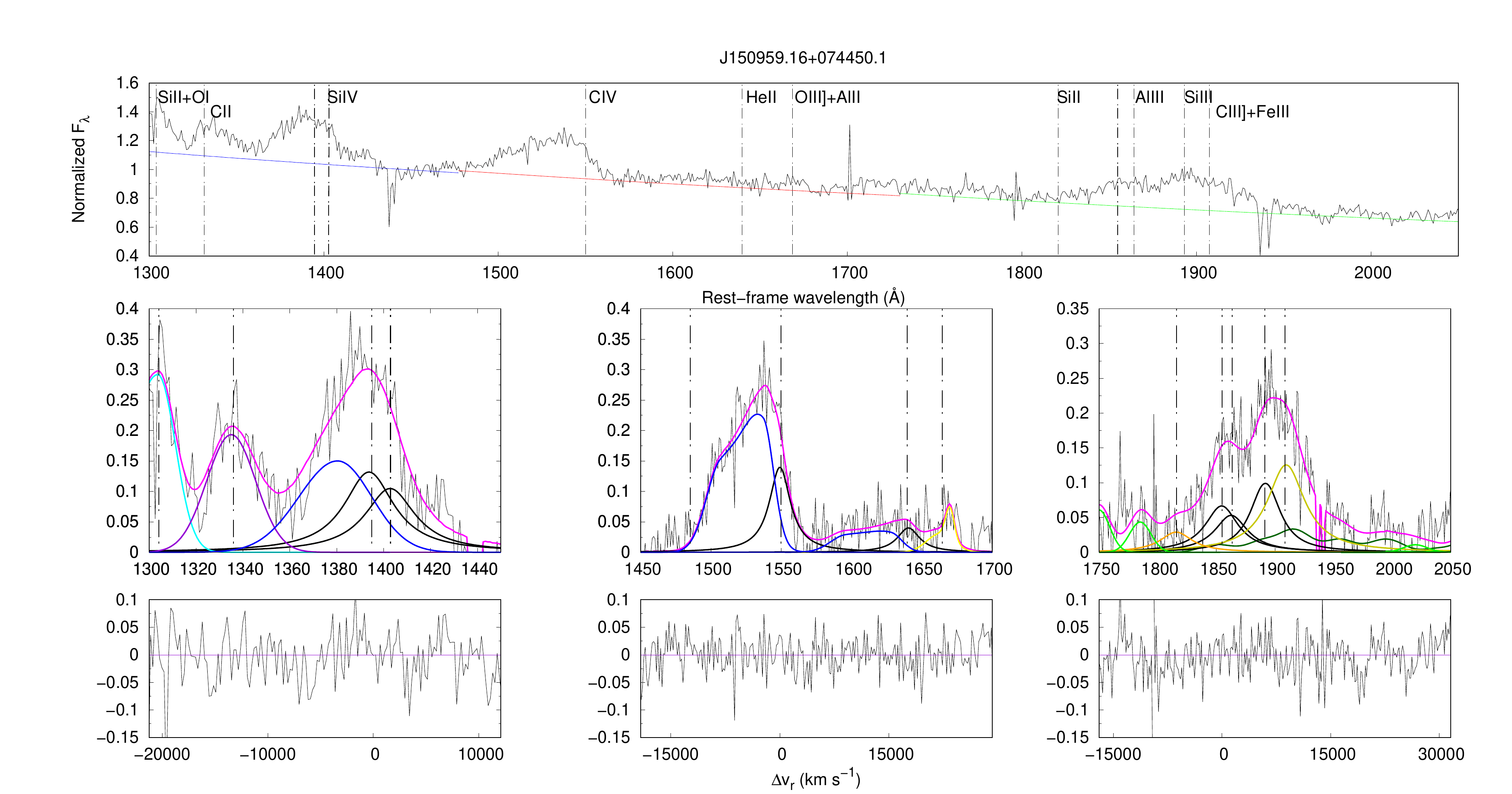}
      \includegraphics[scale=0.37]{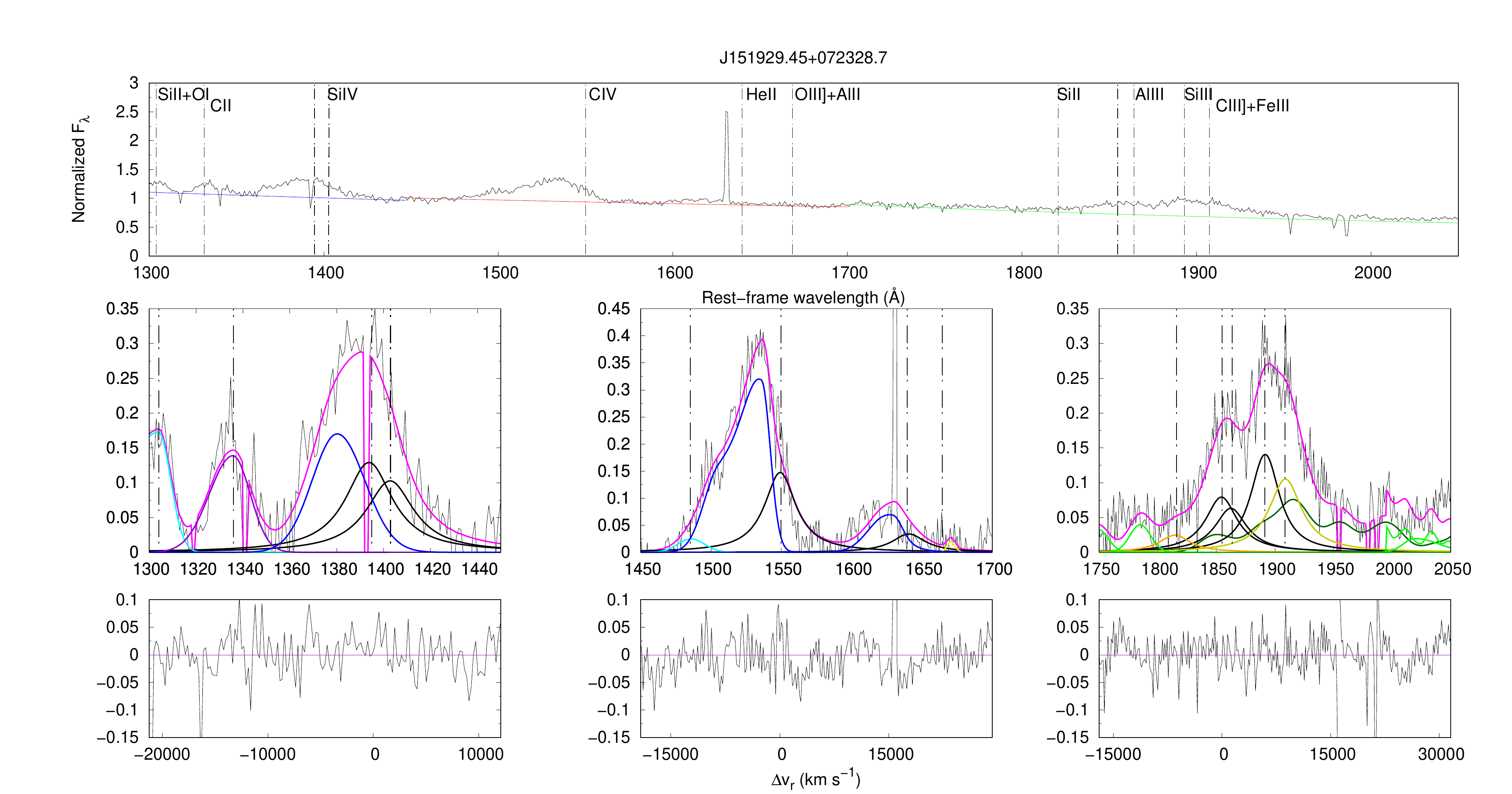}
  \includegraphics[scale=0.37]{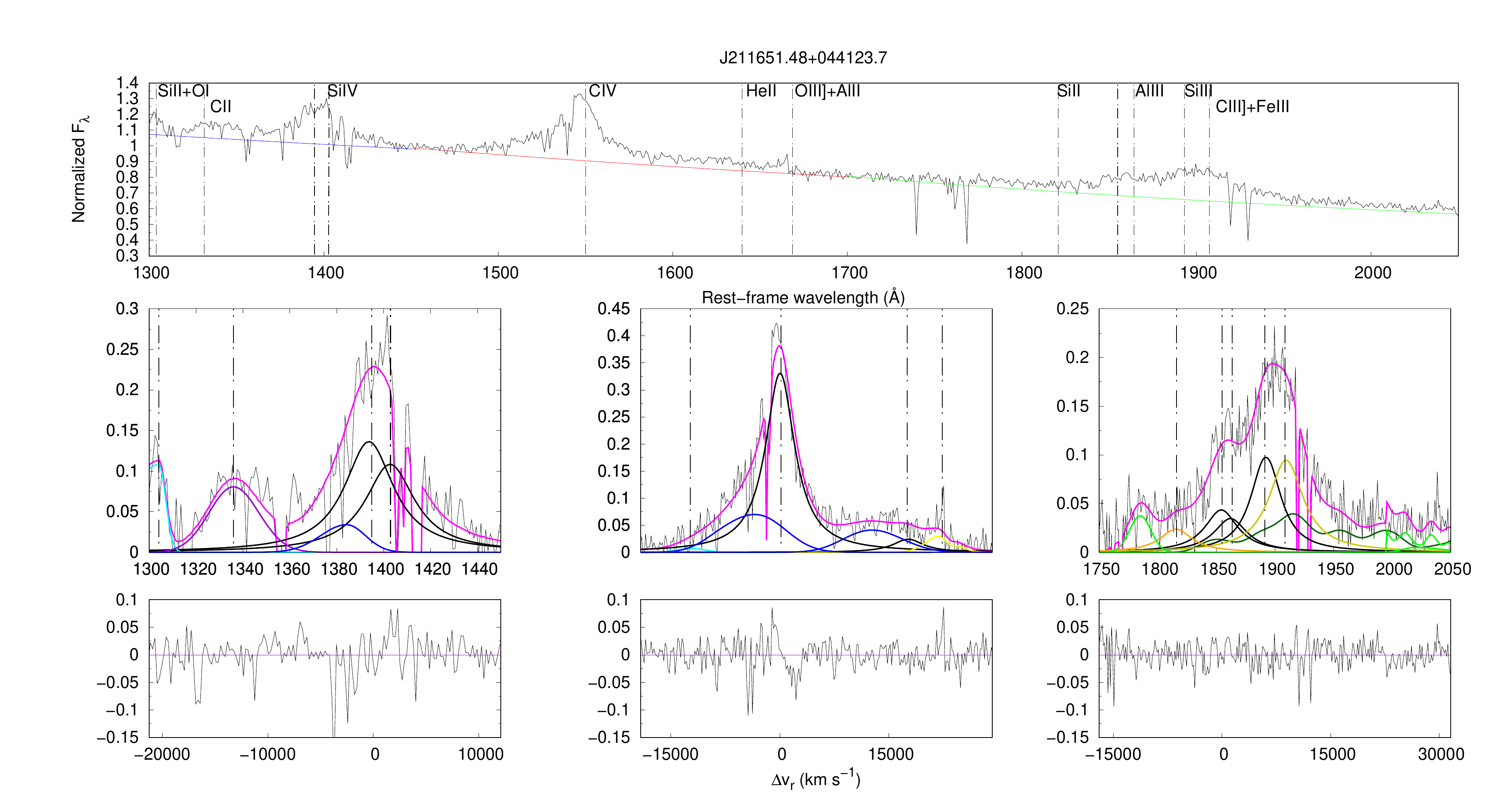}
    \caption{Same of the previous panel, for SDSS J150959.16+074450.1, SDSS J151929.45+072328.7, SDSS J211651.48+044123.7.}
  \end{figure}
%
%
%
%
%
%


\pagebreak\clearpage
\section{Diagnostic intensity ratios in the plane ($U$, \nh) as a function of metallicity}
\label{sec:iso}

The results of the arrays of simulations as a function of \nh, $U$, and $Z$ are shown below, for \nc $=10^{23}$. The SED shape is the same for all simulations ({table$\_$agn}) which corresponds to the SED of \citet{mathewsferland87}. No turbulence was assumed. 
\begin{figure}[ht]
     \includegraphics[scale=.30]{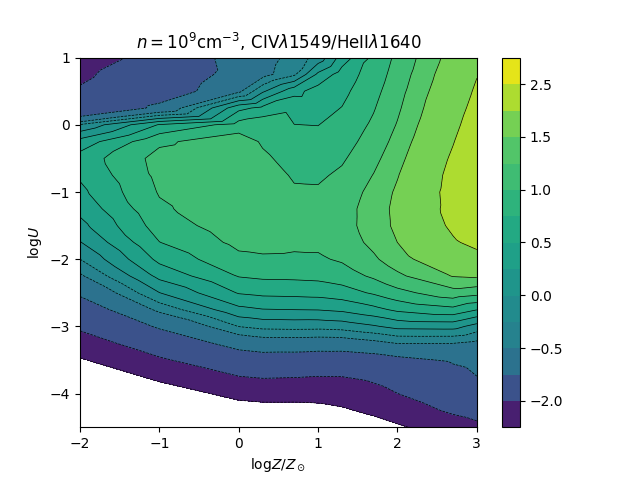}\hspace{-0.75cm}\hspace{-0.05cm}
     \includegraphics[scale=.30]{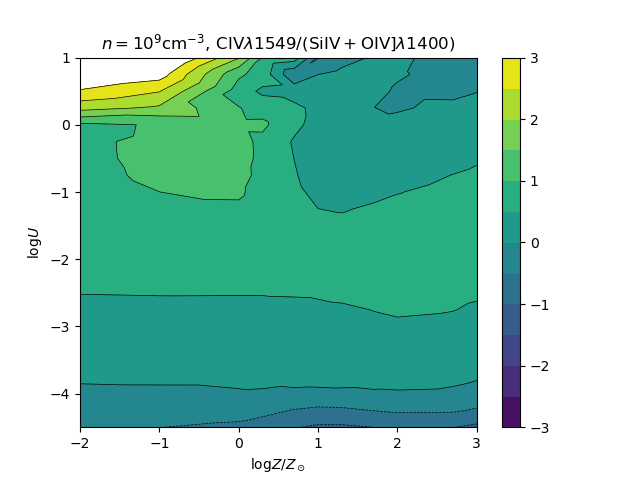}\hspace{-0.75cm}\hspace{-0.05cm}
     \includegraphics[scale=.30]{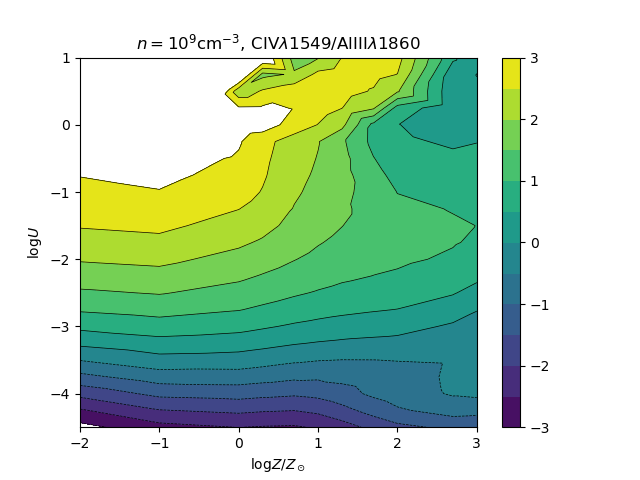}\hspace{-0.75cm}\hspace{-0.05cm}
     \includegraphics[scale=.30]{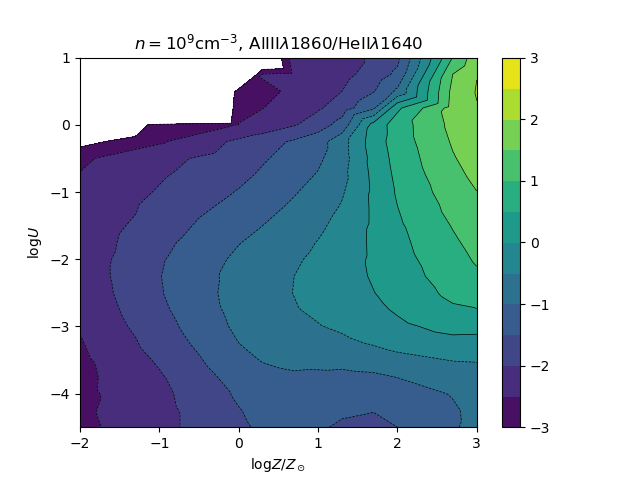}\hspace{-0.75cm}\hspace{-0.05cm}\\
       \includegraphics[scale=.30]{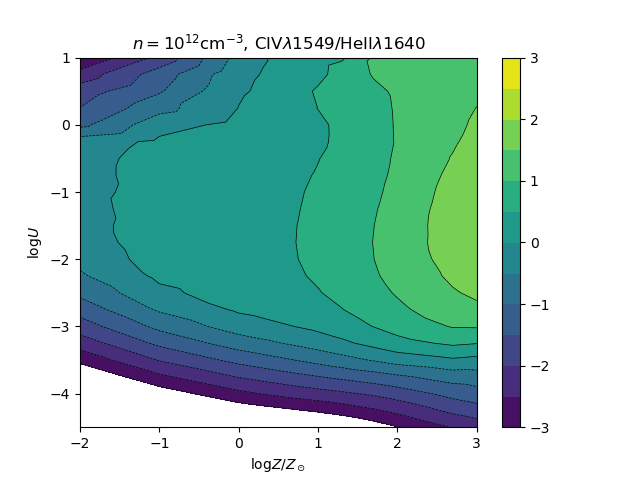}\hspace{-0.75cm}\hspace{-0.05cm}
     \includegraphics[scale=.30]{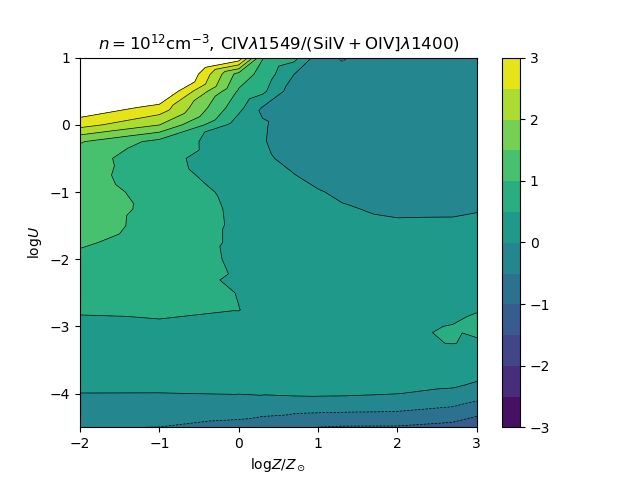}\hspace{-0.75cm}\hspace{-0.05cm}
     \includegraphics[scale=.30]{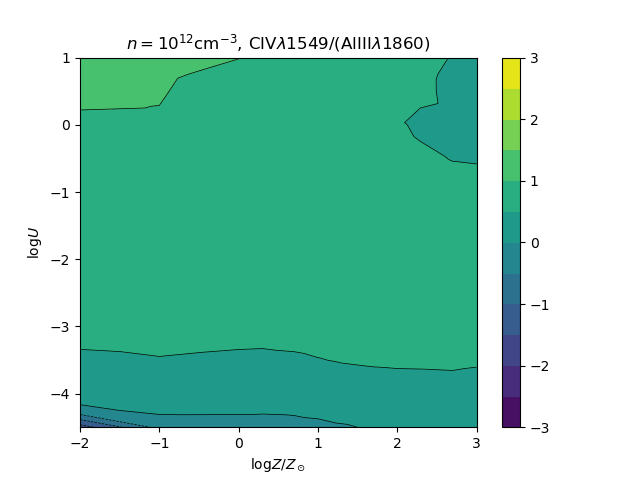}\hspace{-0.75cm}\hspace{-0.05cm}
     \includegraphics[scale=.30]{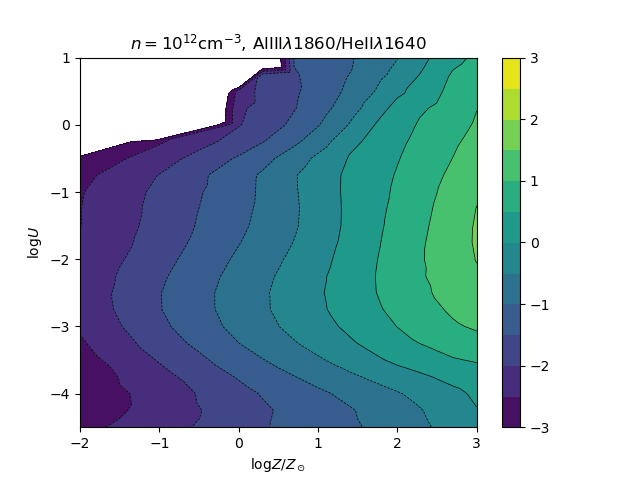}\hspace{-0.75cm}\hspace{-0.05cm}\\
        \includegraphics[scale=.30]{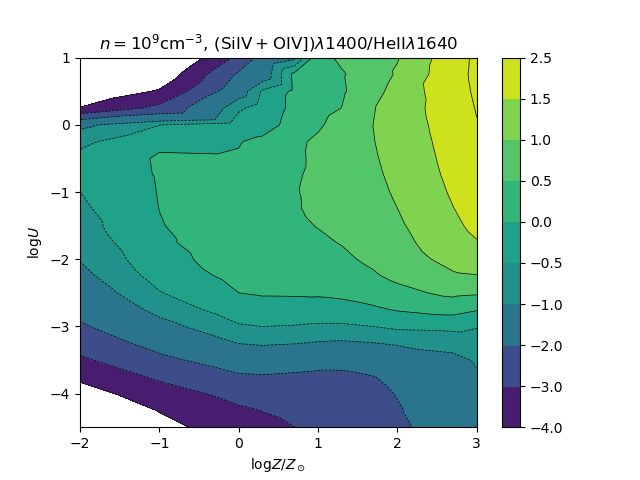}\hspace{-0.75cm}\hspace{-0.05cm}
     \includegraphics[scale=.30]{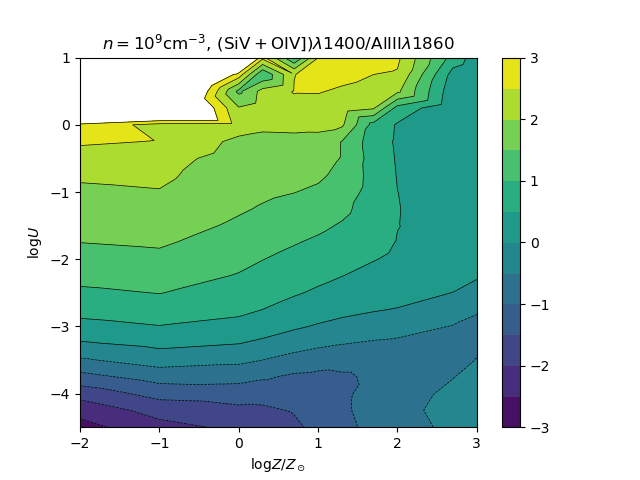}\hspace{-0.75cm}\hspace{-0.05cm}
     \includegraphics[scale=.30]{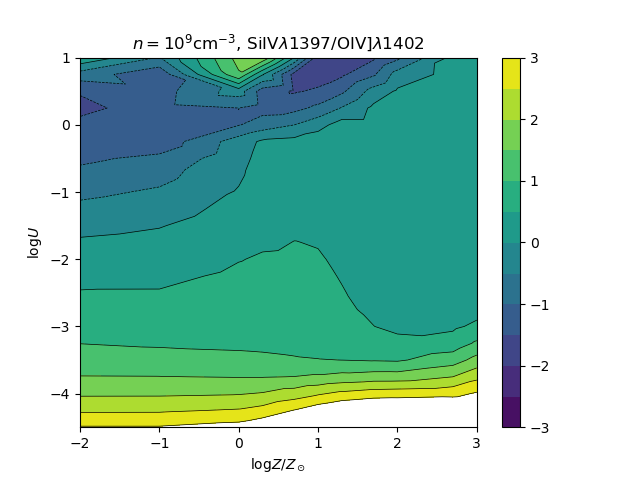}\hspace{-0.75cm}\\
        \includegraphics[scale=.30]{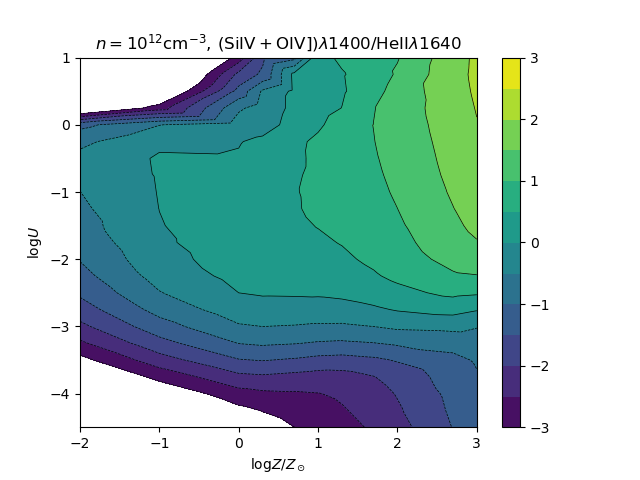}\hspace{-0.75cm}\hspace{-0.05cm}
     \includegraphics[scale=.30]{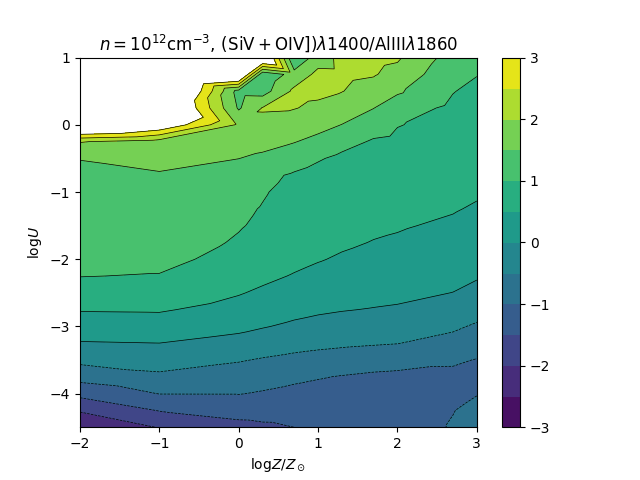}\hspace{-0.75cm}\hspace{-0.05cm}
     \includegraphics[scale=.30]{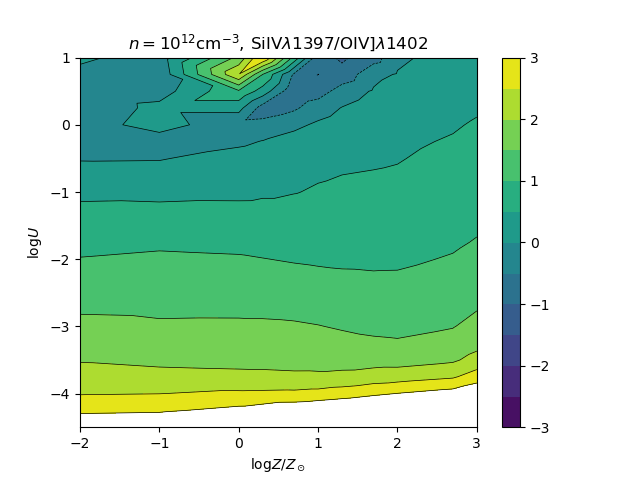}\hspace{-0.75cm}\\
  \caption{ Isophotal contour in the $\log U$ -- $\log Z$ plane of seven diagnostic line intensity ratios,  computed assuming column density $N_\mathrm{c} = 10^{23}$ cm$^{-2}$. 
  Top rows, from left to right: logarithm of \civ/\heiiuv, \civ/(\siiv+\oiv), \civ/\aliii, \aliii/\heii. 
  Bottom rows, from left to right: logarithm (\siiv+\oiv)/\heiiuv, (\siiv+\oiv)/\aliii, \siiv/\oiv. The contour plots are shown for \nh = $10^9$ cm$^{-3}$ (top) and \nh = $10^{12}$ cm$^{-3}$\ (bottom) for each diagnostic ratio.  }
     \label{fig:isoph1}
\end{figure}

\begin{sidewaysfigure}[ht]
     \includegraphics[scale=0.3]{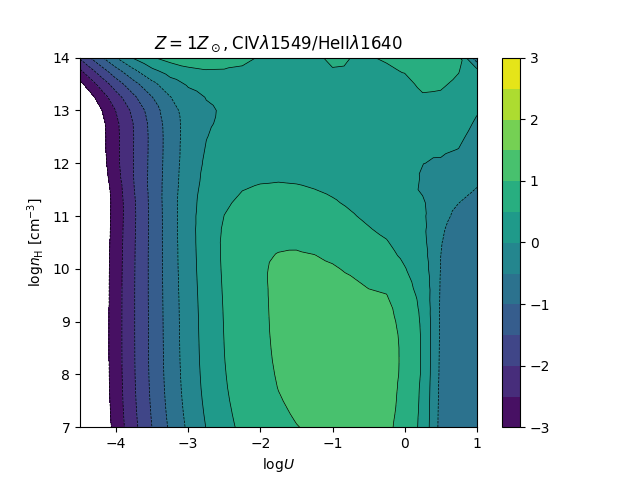}\hspace{-0.75cm}
     \includegraphics[scale=0.3]{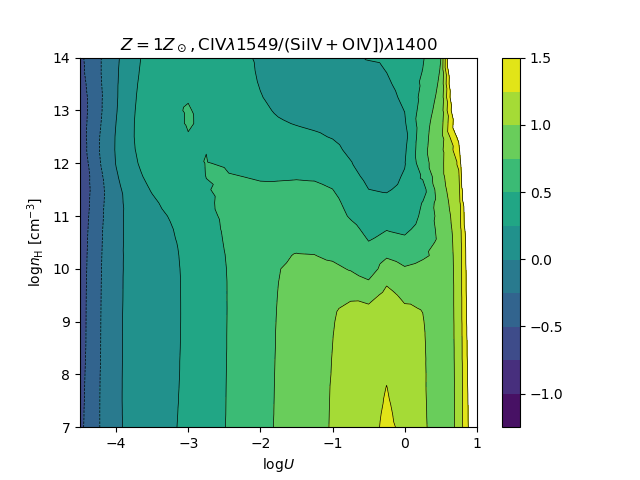}\hspace{-0.75cm}
     \includegraphics[scale=0.3]{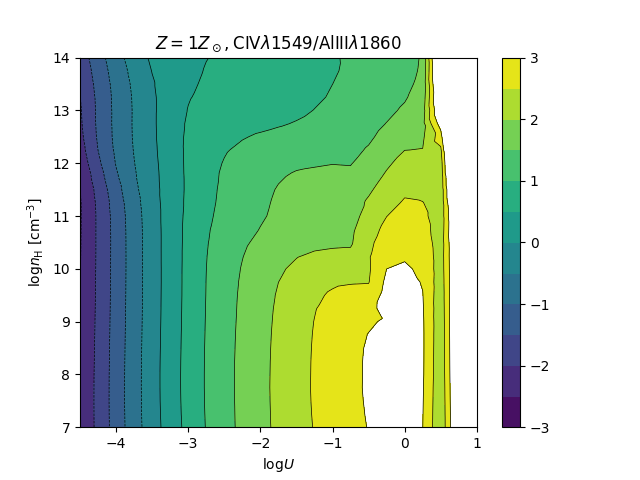}\hspace{-0.75cm}
     \includegraphics[scale=0.3]{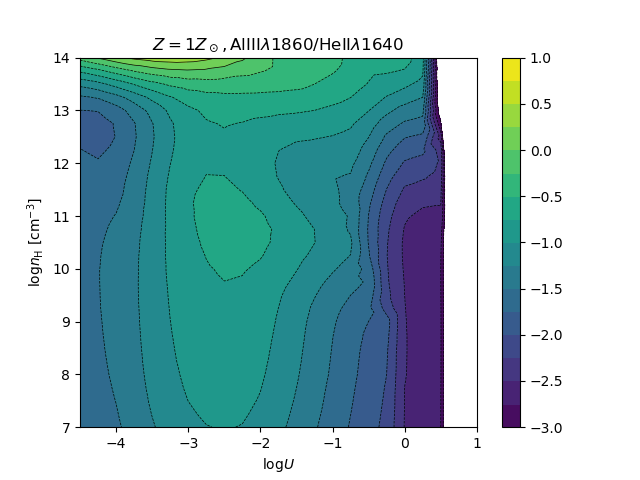}\hspace{-0.75cm}
     \includegraphics[scale=0.3]{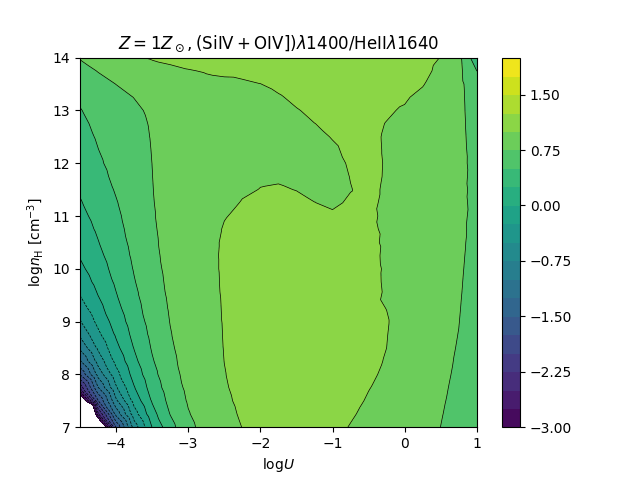}\hspace{-0.75cm}\\
      \includegraphics[scale=0.3]{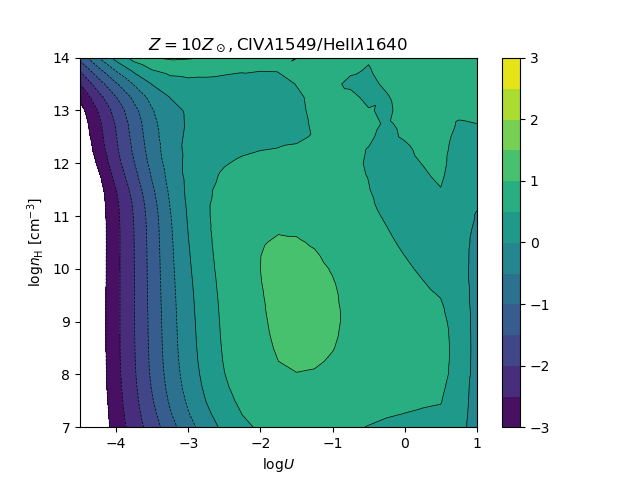}\hspace{-0.75cm}
     \includegraphics[scale=0.3]{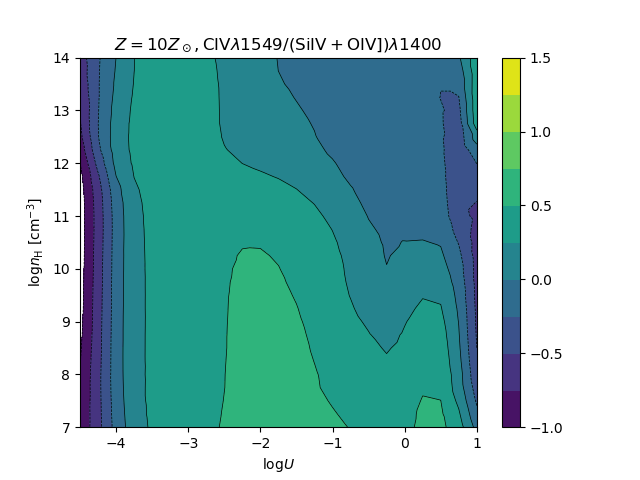}\hspace{-0.75cm}
     \includegraphics[scale=0.3]{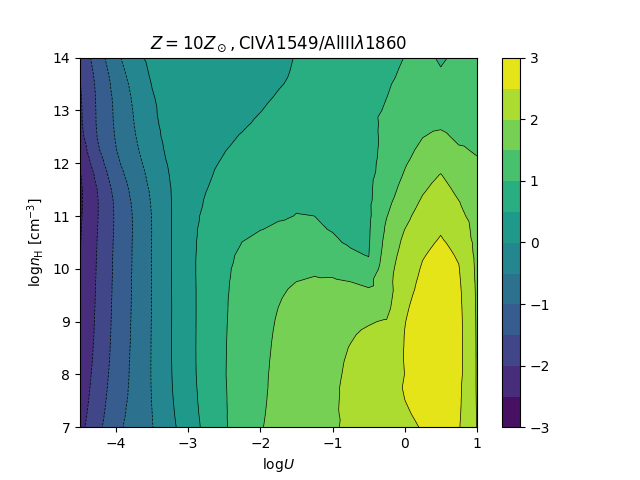}\hspace{-0.75cm}
     \includegraphics[scale=0.3]{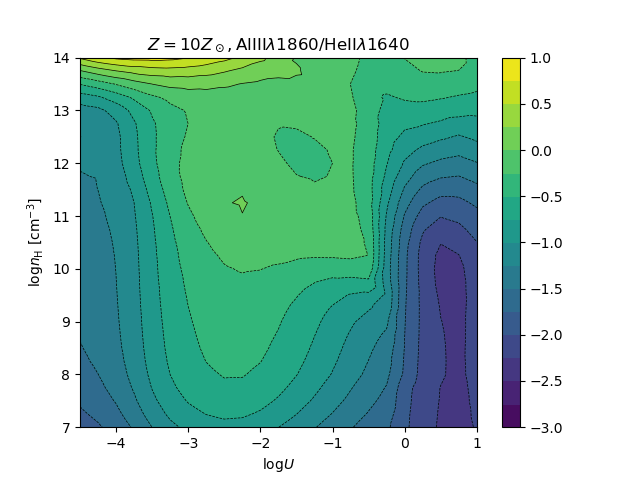}\hspace{-0.75cm}
     \includegraphics[scale=0.3]{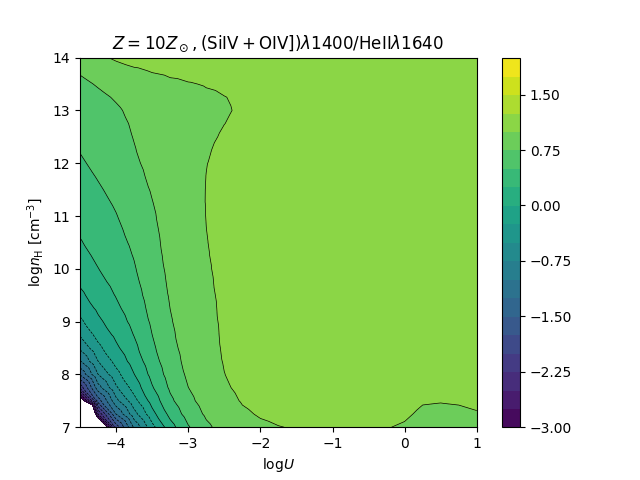}\hspace{-0.75cm}\\
    \includegraphics[scale=0.3]{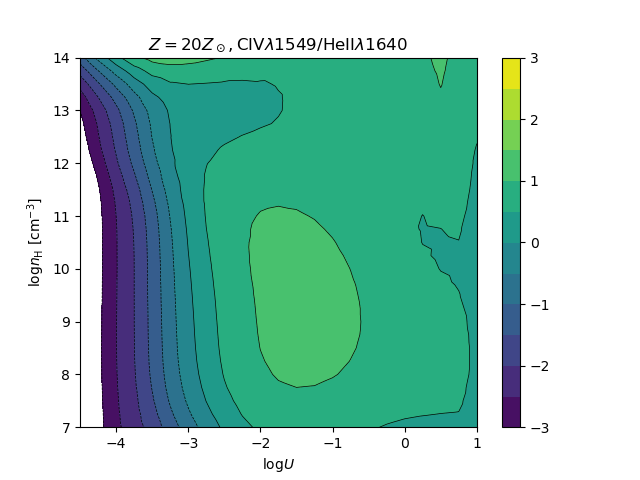}\hspace{-0.75cm}
     \includegraphics[scale=0.3]{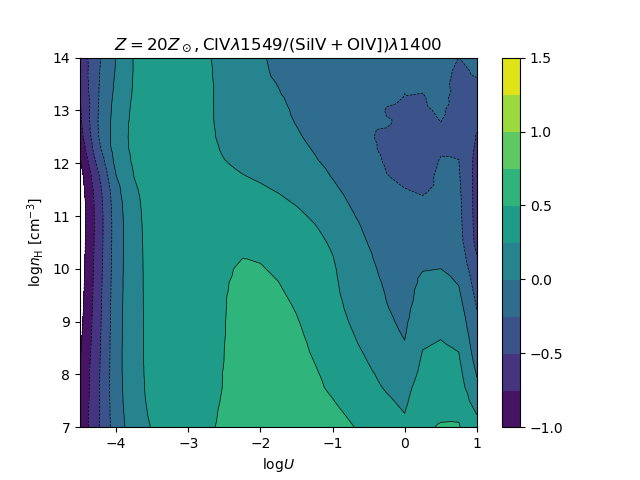}\hspace{-0.75cm}
     \includegraphics[scale=0.3]{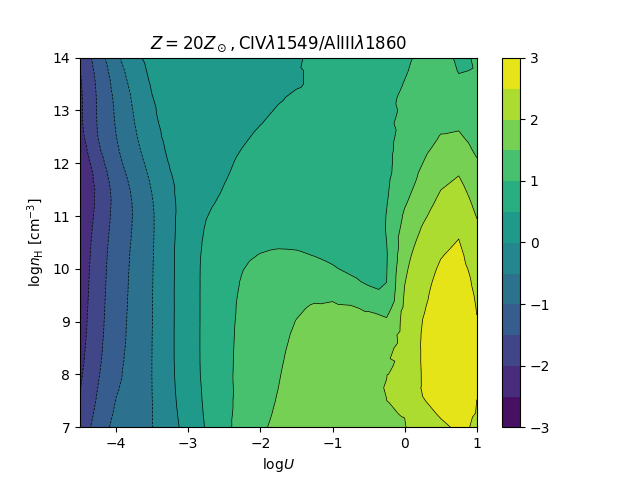}\hspace{-0.75cm}
     \includegraphics[scale=0.3]{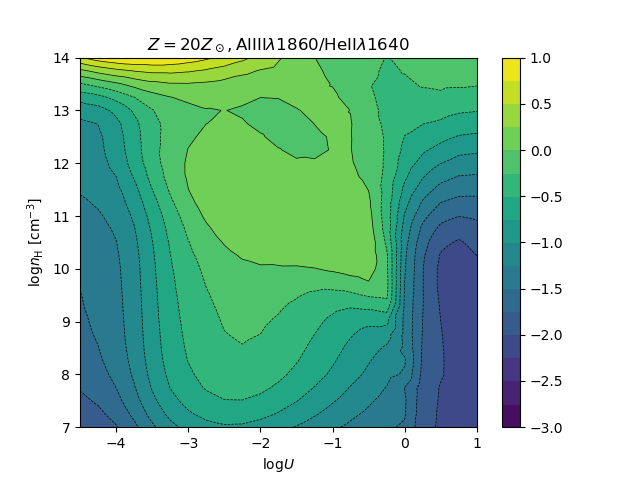}\hspace{-0.75cm}
     \includegraphics[scale=0.3]{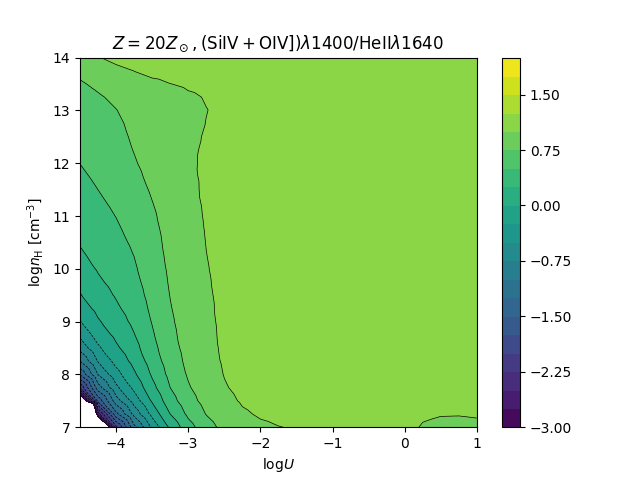}\hspace{-0.75cm}\\
    \includegraphics[scale=0.3]{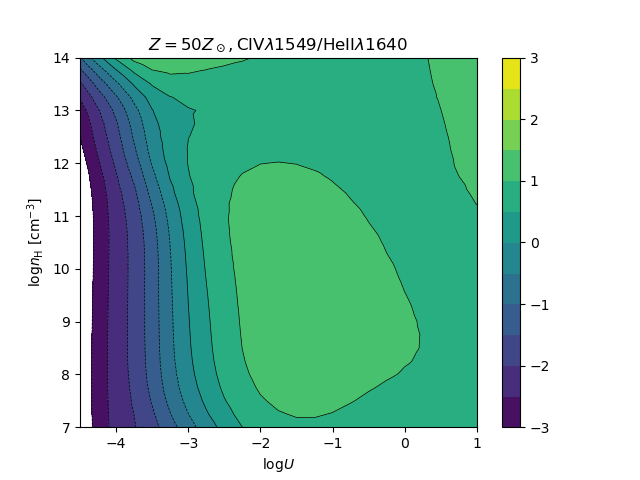}\hspace{-0.75cm}
     \includegraphics[scale=0.3]{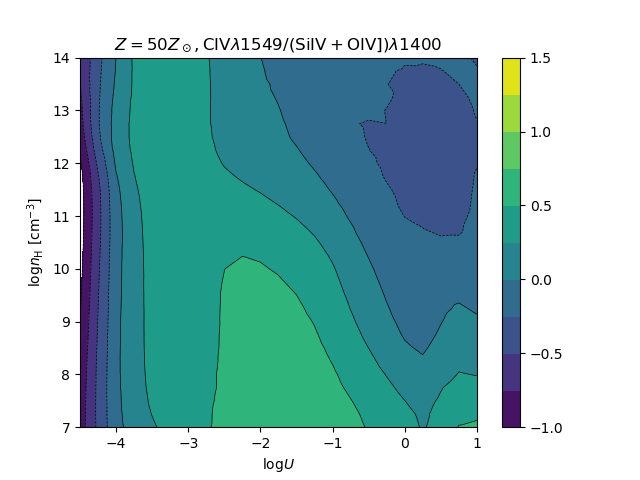}\hspace{-0.75cm}
     \includegraphics[scale=0.3]{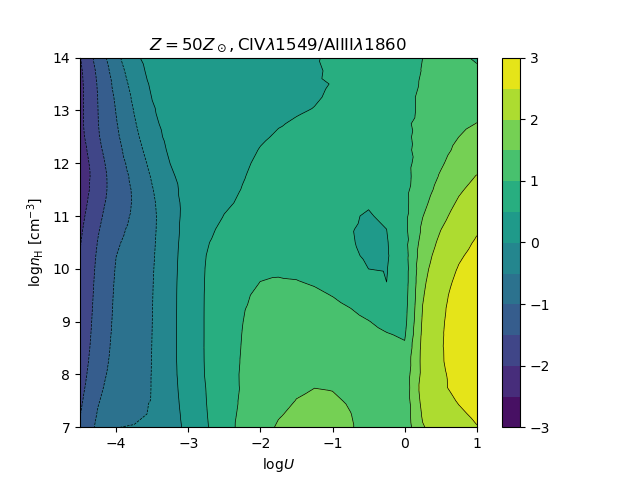}\hspace{-0.75cm}
     \includegraphics[scale=0.3]{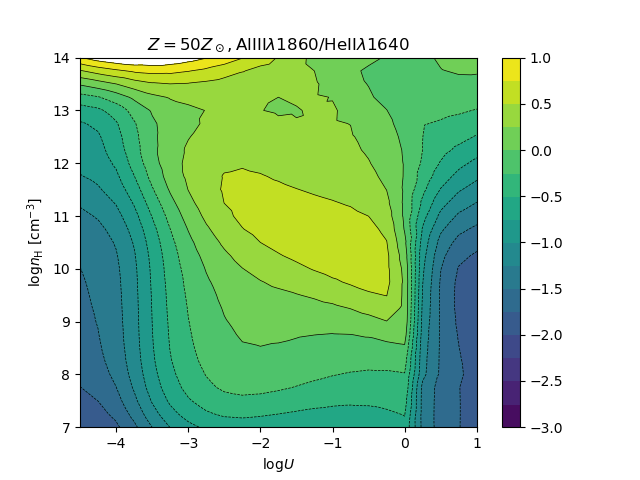}\hspace{-0.75cm}
     \includegraphics[scale=0.3]{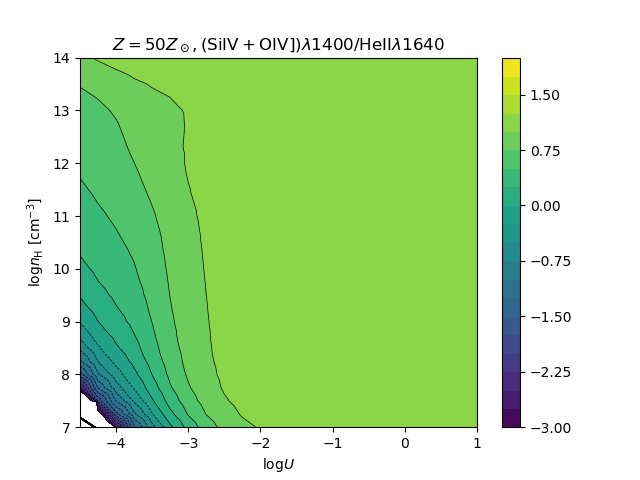}\hspace{-0.75cm}\\
     \includegraphics[scale=0.3]{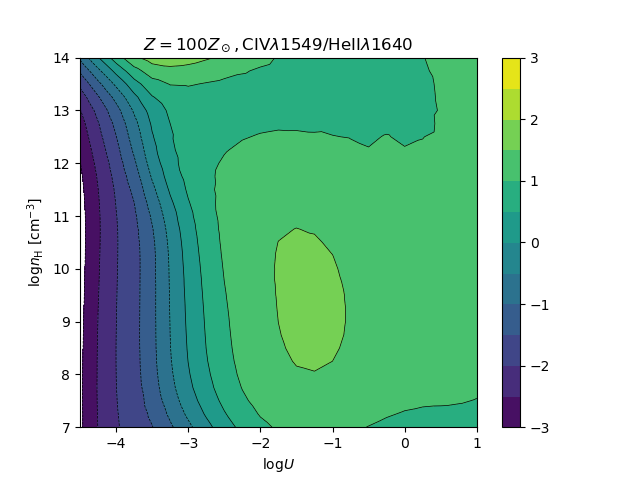}\hspace{-0.75cm}
     \includegraphics[scale=0.3]{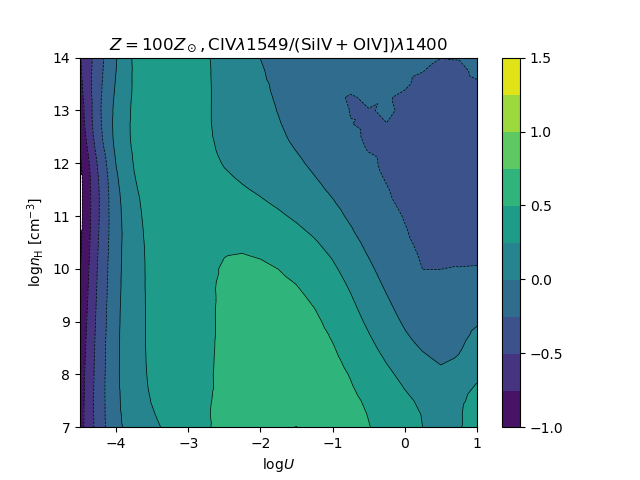}\hspace{-0.75cm}
     \includegraphics[scale=0.3]{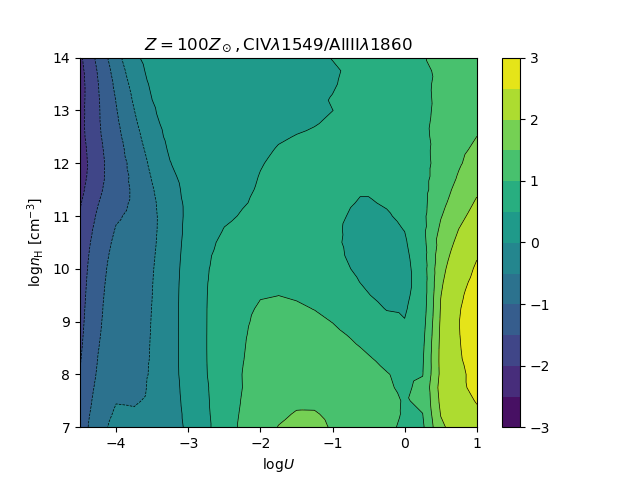}\hspace{-0.75cm}
     \includegraphics[scale=0.3]{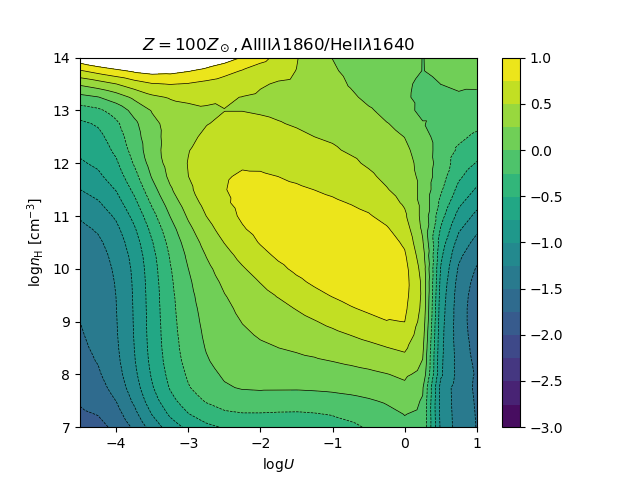}\hspace{-0.75cm}
     \includegraphics[scale=0.3]{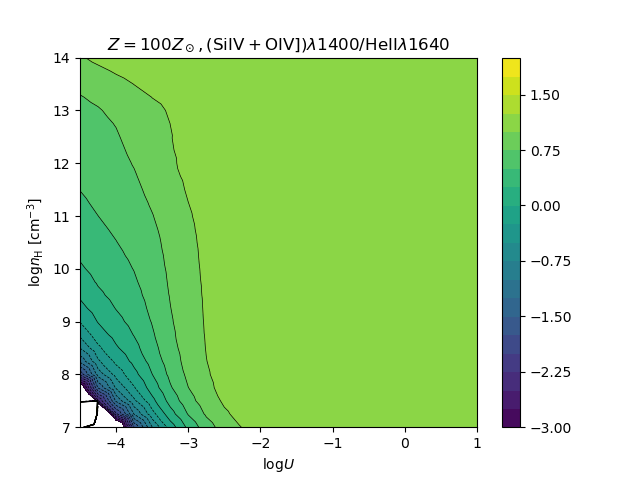}\hspace{-0.75cm}\\
     
     \caption{ Isophotal contour in the $\log U$ -- $\log$ \nh\ for line intensity ratios (from left to right: \civ/\heiiuv, \civ/(\siiv+\oiv), \civ/\aliii, \aliii/\heii, (\siiv+\oiv)/\heii)  as a function of metallicity (from top-to-bottom: $Z$ = 1, 10, 20, 50, 100 $Z_\odot$), for column density $N_\mathrm{c} = 10^{23}$ cm$^{-2}$. }
     \label{fig:isoph2}
\end{sidewaysfigure}

\end{appendix}
\end{document}